\documentclass[a4paper,11pt]{article}
\pdfoutput=1

\usepackage{jheppub}
\usepackage{amsmath,amssymb,mathtools,xspace}
\usepackage{booktabs,multirow,graphicx,tabularx,slashed,multirow}
\usepackage{color,xcolor}

\graphicspath{{figs/}}

\title{O new physics, where art thou?\\ A global search in the top sector} 

\abstract{We provide a comprehensive global analysis of Run II top
  measurements at the LHC in terms of dimension-6 operators. A
  distinctive feature of the top sector as compared to the
  Higgs-electroweak sector is the large number of four-quark
  operators. We discuss in detail how they can be tested and how
  quadratic terms lead to a stable limit on each individual Wilson
  coefficient. Predictions for all observables are computed at NLO in
  QCD. Our SFitter analysis framework features a detailed error
  treatment and shows that theoretical uncertainties are a limiting
  factor.}

\author[1]{Ilaria Brivio,}
\author[1]{Sebastian Bruggisser,}
\author[2,3]{Fabio Maltoni,}
\author[1,4]{Rhea Moutafis,}
\author[1]{\\ Tilman Plehn,}
\author[5]{Eleni Vryonidou,}
\author[1]{Susanne Westhoff,}
\author[6]{and Cen Zhang}

\affiliation[1]{Institut f\"ur Theoretische Physik, Universit\"at Heidelberg, Germany}
\affiliation[2]{CP3, Universit\'e Catholique de Louvain, Louvain-la-Neuve, Belgium}
\affiliation[3]{Dipartimento di Fisica e Astronomia, Universit\`a di Bologna,
        INFN, Sezione di Bologna, Italy}
\affiliation[4]{LAL, IN2P3, CNRS, Orsay, France}
\affiliation[5]{Theoretical Physics Department, CERN, Geneva, Switzerland}
\affiliation[6]{Institute for High Energy Physics and School of Physical Sciences,
        University of Chinese Academy of Sciences, Beijing, China}

\emailAdd{brivio@thphys.uni-heidelberg.de}
\emailAdd{bruggisser@thphys.uni-heidelberg.de}
\emailAdd{fabio.maltoni@uclouvain.be}
\emailAdd{moutafis@thphys.uni-heidelberg.de}
\emailAdd{plehn@uni-heidelberg.de}
\emailAdd{eleni.vryonidou@cern.ch}
\emailAdd{westhoff@thphys.uni-heidelberg.de}
\emailAdd{cenzhang@ihep.ac.cn}

\def\tablename{Table}
\def\figurename{Figure}

\newcommand\one{\leavevmode\hbox{\small1\normalsize\kern-.33em1}}

\newcommand{\lag}{\mathcal{L}}

\newcommand{\unit}[1]{\ensuremath{\, \text{#1}}}

\newcommand{\gev}{\unit{GeV}}
\newcommand{\tev}{\unit{TeV}}

\newcommand{\ifb}{\ensuremath{\text{fb}^{-1}} }

\def\slashchar#1{\setbox0=\hbox{$#1$}           
   \dimen0=\wd0                                 
   \setbox1=\hbox{/} \dimen1=\wd1               
   \ifdim\dimen0>\dimen1                        
      \rlap{\hbox to \dimen0{\hfil/\hfil}}      
      #1                                        
   \else                                        
      \rlap{\hbox to \dimen1{\hfil$#1$\hfil}}   
      /                                         
   \fi}

\newcommand{\ie}{\textsl{i.e.}\;}

\begin{document}
\maketitle
\flushbottom

\clearpage
\section{Introduction}
\label{sec:intro}

Contemporary interpretation frameworks for LHC measurements are driven
by the structural completeness of the Standard Model (SM) with a light
observed Higgs boson and missing hints to the nature of physics beyond
the SM. From a quantum field theory perspective the natural approach
is therefore to consider the SM as an effective field theory (SMEFT).
Here the effects of potential new particles can be systematically included in terms of higher-dimensional operators, suppressed by a sufficiently large matching scale. The framework relies on the idea that new physics affecting LHC
measurements is too heavy to be produced and observed directly. This
is a direct application of the general condition that an effective
Lagrangian is applicable if the additional contributing degrees of
freedom are kinematically decoupled.

Part of the SMEFT
framework~\cite{Buchmuller:1985jz,Leung:1984ni,Degrande:2012wf,Brivio:2017vri,Maltoni:2019aot}
was developed as a gauge-invariant description of anomalous gauge
boson interactions at
LEP~\cite{Hagiwara:1986vm,GonzalezGarcia:1999fq}. Its biggest success
has been the application to Higgs and electroweak boson measurements
at the
LHC~\cite{Butter:2016cvz,deBlas:2016ojx,Ellis:2018gqa,Almeida:2018cld,Biekotter:2018rhp,Kraml:2019sis}.
Most recently, the same
approach~\cite{Degrande:2010kt,Zhang:2010dr,Greiner:2011tt} has been
used to systematically analyze top quark measurements at the
LHC~\cite{Hioki:2013hva,Buckley:2015nca,Buckley:2015lku,Englert:2016aei,Hartland:2019bjb}
and at future colliders~\cite{Durieux:2018tev,Durieux:2019rbz} and their link to bottom sector~\cite{Bissmann:2019gfc}. These efforts pave the
way towards a global SMEFT analysis at the LHC.

Searches for physics beyond the Standard Model in the top sector
reflect three unique aspects of top physics: first, top physics at the
LHC has entered a phase of precision predictions and measurements, a development which many
of us would not have thought to be feasible before the start of the LHC;
second, we might start to doubt available solutions of the hierarchy
problem, but from a field theory perspective it has not lost its
appeal and it singles out the top sector; third, many new physics
scenarios, either weakly or strongly interacting at scales around a few TeV, predict deviations in the top couplings or new top interactions, such as new scalars coupled dominantly to the top
quark~\cite{Morrissey:2009tf}.

In this paper we present a comprehensive analysis of the top sector in
the framework of SMEFT, based on the data collected mostly during the
LHC Run~II. We consider measurements in top pair production, including
associated $t\bar{t}W$ and $t\bar{t}Z$ production, as well as in
single top final states. Since effective interactions typically have a
sizeable impact on kinematic distributions we add a set of kinematic
measurements with a focus on boosted top pair
production~\cite{Aguilar-Saavedra:2014iga,Farina:2018lqo}. Finally, we include charge
asymmetry~\cite{Bauer:2010iq,Degrande:2010kt,Delaunay:2011gv,Rosello:2015sck} and top decay
measurements~\cite{AguilarSaavedra:2010zi,Zhang:2014rja} to shed light on specific sectors of
the effective Lagrangian. To combine all of these measurements in a
coherent picture of the top sector a global SMEFT analysis is without
alternative. Our enlarged set of observables builds on earlier
analyses by the \textsc{TopFitter}~\cite{Buckley:2015nca,Buckley:2015lku} collaboration.

For our simulations we rely on next-to-leading order (NLO) QCD
predictions obtained through \textsc{FeynRules}~\cite{Alwall:2014hca},
\textsc{Nloct}~\cite{Degrande:2014vpa} and
\textsc{Madgraph5\_aMC@NLO}~ \cite{Alwall:2014hca}. NLO predictions
allow us to better control the accuracy and theoretical precision of
our predictions, \emph{i.e.}, the theoretical uncertainties in our
fit. This is especially important in phase space regions where SMEFT
contributions can be large, \emph{e.g.}, in tails of kinematic
distributions. For the global analysis we use the established
\textsc{SFitter} framework~\cite{Lafaye:2007vs,Lafaye:2009vr}.  Our
technical focus is on the consistent treatment of different types of
uncertainties, including correlations of systematic and theoretical
uncertainties. The specific error treatment, a detailed analysis of
several physics aspects not considered before, as well as a slightly
different data set complement other state-of-the-art analyses, such as
that of the \textsc{SMEFiT} collaboration~\cite{Hartland:2019bjb}. In
particular, the \textsc{SFitter} technology allows us to easily study
correlations between Wilson coefficients and to compare the impact of
theoretical and experimental uncertainties.

The paper is organized as follows. We begin with an overview of the
observables used in our fit in Sec.~\ref{sec:eft}, where we derive
analytic expressions for the operator contributions. After
recapitulating the main aspects of the fitting approach in
Sec.~\ref{sec:global} we describe some of the unique aspects of a
SMEFT analysis of top pair production in Sec.~\ref{sec:features}. They
are related to the fact that many features of the set of four-quark
operators are not immediately distinguishable in QCD processes. A
crucial aspect is that while flat directions exist in this sector, the quadratic contributions from dimension-6 operators turn them into
compact circular correlations. This in turn allows us to derive
well-defined limits from profile likelihoods. Finally, we perform a
global fit first of the single top sector in
Sec.~\ref{sec:global_single} and then of the entire top sector in
Sec.~\ref{sec:global_comb}. Details about our choice of operators
and our numerical results are discussed in the Appendix.

\section{Top-quark effective theory}
\label{sec:eft}

In the absence of new resonances, effects of new physics can be
described as effective interactions of SM particles at energies below
a new physics matching scale $\Lambda$. Our goal is to probe effective
interactions with top quarks in LHC
observables~\cite{AguilarSaavedra:2018nen}. The dominant effects are 
parametrized in terms of Wilson coefficients $C_k$ of dimension-6
operators $O_k$ in the effective Lagrangian
\begin{align}
\lag_\text{eff} = \sum_k \left(\frac{C_k}{\Lambda^2}\,^\ddagger O_k + \text{h.c.} \right) 
               + \sum_l \frac{C_l}{\Lambda^2}\,O_l\,,
\end{align}
where the sum runs over all operators that involve top-quarks.
Non-hermitian operators are denoted as $^\ddagger O$. We
neglect operators of mass dimension seven and higher in the EFT
expansion. We focus on CP-conserving extensions of the SM, assuming
that all Wilson coefficients are real and therefore neglecting
CP-violating interactions. Since top-quark observables at the LHC are
largely blind to the flavor of light quarks with the same quantum
numbers, we impose an $U(2)_q\times U(2)_u\times U(2)_d$ flavor
symmetry among quarks of the first and second
generation~\cite{Durieux:2014xla,Degrande:2014tta,AguilarSaavedra:2018nen}.
We use
\begin{align}
q_i & = (u^i_L,d^i_L) \qquad \,u_i = u^i_R,\,d_i=d^i_R \qquad i=1,2 \notag  \\
Q & = (t_L,b_L) \qquad \quad t = t_R,\,b=b_R
\end{align}
to denote left- and right-handed quarks of the first two generations
and the third generation, respectively. Within this framework, we consider 22 independent operators:
\begin{itemize}
\item 8 four-quark operators with $LL$ and $RR$ chiral structure
\begin{align}
  O_{Qq}^{1,8} & = (\bar{Q}\gamma_\mu T^A Q)(\bar{q}_i\gamma^\mu T^A q_i) 
& O_{Qq}^{1,1} & = (\bar{Q}\gamma_\mu Q)(\bar{q}_i\gamma^\mu q_i) \notag \\
  O_{Qq}^{3,8} & = (\bar{Q}\gamma_\mu T^A\tau^I Q)(\bar{q}_i\gamma^\mu T^A \tau^I q_i)
& O_{Qq}^{3,1} & = (\bar{Q}\gamma_\mu\tau^I Q)(\bar{q}_i\gamma^\mu\tau^I q_i) \notag \\
  O_{tu}^8 & = (\bar{t}\gamma_\mu T^A t)(\bar{u}_i\gamma^\mu T^A u_i) 
& O_{tu}^1 & = (\bar{t}\gamma_\mu t)(\bar{u}_i\gamma^\mu u_i) \notag \\
  O_{td}^8 & = (\bar{t}\gamma^\mu T^A t)(\bar{d}_i\gamma_\mu T^A d_i) 
& O_{td}^1 & = (\bar{t}\gamma^\mu t)(\bar{d}_i\gamma_\mu d_i) \; ;
\label{eq:ops_llrr}
\end{align}
\item 6 four-quark operators with $LR$ and $RL$ chiral structure
\begin{align}
  O_{Qu}^8 & = (\bar{Q}\gamma^\mu T^A Q)(\bar{u}_i\gamma_\mu T^A u_i)
& O_{Qu}^1 & = (\bar{Q}\gamma^\mu Q)(\bar{u}_i\gamma_\mu u_i) \notag \\
  O_{Qd}^8 & = (\bar{Q}\gamma^\mu T^A Q)(\bar{d}_i\gamma_\mu T^A d_i)
& O_{Qd}^1 & = (\bar{Q}\gamma^\mu Q)(\bar{d}_i\gamma_\mu d_i) \notag \\
  O_{tq}^8 & = (\bar{q}_i\gamma^\mu T^A q_i)(\bar{t}\gamma_\mu T^A t)
& O_{tq}^1 & = (\bar{q}_i\gamma^\mu q_i)(\bar{t}\gamma_\mu t) \; ;
\label{eq:ops_lr}
\end{align}
\item 8 operators with two heavy quarks and bosons~\cite{Franzosi:2015osa}
\begin{align}
O_{\phi Q}^{1} & = (\phi^\dagger\,i \stackrel{\longleftrightarrow}{D_\mu} \phi)(\bar{Q}\gamma^{\mu}Q) & ^\ddagger O_{tB} & = (\bar{Q}\sigma^{\mu\nu} t)\,\widetilde{\phi}\,B_{\mu\nu} \notag \\
O_{\phi Q}^{3} & = (\phi^\dagger\,i \stackrel{\longleftrightarrow}{D_\mu^I} \phi)(\bar{Q}\gamma^{\mu}\tau^I Q) & ^\ddagger O_{tW} & = (\bar{Q}\sigma^{\mu\nu} t)\,\tau^I\widetilde{\phi}\,W_{\mu\nu}^I \notag \\
O_{\phi t} & = (\phi^\dagger\,i \stackrel{\longleftrightarrow}{D_\mu} \phi)(\bar{t}\gamma^{\mu}t) & ^\ddagger O_{bW} & = (\bar{Q}\sigma^{\mu\nu} b)\,\tau^I\phi \,W_{\mu\nu}^I \notag \\
^\ddagger O_{\phi tb} & = (\widetilde{\phi}^\dagger iD_\mu \phi)(\bar{t}\gamma^{\mu}b) & ^\ddagger O_{tG} & = (\bar{Q}\sigma^{\mu\nu} T^A t)\,\widetilde{\phi}\,G_{\mu\nu}^A \,. 
\label{eq:tboson}
\end{align}
\end{itemize}
The different color structures of the operators will eventually lead
to different color factors in the LHC observables and different limits
on the Wilson coefficients, as we will see later.  In
Appendix~\ref{app:operators}, we list the relations between these
operators and the operators in the Warsaw
basis~\cite{Grzadkowski:2010es}.  Gauge invariance imposes relations
between effective top couplings to gauge bosons. We define
\begin{align}
\label{eq:operator-relations}
C_{\phi Q}^- & \equiv  C_{\phi Q}^1 - C_{\phi Q}^3 & C_{tZ} & \equiv -s_w C_{tB} + c_w C_{tW} \\
C_{\phi Q}^+ & \equiv  C_{\phi Q}^1 + C_{\phi Q}^3 = C_{\phi Q}^- + 2 C_{\phi Q}^3 & \qquad
C_{tA} & \equiv c_w C_{tB} + s_w C_{tW} = \frac{1}{s_w}\big(C_{tW} - c_w C_{tZ}\big)\,, \notag
\end{align}
We choose $C_{\phi Q}^3,\,
C_{\phi Q}^-$ and $C_{tW},\, C_{tZ}$ as degrees of freedom in our
analysis. With this choice, $C_{\phi Q}^-$ and $C_{tZ}$ modify the $t\bar{t} Z$ coupling, $C_{tW}$ modifies the $tbW$ vertex, while $C_{\phi Q}^3$ affects both.

The Wilson coefficients of the
operators in Eqs.~\eqref{eq:ops_llrr}, \eqref{eq:ops_lr}, and
\eqref{eq:tboson} define the 22 parameters in our global
analysis. Further operators either do not leave visible effects in the
observables we have selected (like operators with four heavy quarks)
or are strongly constrained by more sensitive observables (like the
Yukawa operator at dimension six, which is constrained by Higgs
measurements). We therefore do not include them in our analysis, but
mention them whenever they are relevant.

Experimentally, we focus on observables in top pair and electroweak
single top production at the LHC. These processes are precisely
predicted and measured, both at the level of total rates and kinematic
distributions. We also include top pair production in association with
a $W$- or $Z$-boson, which is more sensitive to certain operators than
top pair or single top production and help distinguishing between
operators. Table~\ref{tab:wilson-contributions} shows our set of
Wilson coefficients and their contributions to the various
processes.

\begin{table}[h]
  \centering
  \begin{tabular}{l | ccccccc}
  \hline
\phantom{\Big [}   parameter \  & $t\bar{t}$ & single $t$ & $tW$ & $tZ$ & $t$ decay & $t\bar{t}Z$ & $t\bar{t}W$ \\
  \hline
\phantom{\Big [}   $C_{Qq}^{1,8}$ & $\ \Lambda^{-2}$ & -- & -- & -- & -- & $\Lambda^{-2}$ & $\Lambda^{-2}$ \\
\phantom{\Big [}      $C_{Qq}^{3,8}$ & $\ \Lambda^{-2}$ & $\Lambda^{-4}\ [\Lambda^{-2}]$ & -- & $\Lambda^{-4}\ [\Lambda^{-2}]$ & $\Lambda^{-4}\ [\Lambda^{-2}]$ & $\Lambda^{-2}$ & $\Lambda^{-2}$ \\
\phantom{\Big [}     $C_{tu}^8,\, C_{td}^8$ & $\ \Lambda^{-2}$ & -- & -- & -- & -- & $\Lambda^{-2}$ & -- \\
\phantom{\Big [}     $C_{Qq}^{1,1}$ & $\ \Lambda^{-4}\ [\Lambda^{-2}]$ & -- & -- & -- & -- & $\Lambda^{-4}\ [\Lambda^{-2}]$ & $\Lambda^{-4}\ [\Lambda^{-2}]$ \\
\phantom{\Big [}     $C_{Qq}^{3,1}$ & $\ \Lambda^{-4}\ [\Lambda^{-2}]$ & $\Lambda^{-2}$ & -- & $\Lambda^{-2}$ & $\Lambda^{-2}$ & $\Lambda^{-4}\ [\Lambda^{-2}]$ & $\Lambda^{-4}\ [\Lambda^{-2}]$ \\
\phantom{\Big [}      $C_{tu}^1,\, C_{td}^1$ & $\ \Lambda^{-4}\ [\Lambda^{-2}]$ & -- & -- & -- & -- & $\Lambda^{-4}\ [\Lambda^{-2}]$ & -- \\
  \hline
\phantom{\Big [}      $C_{Qu}^{8}, C_{Qd}^{8}$ & $\ \Lambda^{-2}$ & -- & -- & -- & -- & $\Lambda^{-2}$ & -- \\
\phantom{\Big [}      $C_{tq}^{8}$ & $\ \Lambda^{-2}$ & -- & -- & -- & -- & $\Lambda^{-2}$ & $\Lambda^{-2}$ \\
\phantom{\Big [}      $C_{Qu}^{1}, C_{Qd}^{1}$ & $\ \Lambda^{-4}\ [\Lambda^{-2}]$ & -- & -- & -- & -- & $\Lambda^{-4}\ [\Lambda^{-2}]$ & -- \\
\phantom{\Big [}      $C_{tq}^{1}$ & $\ \Lambda^{-4}\ [\Lambda^{-2}]$ & -- & -- & -- & -- & $\Lambda^{-4}\ [\Lambda^{-2}]$ & $\Lambda^{-4}\ [\Lambda^{-2}]$ \\
  \hline
\phantom{\Big [}       $C_{\phi Q}^-$ & -- & -- & -- & $\Lambda^{-2}$ & -- & $\Lambda^{-2}$ & -- \\
\phantom{\Big [}       $C_{\phi Q}^3$ & -- & $\Lambda^{-2}$ & $\Lambda^{-2}$  & $\Lambda^{-2}$ & $\Lambda^{-2}$  & -- & -- \\
\phantom{\Big [}     $C_{\phi t}$ & -- & -- & -- & $\Lambda^{-2}$  & -- & $\Lambda^{-2}$ & -- \\
\phantom{\Big [}       $C_{\phi tb}$ & -- & $\Lambda^{-4}$ & $\Lambda^{-4}$ & $\Lambda^{-4}$ &  $\Lambda^{-4}$  & -- & -- \\
\phantom{\Big [}     $C_{tZ}$ & -- & -- & -- & $\Lambda^{-2}$ & -- & $\Lambda^{-2}$ & -- \\
\phantom{\Big [}     $C_{tW}$ & -- & $\Lambda^{-2}$ & $\Lambda^{-2}$ & $\Lambda^{-2}$ & $\Lambda^{-2}$  & -- & -- \\
\phantom{\Big [}     $C_{bW}$ & -- & $\Lambda^{-4}$ & $\Lambda^{-4}$ & $\Lambda^{-4}$ & $\Lambda^{-4}$  & -- & -- \\
\phantom{\Big [}       $C_{tG}$ & $\Lambda^{-2}$ & $[\Lambda^{-2}]$ & $\Lambda^{-2}$ &  -- &  $[\Lambda^{-2}]$  & $\Lambda^{-2}$ & $\Lambda^{-2}$\\
  \hline
  \end{tabular}
  \caption{Wilson coefficients in our analysis and their contributions to top-quark observables via SM-interference ($\Lambda^{-2}$) and via dimension-6 squared terms only ($\Lambda^{-4}$). A square bracket indicates that the Wilson coefficient contributes via SM-interference at NLO QCD. All quark masses except $m_t$ are assumed to be zero. `Single $t$' stands for $s-$ and $t-$channel electroweak top production.}
\label{tab:wilson-contributions}
\end{table}

In what follows, we describe in detail how the 22 top operators affect
these processes. We pay special attention to contributions of the
dimension-6 squared terms, which will be crucial for our global
fit. Another important aspect in our discussion is the energy
dependence of operator contributions, which changes the top kinematics
in distributions. For top pair production, we present complete
analytic expressions for four-quark operator contributions at LO, including
both SM-interference and dimension-6 squared terms. We also derive the
structure of operator contributions at NLO QCD.

\subsection{Top pair production}
\label{sec:eft_tbar}

\begin{figure}[t]\centering
\includegraphics[width=.22\textwidth]{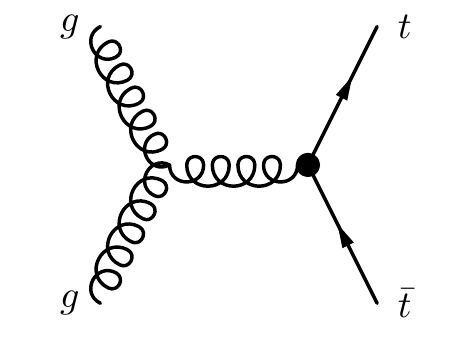}
\includegraphics[width=.22\textwidth]{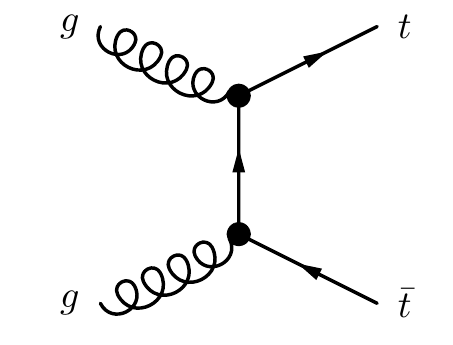}
\includegraphics[width=.22\textwidth]{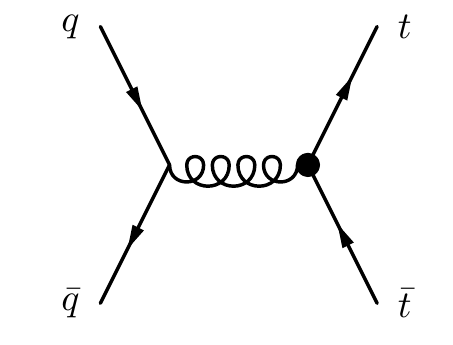}  
\includegraphics[width=.22\textwidth]{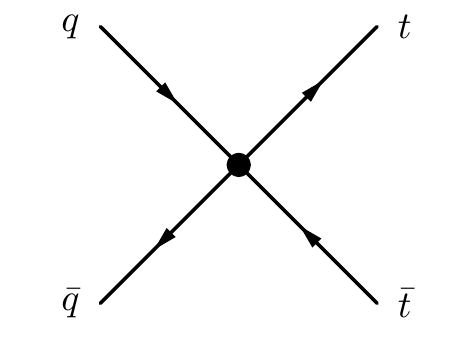}
\caption{Examples of Feynman diagrams contributing to top pair production in SMEFT at leading order. The dots indicate possible insertions of a dimension-6 operator.}\label{fig:ttbar_diagrams}
\end{figure}

Hadronic top pair production involves $gg \to t\bar{t}$ and $q\bar{q}
\to t\bar{t}$ partonic processes. In SMEFT, effective operators
contribute to both processes, as shown in
Fig.~\ref{fig:ttbar_diagrams}. 
 At the LHC, top pair production is
dominated by incoming gluons. At leading order in QCD two operators
contribute to this rate,
\begin{align}
^\ddagger O_{tG}  = (\bar{Q}\sigma^{\mu\nu} T^A t)\,\widetilde{\phi}\,G_{\mu\nu}^A
\qquad \text{and} \qquad
O_{G} = f_{abc} G_{\mu}^{a\nu} G_{\nu}^{b\rho} G_{\rho}^{c\mu}\,.
\end{align}
However, $O_{G}$ is strongly constrained by multi-jet
production~\cite{Krauss:2016ely,Hirschi:2018etq},
\begin{align}
\frac{\Lambda}{\sqrt{g_s |C_{G}|}} > 5.2\,\tev\,.
\end{align}
Since this sensitivity is beyond the reach of top pair production, we
neglect $O_G$ in our analysis. Unfortunately, multi-jet features which
lead to this reach in jet production do not help significantly in the
top sector~\cite{Englert:2018byk,Hirschi:2018etq}.

The contribution of $O_{tG}$ to the partonic differential cross
section is given by~\cite{Zhang:2010dr}\footnote{Notice that in Ref.~\cite{Zhang:2010dr} the top-gluon operator is defined as twice the $O_{tG}$ in Eq.~\eqref{eq:tboson}.}
 \begin{align}
 \frac{d\sigma(g g\to t\bar t)}{d c_t} = \frac{\alpha_s^{3/2}\sqrt{\pi}}{12\sqrt{2}} \frac{ \beta_{t\bar{t}}}{s}\frac{m_t v}{\Lambda^2} \,\frac{7+9\beta_{t\bar{t}}^2\,c_t^2}{1-\beta_{t\bar{t}}^2\,c_t^2}\,C_{tG} + \mathcal{O}\Big( \frac{sv^2}{\Lambda^4}C_{tG}^2\Big)\,,
 \end{align}
where $\beta_{t\bar{t}} = \sqrt{1-4\overline{m}_t^2}$, $\overline{m}_t
= m_t/\sqrt{s}$, and $c_t = \cos\theta_t$ with the scattering angle
$\theta_t$ of the top against one of the incoming gluons in the partonic
center-of-mass (CM) frame. Due to the large gluon luminosity, we
expect a high sensitivity to $O_{tG}$ in inclusive top pair
production. At high energies $\sqrt{s} \gg m_t$, the
$O_{tG}-\text{QCD}$ and $O_{tG}-O_{tG}$ interferences scale as $m_t
v/\Lambda^2$ and $v^2s/\Lambda^4$ relative to the QCD rate,
respectively. In the collinear limit, $c_t \approx 1$, $O_{tG}$
contributions feature the same logarithmic enhancement as QCD. The
kinematics of $O_{tG}$--QCD interference is thus similar to QCD, while
squared $O_{tG}$ contibutions grow with energy relative to the SM. We
will discuss the impact of $O_{tG}$ on kinematic distributions in
detail in Sec.~\ref{sec:ctg}.

Compared to the $gg\to t\bar{t}$ contribution, $q\bar{q}$ scattering
is suppressed by the parton luminosities. However, the quark-antiquark
luminosity is enhanced in the production of boosted tops, where the
partons carry a large fraction of the proton's energy. In this regime,
top pair production is most sensitive to the four-quark operators
introduced in Eqs.~\eqref{eq:ops_llrr} and \eqref{eq:ops_lr} and their
interference with $O_{tG}$.

Contributions of four-quark operators are conveniently classified by
their behavior under top charge conjugation. The vector current $V$ is
odd under charge conjugation, while the axial-vector current $A$ is
even. We define vector and axial-vector combinations of Wilson
coefficients as
\begin{align}
4\,C_{VV}^{u,8} & =  C_{Qq}^{1,8} + C_{Qq}^{3,8} + C_{tu}^8 + C_{tq}^8 + C_{Qu}^8 \notag \\
4\,C_{AA}^{u,8} & = C_{Qq}^{1,8} + C_{Qq}^{3,8} + C_{tu}^8 - C_{tq}^8 - C_{Qu}^8 \notag \\
4\,C_{AV}^{u,8} & = - \big(C_{Qq}^{1,8} + C_{Qq}^{3,8}\big) + C_{tu}^8 + C_{tq}^8 - C_{Qu}^8 \notag \\
4\,C_{VA}^{u,8} & = - \big(C_{Qq}^{1,8} + C_{Qq}^{3,8}\big) + C_{tu}^8 - C_{tq}^8 + C_{Qu}^8 \; .
\label{eq:vv-aa}
\end{align}
Pure $VV$, $AA$, $VA$, or $AV$ currents are obtained for
\begin{align}\label{eq:rel}
|C_{Qq}^{1,8} + C_{Qq}^{3,8}| = |C_{tu}^8| = |C_{tq}^8| = |C_{Qu}^8|\,.
\end{align}
The corresponding combinations with the down-type index $d$ can be
derived by replacing the index $u \to d$ and
$+ C_{Qq}^{3,8} \to - C_{Qq}^{3,8}$ in Eqs.~\eqref{eq:vv-aa} and \eqref{eq:rel}. For color-singlet
coefficients, we define the same relations by changing all indices
$8\to 1$ in Eq.~\eqref{eq:vv-aa}, yielding $C_{VV}^{u,1}$ etc.  
Neglecting electroweak contributions, the
$q\bar{q}\to t\bar{t}$ partonic rate at LO is then given by ($q=u,d$; cf. Refs.~\cite{Zhang:2010dr} and \cite{Ferrario:2008wm})
\begin{align}
\frac{d\sigma(q\bar{q}\to t\bar{t})}{dc_t} & = \frac{2}{9}\frac{\pi \alpha_s^2 \beta_{t\bar{t}}}{2s} \bigg\{ \big(1+\beta_{t\bar{t}}^2c_t^2+4\overline{m}_t^2\big) + \frac{1}{\sqrt{4\pi\alpha_s}} \frac{m_t v}{\Lambda^2} \frac{16}{\sqrt{2}} \, C_{tG} \notag \\
&\quad + \frac{1}{4\pi\alpha_s}\frac{2s}{\Lambda^2} \left[\big(1+\beta_{t\bar{t}}^2c_t^2+4\overline{m}_t^2\big) C_{VV}^{q,8} + 2\beta_{t\bar{t}} c_t\,C_{AA}^{q,8}\right]\notag \\
& \ \ \ + \frac{1}{(4\pi\alpha_s)^2}\frac{s^2}{\Lambda^4}\Big[4\beta_{t\bar{t}} c_t\,\Big(C_{VV}^{q,8}C_{AA}^{q,8} + C_{VA}^{q,8}C_{AV}^{q,8} + \frac{9}{2}\big(C_{VV}^{q,1}C_{AA}^{q,1} + C_{VA}^{q,1}C_{AV}^{q,1}\big)\Big) \notag \\
&  \ \ \  +  \left(1+\beta_{t\bar{t}}^2c_t^2\right)\Big(|C_{V+A}^{q,8}|^2 + \frac{9}{2}|C_{V+A}^{q,1}|^2 \Big) + 4\overline{m}_t^2\Big(|C_{V-A}^{q,8}|^2 + \frac{9}{2}|C_{V-A}^{q,1}|^2 \Big) \Big] \notag \\
& \ \ \ + \frac{1}{(4\pi\alpha_s)^{\frac{3}{2}}}\frac{m_t v s}{\Lambda^4}\frac{4}{\sqrt{2}}\left(C_{VV}^{q,8} + \beta_{t\bar{t}}c_t\, C_{AA}^{q,8}\right) C_{tG} + \mathcal{O}\Big( \frac{v^2s}{\Lambda^4} C_{tG}^2 \Big) \bigg\}\,,
\label{eq:qq_tt_analytic}
\end{align}
with the combinations of color-octet ($\alpha = 8$) and color-singlet
($\alpha = 1$) Wilson coefficients,
\begin{align}
|C_{V+A}^{q,\alpha}|^2 & = |C_{VV}^{q,\alpha}|^2 + |C_{VA}^{q,\alpha}|^2 + |C_{AA}^{q,\alpha}|^2 + |C_{AV}^{q,\alpha}|^2\,, \phantom{\Big]} \notag \\
& \stackrel{q=u}{=} \left(|C_{Qq}^{1,8} + C_{Qq}^{3,8}|^2 + |C_{tu}^8|^2 + |C_{tq}^8|^2 + |C_{Qu}^8|^2\right)/4 \\
|C_{V-A}^{q,\alpha}|^2 & = |C_{VV}^{q,\alpha}|^2 + |C_{VA}^{q,\alpha}|^2 - |C_{AA}^{q,\alpha}|^2 - |C_{AV}^{q,\alpha}|^2 \notag \\
& \stackrel{q=u}{=} \left( (C_{Qq}^{1,8} + C_{Qq}^{3,8}) C_{tq}^8  + C_{tu}^8 C_{Qu}^8 \right) / 2 \,.
\label{eq:chiral_coeff}
\end{align}
To understand the operator effects in kinematic distributions, it is
instructive to explore their behavior at high CM energies
$\sqrt{s}$. The high-energy behavior of the various four-quark
contributions and their interference with $O_{tG}$ is summarized in
Tab.~\ref{tab:he-scaling}.

\begin{table}[t]
\renewcommand{\arraystretch}{1.3}
\begin{center}
\begin{tabular}{c|c|c|c|c|c|c}
\toprule
 & QCD & $\ C_{tG}\ $ & $\ C_{VV}^{q,8}\ $ & $C_{AA}^{q,8}$ & $C_{VV}^{q,1}$ & $C_{AA}^{q,1}$ \\
\hline
 \phantom{\Bigg]} QCD & 1 & $\dfrac{m_t v}{\Lambda^2}$ & $\dfrac{s}{\Lambda^2}$ & $c_t\dfrac{s}{\Lambda^2}$ & -- & -- \\
 \phantom{\bigg]} $C_{tG}$ & \dots & $\dfrac{s v^2}{\Lambda^4}$ & $\dfrac{s m_t v}{\Lambda^4}$ & $c_t\dfrac{s m_t v}{\Lambda^4}$ & -- & -- \\
 \phantom{\bigg]} $C_{VV}^{q,8}$ & \dots & \dots & $\dfrac{s^2}{\Lambda^4}$ & $c_t \dfrac{s^2}{\Lambda^4}$ & -- & -- \\
 \phantom{\bigg]} $C_{AA}^{q,8}$ & \dots & \dots & \dots & $\dfrac{s^2}{\Lambda^4}$ & -- & --\\
 \phantom{\bigg]} $C_{VV}^{q,1}$ & \dots & \dots & \dots & \dots & $\dfrac{s^2}{\Lambda^4}$ & $c_t\dfrac{s^2}{\Lambda^4}$ \\
 \phantom{\bigg]} $C_{AA}^{q,1}$ & \dots & \dots & \dots & \dots &  \dots & $\dfrac{s^2}{\Lambda^4}$ \\
\bottomrule
\end{tabular}
\end{center}
\caption{Relative scaling of operator contributions with respect to QCD in top pair production ($q\bar{q} \to t\bar{t}$) at high energies $\sqrt{s} \gg m_t$.}
\label{tab:he-scaling}
\end{table}

Dipole operators like $O_{tG}$ flip the helicity of the top quark and
require an insertion of the Higgs vacuum expectation value in the
amplitude. Their interference with QCD scales as $m_t v/\Lambda^2$ and
does not feature an enhancement at high energies. Squared $O_{tG}$
contributions, in turn, grow like $sv^2/\Lambda^4$ as in the $gg\to
t\bar{t}$ process. Four-quark operator interferences with QCD and with
$C_{tG}$ grow as $s/\Lambda^2$ and $m_t v s/\Lambda^4$ relative to
QCD, respectively. Squared terms in four-quark operators grow even
stronger with energy, scaling as $s^2/\Lambda^4$. The strong
enhancement at high energies is typical of a four-quark contact
interaction, compared to a short-distance interaction through gluon
exchange. Top pair production is thus most sensitive to four-quark
operators in the tails of kinematic distributions, due to both a
kinematic enhancement and an increased quark-antiquark luminosity.
 
Among the four-quark interactions, only color-octet operators
interfere with QCD and with $C_{tG}$. Color-singlet operators
contribute through interference among themselves. The relative factor
of 9/2 between quadratic contributions of color-singlet and
color-octet operators in Eq.~\eqref{eq:qq_tt_analytic} is due to the
color structure. For color singlets, the rate is proportional to the
number of colors $N_c$ of each quark current, yielding $N_c^2 = 9$.
For color octets, it is proportional to $\sum_{AB}\text{Tr}(T^A T^B)^2
= (N_c^2-1)/4 = 2$.
 
The sensitivity of the $q\bar{q}\to t\bar t$ process to the chirality
of the top quarks is crucial to distinguish between different
four-quark operators. Their interference with QCD or $O_{tG}$ probes
the two combinations $C_{VV}^{q,8}$ and $C_{AA}^{q,8}$, \emph{i.e.},
pure vector and axial-vector currents. Interference of two four-quark
operators introduces the additional chirality structures from
Eq.~\eqref{eq:chiral_coeff}. The impact of chiral operators on
kinematic distributions can be understood by considering
charge-symmetric and -asymmetric differential cross sections
\begin{align}
d\sigma^S & = d\sigma\big(t(p_1)\bar{t}(p_2)\big) + d\sigma\big(t(p_2)\bar{t}(p_1)\big) \notag \\
d\sigma^A & = d\sigma\big(t(p_1)\bar{t}(p_2)\big) - d\sigma\big(t(p_2)\bar{t}(p_1)\big)\,,
\label{eq:sym-asym}
\end{align}
where $p_1$ and $p_2$ are the 4-momenta of of the two tops in the
final state. In Tab.~\ref{tab:tt-ops}, we list the four-quark
coefficients contributing to $\sigma^S$ and $\sigma^A$ in top pair
production at LO
 in QCD.
 At leading order, 5 chiral
combinations of Wilson coefficients per parton contribute. A priori,
they can be distinguished by five measurements of different kinematic
observables. Charge asymmetries play an important role in this
endeavor, since they probe chiral structures that are not accessible
in cross sections or other charge-symmetric
observables~\cite{Rosello:2015sck}. 
These observations will be confronted with data in Sec.~\ref{sec:ttbar_L_vs_R_Ac}, where we show how to use cross sections and asymmetries to gain access to the chirality of the four-quark operators. At NLO, the chiral contributions to $t\bar{t}$ production are modified by QCD corrections, leading to additional kinematic degrees of freedom. In Sec.~\ref{sec:ttbar_L_vs_R_NLO}, we will elaborate more on NLO effects in kinematic distributions.

\begin{table}[t]
\begin{center}
\renewcommand{\arraystretch}{1.3}
\begin{tabular}{c|c}
\toprule
$\ \sigma_k^S$ & $C_{VV}^{q,8}$ \\
\midrule
\multirow{2}{*}{$\ \sigma_{kl}^S\ $}
  & $|C_{V+A}^{q,8}|^2 + \frac{9}{2}\,|C_{V+A}^{q,1}|^2$  \\
  & $|C_{V-A}^{q,8}|^2 + \frac{9}{2}\,|C_{V-A}^{q,1}|^2$  \\
\midrule
$\ \sigma_k^A$ & $C_{AA}^{q,8}$   \\
\midrule
\multirow{1}{*}{$\ \sigma_{kl}^A$}
  & $\ C_{VV}^{q,8} C_{AA}^{q,8} + C_{VA}^{q,8} C_{AV}^{q,8} + \frac{9}{2}\big(C_{VV}^{q,1} C_{AA}^{q,1} + C_{VA}^{q,1} C_{AV}^{q,1}\big)$  \\
\bottomrule
\end{tabular}
\caption{Four-quark contributions to $t\bar{t}$ production in
  SMEFT at LO QCD. We separate SM-interference,
  $\sigma_k^{S,A}$, and dimension-6 squared terms,
  $\sigma_{kl}^{S,A}$ for charge-symmetric and -asymmetric cross sections.\label{tab:tt-ops}}
 \end{center}
\end{table}

\subsection{Single top production and top decay}
\label{sec:eft_single}

\begin{figure}[t]\centering
\parbox{.22\textwidth}{\centering\includegraphics[width=.22\textwidth]{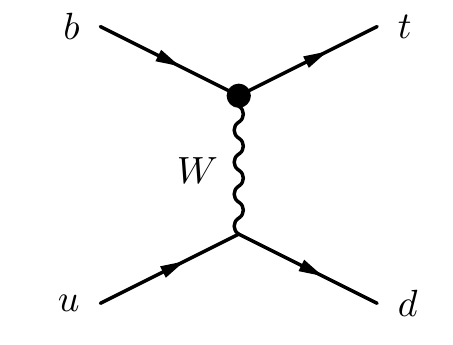}\\($t$-channel)} 
\parbox{.22\textwidth}{\centering\includegraphics[width=.22\textwidth]{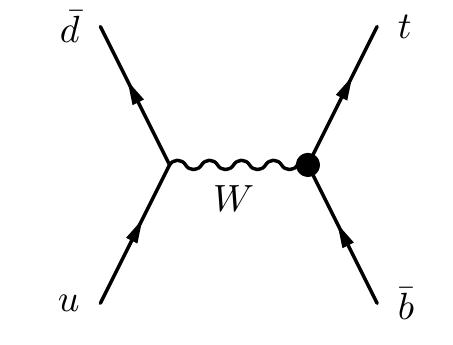}\\($s$-channel)}
\parbox{.22\textwidth}{\centering\includegraphics[width=.22\textwidth]{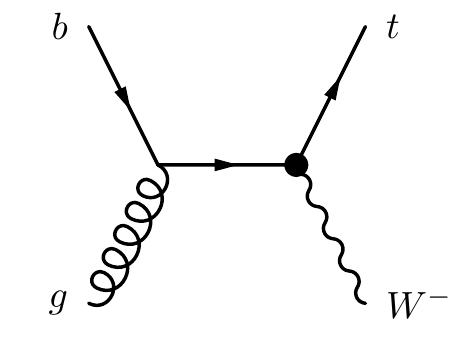}  \\($tW$)}
\parbox{.22\textwidth}{\centering\includegraphics[width=.22\textwidth]{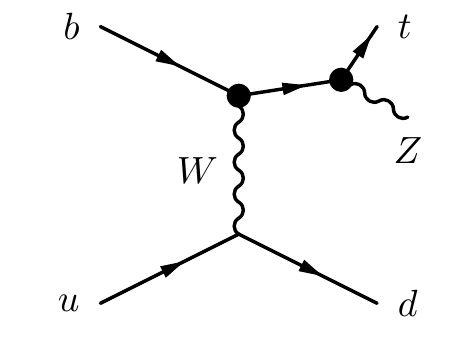}  \\($tZ$)}
\caption{Examples of Feynman diagrams contributing to single top
  production in SMEFT at leading order. The dots indicate possible
  operator insertions.}\label{fig:singletop_diagrams}
\end{figure}

Single top production and top decay are both sensitive to operators
with weak gauge bosons and in this sense complementary to top pair
production. We distinguish $t-$channel and $s-$channel production, as
well as associated $tW$ and $tZ$ production. Examples of Feynman
diagrams for these processes are shown in
Fig.~\ref{fig:singletop_diagrams}. 

\paragraph{$t-$ and $s-$channel production} 
probe the same set of operators, because the underlying partonic
processes $ub\to dt$ and $u\bar{d} \to t\bar{b}$ are related by
a crossing symmetry. At the level of SM-interference 
three dimension-6 operators contribute,
\begin{align}
O_{Qq}^{3,1} & = (\bar{Q}\gamma_\mu\tau^I Q)(\bar{q}_i\gamma^\mu\tau^I
q_i) & ^\ddagger O_{tW} & = ( \bar{Q} \sigma^{\mu\nu} \tau^I t)
\tilde{\phi} W_{\mu\nu}^I \notag \\ O_{\phi Q}^3 & = (\phi^\dagger
\stackrel{\longleftrightarrow}{iD_\mu^I}
\phi)(\bar{Q}\gamma^{\mu}\tau^I Q)\,.
\end{align}
Since the kinematics of the two channels is different, we also expect
a different sensitivity to these operators. The dominant partonic
cross sections for $t-$ and $s-$channel production are given by~\cite{Zhang:2010dr}
\begin{align}
\frac{d\sigma_t(ub \to dt)}{dc_{tu}} = & \frac{G_F^2 m_W^4 \beta_t^2}{\pi s \big(2\overline{m}_W^2 + \beta_t(1 + c_{tu})\big)^2} \bigg\{ 1 + 2\frac{v^2}{\Lambda^2} C_{\phi Q}^3 + \sqrt{2}\frac{v^2}{\Lambda^2} C_{tW}\frac{m_t}{m_W}(1 + c_{tu}) \notag \\
& - \frac{v^2}{\Lambda^2} \frac{s}{m_W^2}\,C_{Qq}^{3,1} \big(2\overline{m}_W^2 + \beta_t(1+ c_{tu})\big)\bigg\} \notag \\
\frac{d\sigma_s(u\bar{d} \to t\bar{b})}{dc_{tu}} = & \frac{G_F^2 m_W^4 \beta_t^2}{16\pi s(1-\overline{m}_W^2)^2}(1+c_{tu})(1+ \beta_t c_{tu} + \overline{m}_t^2) \bigg\{ 1 + 2\frac{v^2}{\Lambda^2}\,C_{\phi Q}^3 \notag \\
& + 4\sqrt{2}\frac{v^2}{\Lambda^2}\, C_{tW} \frac{m_t}{m_W}\frac{1}{1+ \beta_t c_{tu} + \overline{m}_t^2} + 2\frac{v^2}{\Lambda^2}\frac{s}{m_W^2}\,C_{Qq}^{3,1} (1-\overline{m}_W^2)\bigg\}\,.
\end{align}
Here $\beta_t = 1-\overline{m}_t^2$, $\overline{m}_W^2 = m_W^2/s$, and
$c_{tu} = \cos\theta_{tu}$ is the cosine of the angle between the top
and the incoming up quark in the CM system. We set $V_{tb} = 1 =
V_{ud}$ and neglect all quark masses except $m_t$. In $t-$channel production, the process $\bar{d}b \to \bar{u}t$ also contributes, but is subdominant due to the smaller parton luminosity. We do not provide analytic expressions for this channel, but include it in the numerical analysis.

The operator $O_{\phi Q}^3$ has the same Lorentz structure as the $tbW$ coupling in
weak interactions, so its interference with the SM causes a constant
shift in the rate. In turn, the contributions of $O_{Qq}^{3,1}$ is logarithmically enhanced at high energies, while $O_{tW}$ scales as $\ln(s/m_W^2)/s$. In $s-$channel production the logarithmic enhancement is absent. In
Tab.~\ref{tab:he-scaling-singletop}, we summarize the relative scaling of operators with respect to the SM for $t-$ and $s-$channel production.

\begin{table}[t]
\renewcommand{\arraystretch}{1.3}
\begin{center}
\begin{tabular}{c|c|c|c|c|c|c|c}
\toprule
 & $\,\text{SM}\,$ & $C_{Qq}^{3,1}$ & $C_{\phi Q}^3$ & $C_{tW}$ & $\phantom{\Big]} C_{Qq}^{3,8}\ $ & $\ C_{\phi t b}\ $ & $C_{bW}$ \\
\hline
 $\ \text{SM}$ & $1$ & $\dfrac{m_W^2}{\Lambda^2}\ln \dfrac{s}{m_W^2}$ & $\dfrac{v^2}{\Lambda^2}$ & $\dfrac{m_t v}{\Lambda^2}\dfrac{m_W^2}{s} \ln \dfrac{s}{m_W^2} $ & -- & $\propto m_b$ & $\propto m_b$ \\
 \phantom{\bigg]} $C_{Qq}^{3,1}\ $  & \dots &  $\dfrac{s m_W^2}{\Lambda^4}$ & $\dfrac{v^2 m_W^2}{\Lambda^4} \ln \dfrac{s}{m_W^2} $ & $\dfrac{m_t v m_W^2}{\Lambda^4}$ & -- & $\propto m_b$ & $\propto m_b$ \\
 \phantom{\bigg]} $C_{\phi Q}^3\ $ & \dots & \dots & $\dfrac{v^4}{\Lambda^4}$ & $\dfrac{m_t v^3}{\Lambda^4}\dfrac{m_W^2}{s} \ln \dfrac{s}{m_W^2} $ & -- & $\propto m_b$ & $\propto m_b$\\
 \phantom{\bigg]} $C_{tW}\ $ & \dots & \dots & \dots & $\dfrac{v^2 m_W^2}{\Lambda^4}$ & -- & $\propto m_b$ & $\propto m_b$\\
 \phantom{\bigg]} $C_{Qq}^{3,8}\ $ & \dots & \dots & \dots & \dots & $\dfrac{s m_W^2}{\Lambda^4}$ & -- & -- \\
 \phantom{\bigg]} $C_{\phi t b}\ $ & \dots & \dots & \dots & \dots & \dots & $\dfrac{v^4}{\Lambda^4}$ & $\dfrac{m_t v^3}{\Lambda^4}\dfrac{m_W^2}{s} \ln \dfrac{s}{m_W^2}$\\
 \phantom{\bigg]} $C_{bW}\ $ &\dots  & \dots & \dots & \dots & \dots & \dots & $\dfrac{v^2 m_W^2}{\Lambda^4}$ \\
\bottomrule
\end{tabular}
\end{center}
\caption{Relative scaling of operator contributions in $t-$channel single top production at high energies $\sqrt{s} \gg m_t$. The scaling for $s-$channel production is obtained by replacing $m_W^2\to s$ and $\ln(\dots) \to 1$. The SM contribution scales as $1/m_W^2$ in $t-$channel production and as $1/s$ in $s-$channel production. We denote a negligible bottom mass insertion as $\propto m_b$.}
\label{tab:he-scaling-singletop}
\end{table}

Once we include dimension-6 squared terms, three additional operators
$O_{Qq}^{3,8}$, $O_{\phi tb}$ and $O_{bW}$ contribute to single top
production. The operator $O_{Qq}^{3,8}$ does
not interfere because of its color structure. The interference of
$O_{\phi tb}$ and $O_{bW}$ with the SM and all other operators is
suppressed by the bottom mass $m_b$, because $O_{\phi tb}$ and
$O_{bW}$ involve right-handed bottom quarks. This means that their
interference with left-handed currents is helicity-suppressed. The
interference between $O_{\phi tb}$ and $O_{bW}$ are not
$m_b-$suppressed and thus much larger than their interference with the
SM amplitude.

\paragraph{Top decay} is sensitive to the operators $O_{Qq}^{3,1}$ (or operators with two heavy quarks and two leptons in semi-leptonic top decays), $O_{\phi Q}^3$ and $O_{tW}$ in SM-interference. Since
the $W$-boson in $t\to bW \to b q\bar{q}'$ decays is on-shell in the observables we consider, the
contribution of four-quark operators is negligible. As in single top
production, $O_{\phi Q}^3$ re-scales the SM rate by a factor $(1+
2C_{\phi Q}^3v^2/\Lambda^2)$. With the current experimental situation,
we expect a higher sensitivity to $O_{\phi Q}^3$ in $t$-channel single
top production than in the top width $\Gamma_t$. The dipole operator
$O_{tW}$ in turn modifies the top decay kinematics. In particular,
the helicity fractions $F_i = \Gamma_i/\Gamma_t$ of the $W$-boson are
very sensitive to this operator. In our analysis, we
consider~\cite{Zhang:2010dr}
\begin{align}\label{eq:w-hel}
F_L & = \frac{m_t^2}{m_t^2 + 2m_W^2} -
4\sqrt{2}\frac{v^2}{\Lambda^2}C_{tW} \frac{m_t m_W(m_t^2 -
  m_W^2)}{(m_t^2 + 2m_W^2)^2} \notag \\ F_0 & = \frac{m_t^2}{m_t^2 +
  2m_W^2} + 4\sqrt{2}\frac{v^2}{\Lambda^2}C_{tW} \frac{m_t m_W(m_t^2 -
  m_W^2)}{(m_t^2 + 2m_W^2)^2}\,,
\end{align}
where $F_L$ and $F_0$ denote the $t\to bW$ branching ratios into $W$
bosons with negative $(L)$ and zero $(0)$ helicity, respectively. At
the level of dimension-6 contributions squared, $O_{bW}$ and $O_{\phi
  tb}$ contribute to the helicity fractions. Numerically their
contribution is at the permille level, similar to $t$-channel
production. In a global analysis, top decays are thus relevant in
probing these operators.

\paragraph{Associated $tW$ production} 
probes $O_{\phi Q}^3$, $O_{tW}$ and $O_{tG}$ interfering with the SM
amplitudes. Obviously, its sensitivity to $O_{tG}$ is much smaller
than for top pair production. The operator $O_{\phi Q}^3$ is also
probed in $t$-channel and $s$-channel production, and $O_{tW}$ is best
probed in top decays. We therefore do not expect much
additional information on SMEFT operators from $tW$ production. 

\paragraph{Associated $tZ$ production} 
is essentially $t-$channel single top production with an additional
$Z$-boson in the final state. It probes all operators that contribute
to $t$-channel production, as shown in
Tab.~\ref{tab:wilson-contributions}. The contribution of an operator
relative to the SM, however, is different for the two processes. In
general, $tZ$ production probes operators at higher energies than
$t$-channel production, leading to enhanced effects of operators that
grow with energy~\cite{Degrande:2018fog}. A larger theoretical
sensitivity in $tZ$ production can thus compensate for the lower
experimental sensitivity.

In addition, $tZ$ production probes operators that modify the top
coupling to the $Z$-boson, namely $O_{\phi Q}^-$, $O_{\phi t}$, and
$O_{tZ}$. All three operators interfere with the SM. Their scaling at
high energies depends on the polarization of the $Z$-boson and has
been studied in detail in
Refs.~\cite{Degrande:2018fog,Maltoni:2019aot}. The operators $O_{\phi
  Q}^-$ and $O_{\phi t}$ modify the SM $Z$ couplings with left- and
right-handed tops, respectively. The main difference is observed in
the longitudinal $Z$-mode. While the $O_{\phi Q}^- -$SM interference
grows with energy, the $O_{\phi t} -$SM interference requires a
helicity flip of the top quark and thus does not feature this
growth. Similar considerations apply at the level of $O_{\phi Q}^- -
O_{\phi Q}^-$ and $O_{\phi t} - O_{\phi t}$ interference. We therefore
expect a higher sensitivity to $O_{\phi Q}^-$ than to $O_{\phi t}$,
which we will confirm numerically in
Sec.~\ref{sec:global_single}. The dipole operator $O_{tZ}$ features
a similar energy growth in longitudinal $Z$-production as $O_{\phi
  Q}^-$, resulting in a sizeable contribution to $tZ$
production.

Associated single top production with a Higgs boson is an interesting
channel that complements $tW$ and $tZ$ production~\cite{Degrande:2018fog}. In addition to the gauge operators $tH$ production probes
modifications of the top Yukawa coupling. We do not consider $tH$
production in our fit here, but plan to include it in a combined
analysis with Higgs observables.

NLO QCD corrections can modify the kinematics of operator
contributions in single top production. However, the number of single
top observables is large enough to distinguish between all
contributing operators already at leading order. We therefore do not
investigate NLO contributions in detail here, but include them in our
numerical analysis. For $t-$channel single top production and top
decay, NLO corrections in SMEFT have been calculated in Refs.~\cite{Zhang:2016omx,Boughezal:2019xpp}.

\subsection{Associated $t\bar{t}W$ and $t\bar{t}Z$ production}
\label{sec:eft_ttW_ttZ}

Compared with top pair and single top production, associated
$t\bar{t}W$ and $t\bar{t}Z$ production do not bring us sensitivity to
new operators, but probe them in a different context. The main purpose
of including these two processes in our global analysis is to resolve
blind directions and to better probe operators that are difficult to
access in other channels. Just as for the $tH$ production channel we leave
$t\bar{t}H$ production~\cite{Maltoni:2016yxb} to a future combination
with global Higgs analyses.

Associated $t\bar{t}W$ production is sensitive to a subset of the
possible four-quark operators. Since the $W$ can only be radiated from
the initial state in the SM, only operators with left-handed initial quarks contribute while $RR$
and $LR$ contributions are absent if we neglect light-quark
masses. The non-trivial electroweak structure of $t\bar{t}W$
production affects the relative contributions of the weak singlet and
triplet operators $O_{Qq}^{1,8}$ and $O_{Qq}^{3,8}$. We will use this
effect to distinguish between these two operators.  In the SM, the
total rate of $t\bar{t}W$ production is dominated by quark-antiquark
interactions, while inclusive $t\bar{t}$ production is
gluon-dominated. This means that relative to the SM contribution
four-quark operators thus give sizeable effects in the $t\bar{t}W$
rate and we can hope for a good sensitivity to $LL$ and $RL$
operators.

Associated $t\bar{t}Z$ probes the same four-quark operators that enter
$t\bar{t}$ production. Similarly to $t\bar{t}W$ production, the
emission of the boson changes the relative contributions of four-quark
operators with different weak gauge structure. In addition,
$t\bar{t}Z$ is sensitive to operators with right-handed up versus down
quarks, namely $O_{tu}^8$ and $O_{td}^8$ or $O_{Qu}^8$ and
$O_{Qd}^8$. In Sec.~\ref{sec:ttbar_u_vs_d}, we will disentangle these
operators by combining $t\bar{t}$ distributions with $t\bar{t}W$ and
$t\bar{t}Z$ production in a global analysis.  In addition, we use the
$t\bar{t}Z$ process to probe $O_{\phi Q}^-$, $O_{\phi t}$ and
$O_{tZ}$, which are so far constrained by $tZ$ production. In
$t\bar{t}Z$ production, all three operators interfere with the SM. The
contributions of $O_{\phi Q}^-$ and $O_{\phi t}$ re-scale the SM
$Z$-couplings to left- and right-handed tops, but do not change the
process kinematics. The dipole operator $O_{tZ}$ changes the kinematic
distributions mildly, but its overall effect on the rate is
modest~\cite{Bylund:2016phk}. Combining searches for $O_{tA}$ in
$t\bar{t}\gamma$ and $O_{tW}$ in top decays is an alternative way to
get access to $C_{tZ}$, see Eq.~\eqref{eq:operator-relations} and
Ref.~\cite{Schulze:2016qas}.

\section{Global analysis setup}
\label{sec:global}

\subsection{Data set}
\label{sec:data_set}

\begin{table}[t]\centering
\setlength{\tabcolsep}{2pt}
\renewcommand{\arraystretch}{1.2}
\begin{small} \begin{tabular}{lcc>{$}l<{$}>{$}l<{$}c|*4{c}}
\toprule
experiment& $\sqrt{S}$ (TeV)& $\mathcal{L}$ (fb$^{-1}$)& \text{channel}&
\text{observable \& $K$-factor}& \#bins &
\ R& M& D& A \\
\midrule[1pt]

$p p \to t\bar t$ & & & & & & & & \\

CMS\hfill~\cite{Khachatryan:2016mqs}& 8& 19.7& e\mu& \sigma_{t\bar t}\hfill \text{\cite{Czakon:2011xx}}& & \ \checkmark& \checkmark & . & . \\

ATLAS\hfill~\cite{Aaboud:2017cgs}& 8& 20.02& lj& \sigma_{t\bar t}\hfill\text{\cite{Czakon:2011xx}}& &\  \checkmark& \checkmark & . & . \\

CMS\hfill~\cite{Sirunyan:2017uhy}& 13& 2.3& lj& \sigma_{t\bar t}\hfill\text{\cite{Czakon:2011xx}}& & \ \checkmark& \checkmark & . & . \\

CMS\hfill~\cite{Sirunyan:2018goh}& 13& 3.2& ll& \sigma_{t\bar t}\hfill\text{\cite{Czakon:2011xx}}& & \ \checkmark & \checkmark& . & . \\

ATLAS\hfill~\cite{Aaboud:2016pbd}& 13& 3.2& e\mu&
\sigma_{t\bar t}\hfill\text{\cite{Czakon:2011xx}}& & \ \checkmark& \checkmark & . & . \\ 

\cmidrule[0.1pt]{1-10}

ATLAS\hfill~\cite{Aad:2015mbv}& 8& 20.3& lj& \sigma^{-1}(d\sigma/dm_{t\bar t})\hfill \text{\cite{Czakon:2017dip,Czakon:2015owf,Czakon:2016dgf}}& 7 &. &\checkmark& \checkmark & . \\

CMS\hfill~\cite{Khachatryan:2015oqa}& 8& 19.7& lj& \sigma^{-1}(d\sigma/dp_{T,t})\hfill\text{\cite{Czakon:2017dip,Czakon:2015owf,Czakon:2016dgf}}& 7 & . &. &\checkmark & . \\

& & & ll& \sigma^{-1}(d\sigma/d p_{T,1})& 5 & . &. &\checkmark& . \\

CMS\hfill~\cite{Sirunyan:2017azo}& 8& 19.7& e\mu&
\sigma^{-1}(d^2\sigma/dm_{t\bar t}dy_{t\bar t})\hfill\text{\cite{Czakon:DoubleDiff}}& 16 & . &. &. & . \\

CMS\hfill~\cite{Khachatryan:2016gxp}& 8& 19.7 & lj \text{ high }p_T& d\sigma/dp_{T,t} &5 &. & . & . & . \\

CMS\hfill~\cite{Khachatryan:2016mnb}& 13& 2.3& lj& \sigma^{-1}(d\sigma/dm_{t\bar t})& 8 & . &\checkmark& \checkmark & . \\

CMS\hfill~\cite{Sirunyan:2018wem}& 13& 35.8& lj& \sigma^{-1}(d\sigma/dp_{T}(t_h))\ \hfill\text{\cite{Czakon:2017dip,Czakon:2015owf,Czakon:2016dgf}}& 12 &. &  . & \checkmark & . \\

CMS\hfill~\cite{Sirunyan:2017mzl}& 13& 2.1& ll& \sigma^{-1}(d\sigma/dp_{T,t})\hfill\text{\cite{Czakon:2017dip,Czakon:2015owf,Czakon:2016dgf}}& 6 &. & . & \checkmark & . \\

CMS\hfill~\cite{Sirunyan:2018ucr}& 13& 35.9& ll& \sigma^{-1}(d\sigma/d\Delta y_{t\bar t})\hfill\text{\cite{Czakon:2017dip,Czakon:2015owf,Czakon:2016dgf}}& 8 &. & . & . & \checkmark \\

ATLAS\hfill~\cite{Aaboud:2018eqg}& 13& 36.1& aj \text{ high } p_T& \sigma^{-1}(d\sigma/dm_{t\bar t})& 8& . &. & . & . \\
\cmidrule[0.1pt]{1-10}

CMS\hfill~\cite{Khachatryan:2015oga}& 8& 19.7& lj& A_C\hfill\text{ \cite{Czakon:2017lgo}}& & . & . &. & \checkmark \\

CMS\hfill~\cite{Khachatryan:2016ysn}& 8 & 19.7& ll& A_C\hfill\text{ \cite{Czakon:2017lgo}}&  & . & . &. & \checkmark \\

ATLAS\hfill~\cite{Aad:2015noh}& 8& 20.3& lj& A_C\hfill\text{ \cite{Czakon:2017lgo}}& & . & . &. & \checkmark \\
 
ATLAS\hfill~\cite{Aad:2016ove}& 8& 20.3& ll& A_C\hfill\text{ \cite{Czakon:2017lgo}}& & . & . &. & \checkmark \\

ATLAS\hfill~\cite{ATLAS-CONF-2019-026}& 13& 139& lj& A_C\hfill\text{ \cite{Czakon:2017lgo}}& &. & . &. & \checkmark  \\

\midrule[1pt]
\multicolumn{2}{l}{$pp \to t\bar{t}Z$} & & & & & & & & \\
CMS\hfill~\cite{CMS:2019too}& 13& 77.5& \text{multi lept.} & \sigma_{t\bar{t}Z} \hfill\text{\cite{Kulesza:2018tqz}} & & . & . &. &  .\\

ATLAS\hfill~\cite{Aaboud:2016xve}& 13& 3.2& \text{multi lept.} & \sigma_{t\bar{t}Z}\hfill\text{\cite{Kulesza:2018tqz}} & & . & . &. & .\\

\midrule[0.1pt]
\multicolumn{2}{l}{$pp \to t\bar{t}W$} & & & & & & & & \\
CMS\hfill~\cite{Sirunyan:2017uzs}& 13& 35.9& \text{multi lept.} & \sigma_{t\bar{t}W}\hfill\text{\cite{Kulesza:2018tqz}} & & . & . &. & . \\

ATLAS\hfill~\cite{Aaboud:2016xve}& 13& 3.2& \text{multi lept.} & \sigma_{t\bar{t}W}\hfill\text{\cite{Kulesza:2018tqz}} & & . & . &. & . \\

\bottomrule[1pt]
\end{tabular} \end{small}
\caption{Top-pair observables included in our global analysis. The labels R, M,
  D, A define four different data
  sets with rates, rates and invariant mass distributions, distributions only, and asymmetries, used in the numerical analysis of Section~\ref{sec:features}.}
\label{tab:data1}
\end{table}

\begin{table}[t]\centering
\setlength{\tabcolsep}{5pt}
\renewcommand{\arraystretch}{1.2}
\begin{small} \begin{tabular}{lcc>{$}l<{$}>{$}l<{$}}
\toprule
experiment& $\sqrt{S}$ (TeV)& $\mathcal{L}$ (fb$^{-1}$)& \text{channel}&
\text{observable \& $K$-factor }
\\
\midrule[1pt]
$t-$channel & & & &  \\

CMS\hfill~\cite{Chatrchyan:2012ep}& 7& 1.17 ($\mu$), 1.56 ($e$)& e+\mu& \sigma_{tq + \bar t q} \\

ATLAS\hfill~\cite{Aad:1712656}& 7& 4.59& e+\mu& \sigma_{tq + \bar t q} \\

ATLAS\hfill~\cite{Aaboud:2017pdi}& 8& 20.2& e+\mu& \sigma_{tq}, \sigma_{\bar t q} \\

CMS\hfill~\cite{Khachatryan:2014iya}& 8& 19.7& e+\mu& \sigma_{tq}, \sigma_{\bar t q} \\

ATLAS\hfill~\cite{Aaboud:2215404}& 13& 3.2& e+\mu& \sigma_{tq}, \sigma_{\bar t q} \hfill\text{\cite{Berger:2016oht}} \\

CMS\hfill~\cite{CMS-PAS-TOP-16-003}& 13& 2.3& \mu& \sigma_{tq}, \sigma_{\bar t q} \hfill\text{\cite{Berger:2016oht}} \\

\midrule[1pt]
$s-$channel & & & &  \\

CMS\hfill~\cite{Khachatryan:2137611}& 7& 5.1& \mu& \sigma_{t\bar b+ \bar t b} \\
& 8& 19.7& e+\mu& \sigma_{t\bar b+ \bar t b} \\

ATLAS\hfill~\cite{Aad:2102936}& 8& 20.3& e+\mu& \sigma_{t\bar b+ \bar t b} \\

\midrule[1pt]
$tW$ channel & & & & \\

ATLAS\hfill~\cite{Aad:2012xca}& 7& 2.05& 2lj& \sigma_{tW+\bar tW} \\

CMS\hfill~\cite{Chatrchyan:2012zca}& 7& 4.9& 2lj& \sigma_{tW+\bar tW} \\

ATLAS\hfill~\cite{Aad:2015eto}& 8& 20.3& 2lj& \sigma_{tW+\bar tW}  \\

CMS\hfill~\cite{Chatrchyan:2014tua}& 8& 12.2& 2lj& \sigma_{tW+\bar tW} \\

ATLAS\hfill~\cite{Aaboud:2016lpj}& 13& 3.2& 2lj& \sigma_{tW+\bar tW} \\

CMS\hfill~\cite{Sirunyan:2018lcp}& 13& 35.9& e\mu j& \sigma_{tW+\bar tW}  \\

\midrule[1pt]
$tZ$ channel & & & &  \\

ATLAS\hfill~\cite{Aaboud:2017ylb}& 13& 36.1& 3l2j& \sigma_{tZq} \\

\midrule[1pt]
\multicolumn{2}{l}{$W$ helicities in top decays} & & & \\

ATLAS\hfill~\cite{Aad:2012ky}& 7& 1.04& & F_0,\, F_L \\

CMS\hfill~\cite{Chatrchyan:2013jna}& 13& 5.0& & F_0,\, F_L \\

ATLAS\hfill~\cite{Aaboud:2016hsq}& 8& 20.2& & F_0,\, F_L \\

CMS\hfill~\cite{Khachatryan:2016fky}& 8& 19.8& & F_0,\, F_L \\
\bottomrule[1pt]
\end{tabular} \end{small}
\caption{Observables included in the single top fit, in analogy to
  Tab.~\ref{tab:data1} .}
\label{tab:data2}
\end{table}

The key to any global analysis is the availability of enough
measurements to constrain the model parameters. In case of the top
sector we confront 22 dimension-6 operators with a much larger number
of available measurements shown in Tabs.~\ref{tab:data1}
and~\ref{tab:data2}.  The data forms two main parts, measurements of
the leading top pair production process mediated by QCD couplings and
measurements of processes including a weak coupling. The latter
include single top production as well as associated top pair
production with electroweak bosons. Because all our measurements are
unfolded to the level of stable top quarks, and because there is
essentially only one top decay channel leading to a universal
branching ratio of one, we can assume SM-like top decays for all
measurements except for the $W$ helicity fractions in top
decays. Observables combining top production and decay play a special
role in the SMEFT interpretation, because they probe features of
operators not accessible in top production
alone~\cite{AguilarSaavedra:2010zi,Aguilar-Saavedra:2017nik,deBeurs:2018pvs,Neumann:2019kvk}. In
our analysis, the charge asymmetry described in
Sec.~\ref{sec:eft_tbar} plays a similar role in probing operators, even though it is based
on kinematics of fully reconstructed top quarks.

In terms of the Wilson coefficients of Sec.~\ref{sec:eft} all
our rate observables have the form
\begin{align}
\sigma = \sigma_\text{SM} + \sum_k \frac{C_k}{\Lambda^2} \; \sigma_k + \sum_{k,l} \frac{C_k C_l}{\Lambda^4} \; \sigma_{kl} \,,
\label{eq:fit_input}
\end{align}
where $\sigma_\text{SM}$ is the SM prediction, $\sigma_k$ are
contributions arising from the interference of a single dimension-6
operator with the SM, and $\sigma_{kl}$ arise from the interference of two diagrams containing one operator each.  Technically, $\sigma_k$
and $\sigma_{kl}$ are the theory input which
\textsc{Madgraph5\_aMC@NLO}~\cite{Alwall:2014hca} provides at NLO QCD
accuracy.

The SMEFT Lagrangian leads to two main kinds of
corrections, as illustrated for top pair production in Tab.~\ref{tab:he-scaling}: operators which
change the high-energy behavior of the process through an additional
energy dependence of the kind $s/\Lambda^2$ and those which scale
merely like $v^2/\Lambda^2$ or $m_tv/\Lambda^2 \approx y_t v^2/\Lambda^2$ compared to the SM. For
the latter the leading observables are rate measurements or the total
cross section, because they offer the best statistics and often minimize
theoretical uncertainties. From global Higgs-electroweak analyses we
know that a modified momentum dependence can be constrained most
efficiently by high-energy tails of kinematic distributions or
 simplified template cross sections~\cite{Butter:2016cvz,deBlas:2016ojx,Ellis:2018gqa,Almeida:2018cld,Biekotter:2018rhp,Kraml:2019sis}.~\footnote{Note
  that for a distribution to constrain a dimension-6 contribution in
  this phase space region it is not necessary that we actually observe
  the SM process in the same phase space
  region~\cite{Biekotter:2018rhp}.} Similarly, we know that for many
kinematic distributions the few bins with the highest momentum
transfer include the relevant information on individual operators, whereas
for several operators with a different high-energy behavior there
often exist several relevant regimes~\cite{Brehmer:2019gmn}.

Unlike in the Higgs sector, cross section measurements in the
top sector are reported such that we can easily compare them to 
parton-level predictions. Kinematic distributions are typically
reported as normalized distributions, \ie they integrate to one and
can be combined with total rate measurements without
double-counting information. A problem arises when we include operator contributions to the distribution in the numerator and
to the rate in the denominator. In this case the normalized bin
entries entering our fit become correlated and develop a distinct
non-linear behavior.

\subsection{SFitter analysis}
\label{sec:sfitter}

For our global LHC analysis we use the \textsc{SFitter}
framework~\cite{Lafaye:2007vs,Lafaye:2009vr}, which focuses on a proper treatment of uncertainties in a conservative frequentist approach. We extract the statistical uncertainties and a leading set of up to $\sim20$
 systematic uncertainties for each experiment and simulate a Gaussian
shape of the completely exclusive likelihood for statistics and
systematics. For the systematic uncertainties we also allow for
correlations within the same experiment, collider energy scale, and
top signature. This applies for example to jet uncertainties like the
jet energy scale or the jet efficiencies. An exception is the
uncertainty on the luminosity, which we correlate for all channels and
both experiments.
In order to simplify the treatment of uncertainty correlations, we fit only one observable from each experimental analysis, and we never take two measurements of the same observable at the same energy scale. Moreover, for total rates and charge asymmetries we fit only two observables, one for each collider energy, obtained with weighted averages of the measurements performed with different experiments and datasets. 

In addition to the experimental sources of uncertainties, theoretical
error bars reflect missing higher orders in the perturbative
series. Precise predictions are crucial to extract any Lagrangian
parameter from LHC rate measurements. We rely on NLO predictions for
$t \bar{t}$ and single top observables in SMEFT using
\textsc{Madgraph5\_aMC@NLO}, while we use LO QCD predictions for the
statistics-limited $t\bar{t}V$ rates. For the central values we add
NNLO $K$-factors in the SM, whenever available, to account for the most precise
predictions available (see Tabs.~\ref{tab:data1},
\ref{tab:data2}). This means we assume that the operator
contributions scale like the SM rate beyond NLO.  The only exception
is the charge asymmetry, that does not scale multiplicatively with higher-order corrections. In this case we fit the sum of the most precise available SM prediction and the new physics corrections at (N)LO from \textsc{Madgraph5\_aMC@NLO}.

The theoretical uncertainties are obtained by varying the
renormalization and factorization scales by a factor of two around the
respective central scale choices. Since technically we cannot distinguish the uncertainties due to operator effects, we use the scale uncertainties on the SM prediction from our NLO simulations as an overall theory uncertainty on the observable. This gives for instance a 12\%
uncertainty for the combined SM and dimension-6 $t\bar{t}$ and
$t\bar{t}V$ rates. Regarding higher-order corrections, our estimate based on the NLO scale dependence in
the presence of dimension-6 effects is likely to be conservative
for small new physics effects~\cite{Czakon:2017wor}. Since NLO corrections to the non-SM contributions are included in our simulations, we generally expect QCD effects beyond NLO to be moderate. Exceptions occur in single top production, where SM QCD effects first occur beyond tree level, or in bins of kinematic distributions near the endpoints of the spectrum. Whenever the scale uncertainty in the NLO simulation for such rate measurements happens to be very small we replace it by a minimum of 10\%.  Similarly, when strong cancellations of scale uncertainties occur in normalized kinematic distributions, we replace the theoretical uncertainty in
each bin by 2\% whenever the scale variation drops below this
level~\cite{Czakon:2016ckf,Czakon:2018nun}. 

In \textsc{SFitter} all theoretical uncertainties are modelled as a
flat likelihood within the quoted error band. This applies to the
theoretical uncertainties on the signal as well as the theoretical
uncertainties on the background, quoted in the experimental
analyses. If we combine them with Gaussian experimental uncertainties
in a profile likelihood this leads to the \textsc{RFit}
scheme~\cite{Hocker:2001xe}.  Theoretical uncertainties are generally
uncorrelated unless they describe the same fiducial volume at the same
collider energy. This also includes the theoretical errors for
individual bins in a kinematic distribution, which we
assume to be uncorrelated.  Uncertainties from the limited precision of parton
densities are evaluated in analogy to the theoretical uncertainties
reflecting the missing higher orders in the hard process. We evaluate
them using a set of 209 predictions from the \textsc{Nnpdf3.0} NLO
set with $\alpha_s(m_Z) = 0.118$~\cite{Ball:2017nwa}, 
the \textsc{Mmht2014Nlo} set at 68\%~CL~\cite{Harland-Lang:2014zoa},  
and the \textsc{Ct14Nlo} set~\cite{Dulat:2015mca}. 
A typical error bar from the parton distribution functions (PDFs) is 6\% for the $t\bar{t}$ or $t\bar{t}V$ rates.
Because we assume a flat likelihood for these PDF uncertainties, the
profile likelihood combination of higher-order and PDF uncertainties
adds the two error bars linearly.

To probe the parameter space we rely on Markov chains, similar to
Ref.~\cite{Butter:2016cvz}, rather than the numerically more complex
toy measurements used in Ref.~\cite{Biekotter:2018rhp}. To cover the
full 22-dimensional parameter space we use up to 2000 Markov chains
giving up to 400 million parameter points. This defines our fully
exclusive likelihood which we then profile down to two and one
relevant dimensions.

\section{Top pair features}
\label{sec:features}

Before entering a global analysis of the top sector we study
some of the underlying features in detail. This is essential for the
top-pair side of the analysis. Its unique challenges are very
different from the electroweak-Higgs
sector~\cite{Butter:2016cvz,Ellis:2018gqa,Almeida:2018cld,Biekotter:2018rhp}
and the single top sector discussed in Sec.~\ref{sec:global_single}.

In top pair production the operator $O_{tG}$ induces large corrections to the total
and differential rates, as it is the only
operator modifying the gluon-induced production process. We discuss
its known and expected behavior in Sec.~\ref{sec:ctg} and roughly
estimate the expected sensitivity of our global fit.

The new feature in top pair production is the large set of
four-quark operators affecting the partonic process $q\bar q\to
t\bar t$.  Fourteen such operators, different in their QCD and
electroweak structure, contribute to one and the same process. Since top
pair production is a QCD process, most of its observables average
or sum over the electroweak properties of the external particles. To distinguish these operators we rely on the observables
\begin{align}
\left\{ \; \sigma_\text{tot} , 
      \frac{d \sigma}{d m_{t\bar t}} , 
      \frac{d \sigma}{d p_{T,t}} ,
      \frac{d \sigma}{d \Delta y_{t\bar{t}}} , 
      A_C
\; \right\} 
\end{align}
supplemented by $p_{T,t}$ and $m_{t\bar t}$ distributions
in the boosted region. In Secs.~\ref{sec:ttbar_u_vs_d}
to~\ref{sec:ttbar_L_vs_R_NLO} we will study how the gauge and chiral
structure of four-quark operators can be resolved by dedicated
$t\bar{t}$ measurements. This allows us to break some of the flat
directions in model space already at LO in the EFT analysis, where only the tree-level interference between the SM
and the dimension-6 operators are considered.

Finally, in Sec.~\ref{sec:quadratic} we will study the effect of
dimension-6-squared contributions on the sensitivity to operators. We  will see that the flat directions turn into compact circles which allow us to derive
more stringent limits on individual operators.

\subsection{Event kinematics}
\label{sec:ctg}

Before we study the effects of specific dimension-6 operators on top
pair production, we roughly estimate the reach of our analysis for
operators affecting the total rate and for operators affecting the
event kinematics.  The top-gluon dipole operator $O_{tG}$ is the only
top EFT contribution to the leading partonic process $gg\to t\bar{t}$.
We therefore expect a high sensitivity to $O_{tG}$ in inclusive top
pair production. In contrast, four-quark operators contribute only to
the $q\bar{q} \to t\bar{t}$ process, which is subleading, but enhanced
at high energies. We thus expect the best sensitivity to four-quark
operators in tails of kinematic distributions.

In our analysis, all distributions are normalized to the total
rate. These normalized distributions are direct probes of the dynamics
of operator contributions relative to the SM. To illustrate this
important point, we compare the event kinematics of the dipole operator
$O_{tG}$ with the four-quark operator $O_{tu}^8$. The normalized
$m_{t\bar{t}}$ distribution depends on these two operators
as (neglecting $O_{tG}-O_{tu}^8$ interference)
\begin{align}\label{eq:mtt-dist}
\frac{1}{\sigma}\frac{d \sigma}{d m_{t\bar{t}}} & \approx \frac{\sigma_{\rm SM}(m_{t\bar{t}})}{\sigma_{\rm SM}(2m_t)} \Bigg(1 + \mathcal{O}\left(m_t v - m_t v \right)\frac{C_{tG}}{\Lambda^2} +  \mathcal{O}\left(m_{t\bar{t}}^2 - (2m_t)^2\right) v^2 \frac{|C_{tG}|^2}{\Lambda^4}\\\nonumber
& \hspace*{2.8cm} + \mathcal{O}\left(m_{t\bar{t}}^2 - (2m_t)^2\right)\frac{C_{tu}^8}{\Lambda^2} 
 + \mathcal{O}\left(m_{t\bar{t}}^4 - (2m_t)^4\right)\frac{|C_{tu}^8|^2}{\Lambda^4} \Bigg)\,.
\end{align}
Here $2m_t$ denotes the invariant mass close to the
production threshold, which dominates in the total cross
section, while $m_{t\bar{t}}$ can be much higher in differential distributions. Since the $O_{tG}-$QCD interference does not feature an
energy enhancement, it cancels almost completely in normalized
distributions. Kinematic distributions are therefore expected to lead
to relatively weak constraints driven by the $|C_{tG}|^2$ term. In
contrast, the four-quark contribution of $O_{tu}^8$ features an energy
enhancement already at $\mathcal{O}(\Lambda^{-2})$. This leads to a
good sensitivity at high energies, despite the relative suppression by
the parton luminosity. Total rates and distributions are thus
complementary in probing dipole operators and four-quark operators. Notice that in our numerical analysis we keep the full operator contributions in the normalization of distributions.

\begin{figure}[t]
\centering
\includegraphics[width=.49\textwidth]{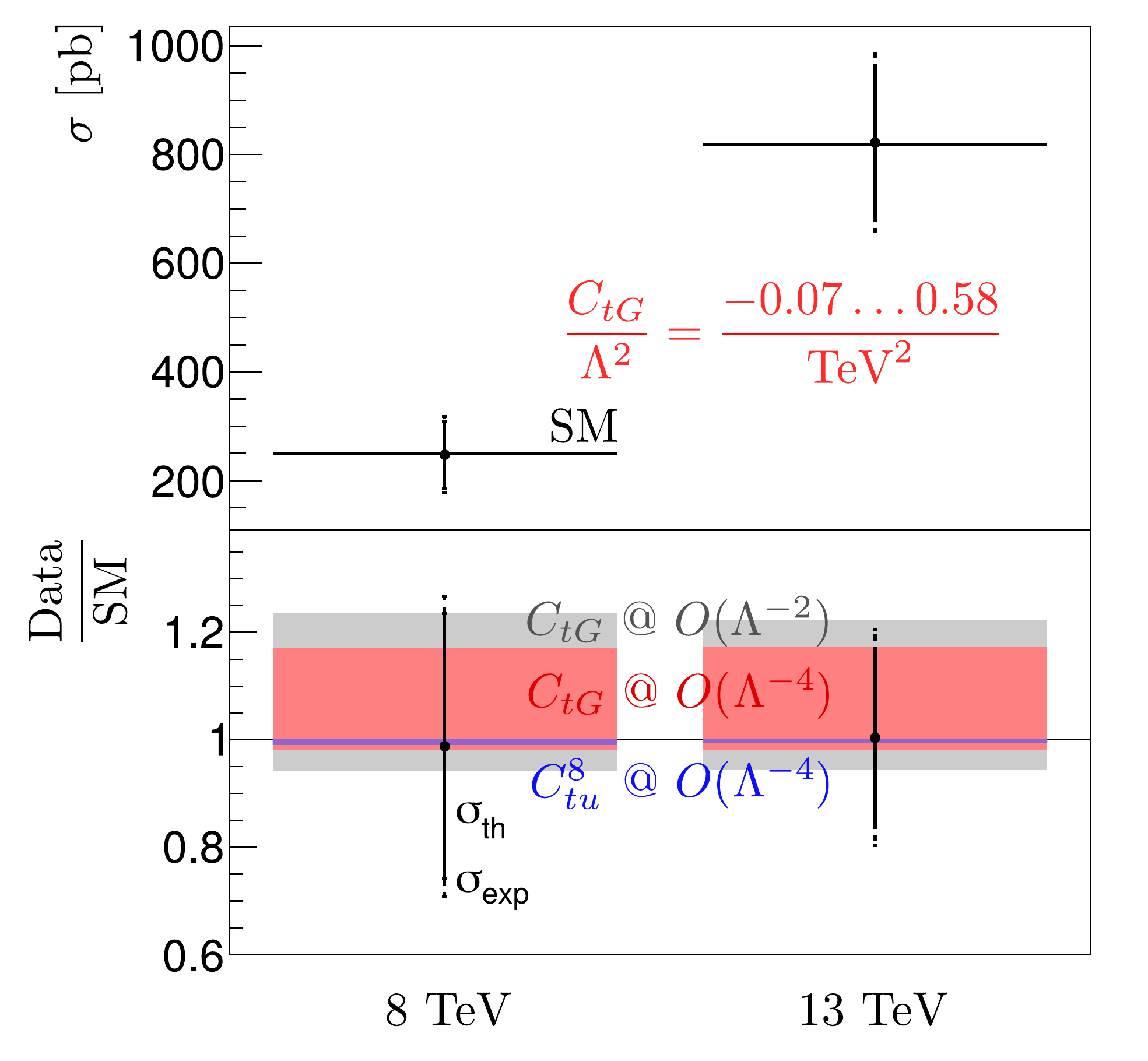}
\includegraphics[width=.49\textwidth]{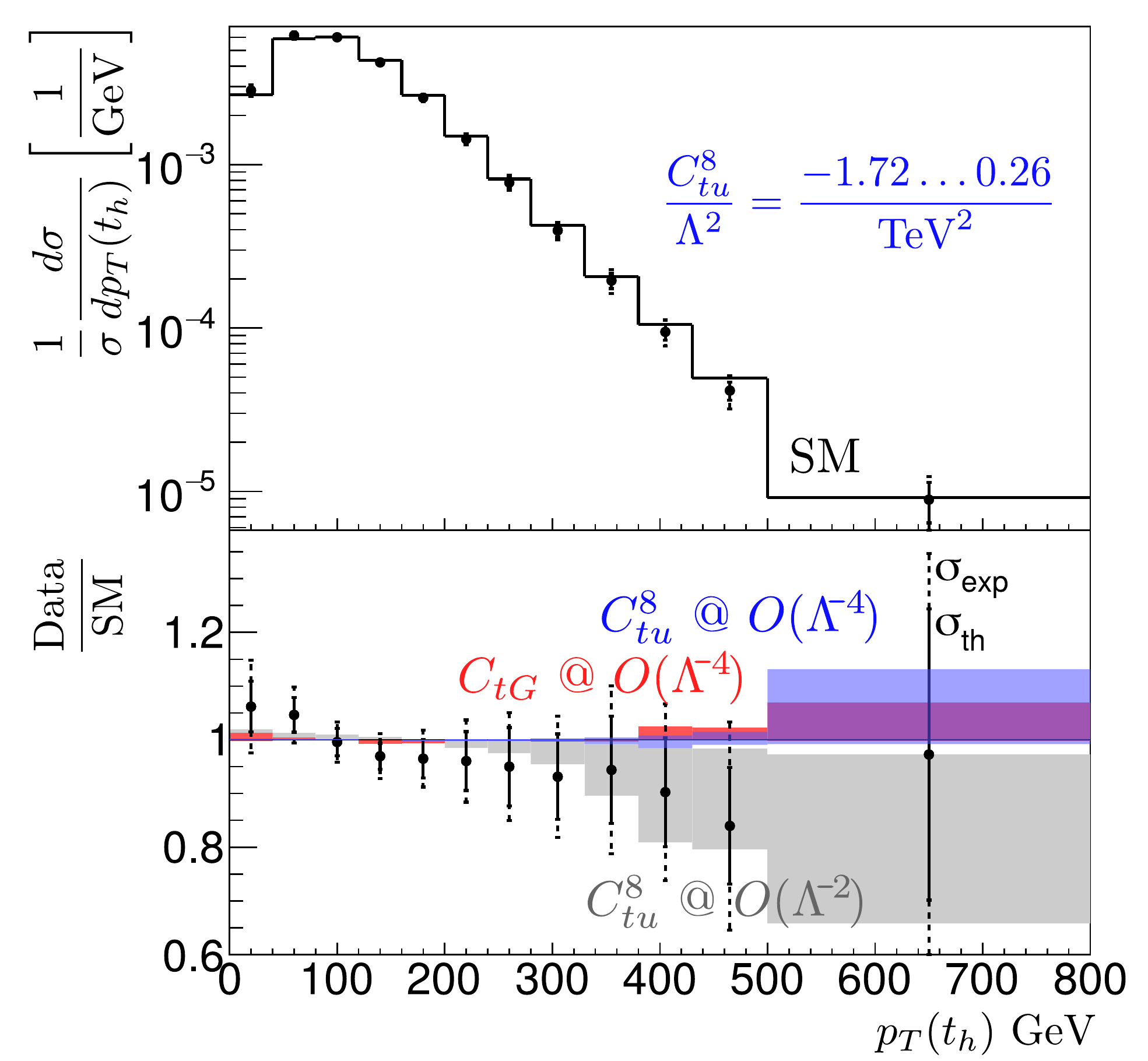}
\caption{Contribution of $O_{tG}$ and $O_{tu}^8$ to total $t\bar{t}$
  rates (left) and the normalized $p_{T}$ distribution of a hadronically decaying top (right). The
  shaded regions correspond to the $68\%$ CL from a simultaneous fit
  to the two rates and the distribution. The grey shaded regions show
  the contribution from $O_{tG}$ (rates) and $O_{tu}^8$ (distribution)
  at order $\Lambda^{-2}$.  The red and blue shaded regions show the
  contribution from $O_{tG}$ and $O_{tu}^8$ to order $\Lambda^{-4}$,
  respectively.}
\label{fig:dist}
\end{figure}

The reach of measurements of total cross sections at
8~TeV and 13~TeV and a $p_{T}$ distribution at 13~TeV is estimated for $O_{tG}$ and $O_{tu}^8$ in
Fig.~\ref{fig:dist}.
%

The upper-left panel shows the averaged cross section
measurements with their combined uncertainties. Similarly, the
upper-right panel shows the normalized $p_T$ distribution for the
hadronically decaying top at 13~TeV from Ref~\cite{Sirunyan:2018wem} (see Tab.~\ref{tab:data1}). The two lower panels show the
relative deviations from the SM prediction and the 68\% CL limits from
a combined analysis of $C_{tG}$ and $C_{tu}^8$ to the small data set consisting of only the observables shown in Fig.~\ref{fig:dist}. The grey panels show the
result from the new physics interference at order $\Lambda^{-2}$ for
$C_{tG}$ in terms of the total rates and for $C_{tu}^8$ in terms of the
kinematic distribution, corresponding to the 68\% CL ranges
\begin{align}
C_{tG}/\Lambda^2 \in [-0.19,0.78]/\tev^2
\qquad
C_{tu}^8/\Lambda^2 \in [-6.77,-0.57]/\tev^2\;.
\end{align}
The limits are slightly asymmetric, because the top quarks in the
normalized distribution are softer than in the SM expectation.

For the red ($O_{tG}$) and blue ($O_{tu}^8$) shaded regions we also
include the contributions to order $\Lambda^{-4}$ and find at 68\% CL
\begin{align}
C_{tG}/\Lambda^2 \in [-0.07,0.58]/\tev^2
\qquad
C_{tu}^8/\Lambda^2 \in [-1.72,0.26]/\tev^2\;.
\label{eq:limit_estimate}
\end{align}
While these limits are just based on a small fit to three observables,
they give us an intuition of what to expect from our fit. For $C_{tG}$
an expected range around $\Lambda/\sqrt{|C_{tG}|} = 1.3$~TeV saturates
the error bars of the leading total rate measurement, while for the
four-quark operator $C_{tu}^8$ values around $\Lambda/\sqrt{|C_{tu}^8|}
= 0.7$~TeV can be expected from this one kinematic
distribution. Comparing these limits to the kinematic range probed by
the $p_T$ distribution in Fig.~\ref{fig:dist}, we see that the effective theory interpretation is valid for an underlying theory that does not predict propagating new states at the LHC and is not too strongly coupled.

\begin{figure}[t]\centering
\includegraphics[width=.45\textwidth]{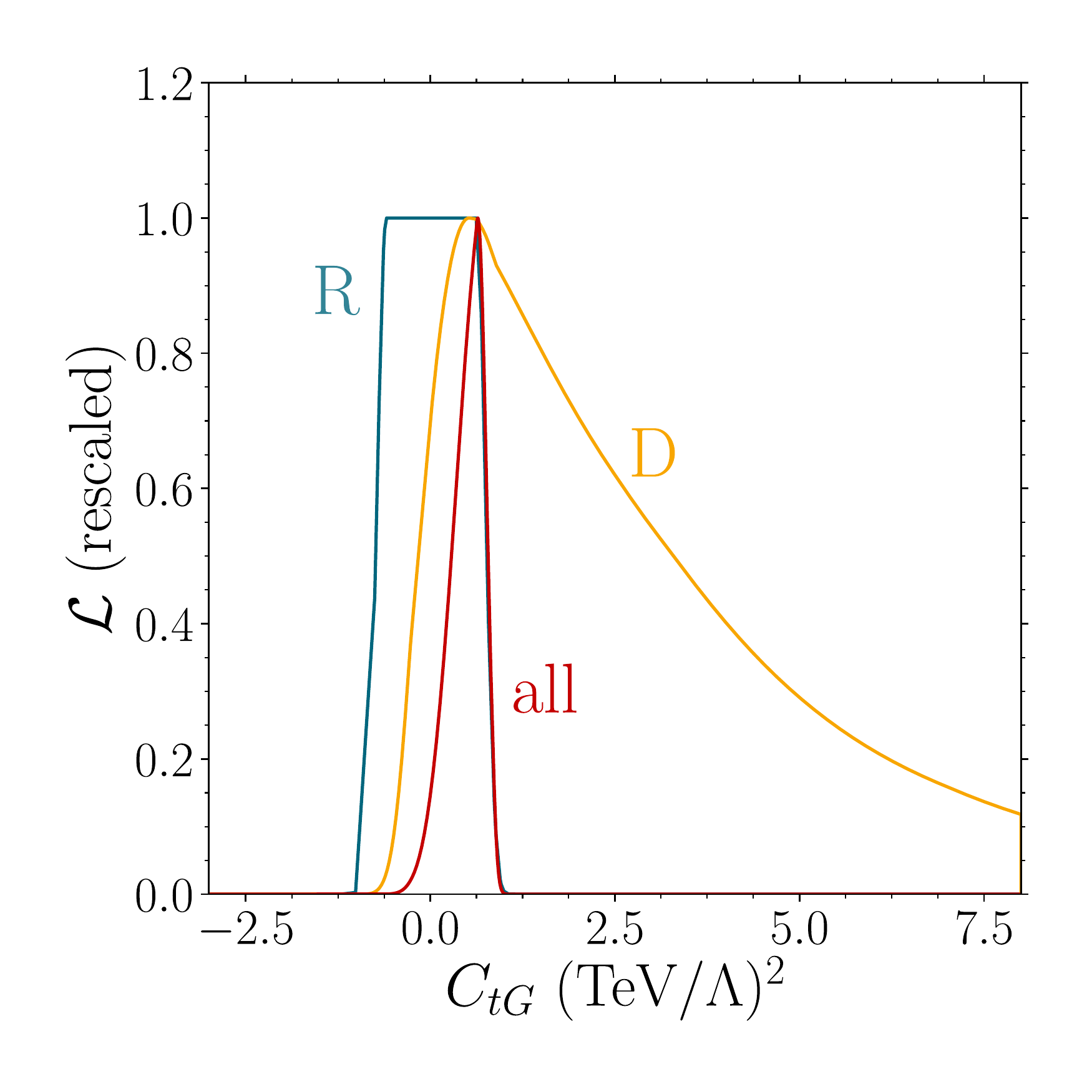}
\includegraphics[width=.45\textwidth]{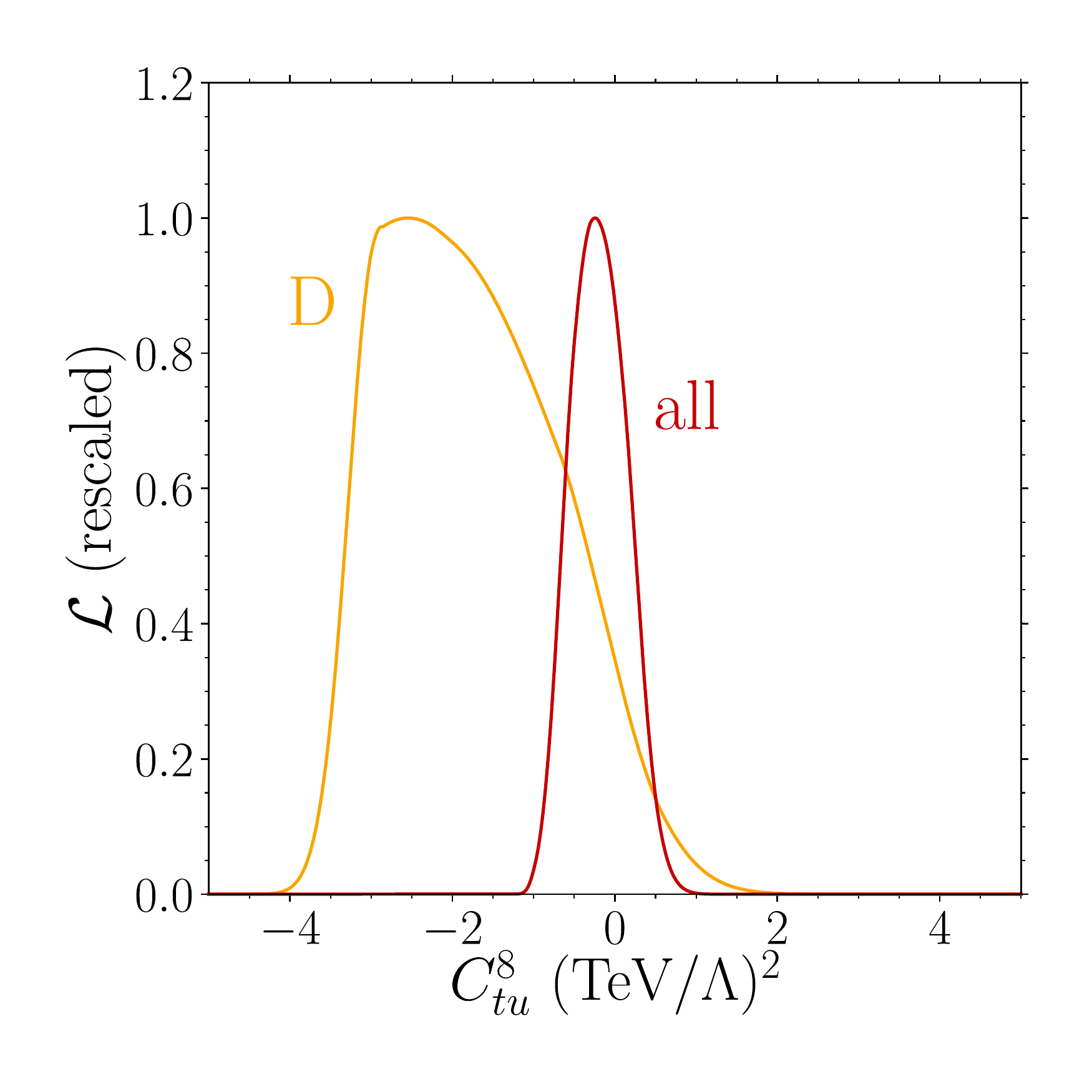}
\caption{Normalized likelihood as a function of $C_{tG}$ (left) and a
  $C_{tu}^8$ (right) in individual fits at LO to order
  $\Lambda^{-2}$. We show fits to the R (blue) and D (yellow) observable sets of Tab.~\ref{tab:data1}, and to all $t\bar t$ observables in Tab.~\ref{tab:data1} (red). In the right panel, we do not show the R likelihood, as it is 1 for all the values of $C_{tu}^8$ in the displayed range. }
\label{fig:ctG}
\end{figure}

In Fig.~\ref{fig:ctG} we show how total rates and normalized kinematic
distributions lead to very different likelihood distributions.
%
%
 First, rate measurements alone have a strong constraining power on $C_{tG}$
compared to four-quark operators, due to the SMEFT correction being
relatively large. The likelihood from normalized distribution is strongly asymmetric: negative values of $C_{tG}$ are strongly limited by the physical requirement that the bin content of all the measured
distributions remains positive. Positive values of $C_{tG}$, on the other hand, are less constrained as discussed in the previous section.
More specifically, let  $n_k$ be the number of entries in $k$-th bin of a normalized distribution. As a function of the SM ($n_k^{SM}$) and SM-$C_{tG}$ interference ($n_k^{\rm int}$) contributions, it scales as
\begin{align}
\frac{n_k^\text{SM} + n_k^\text{int} C_{tG}/\Lambda^2}
     {\sum_l \left( n_l^\text{SM} +  n_l^\text{int} C_{tG}/\Lambda^2 \right)} 
\stackrel{C_{tG}\to\infty}{\longrightarrow}
\frac{n_k^\text{int}}
     {\sum_l n_l^\text{int}}  \; .
\end{align}
For large values of $C_{tG}/\Lambda^2\to\infty$ the
normalized bin content becomes a constant. As noted in Eq.~\ref{eq:mtt-dist}, $C_{tG}$ is characterized by a kinematic behavior very similar to that of the SM, which leads to values $n_k^{\rm int}/\sum_l n_l^{\rm int}$ generally compatible with $n_k^{\rm SM}/\sum_l n_l^{\rm SM}$.   As a consequence the
corresponding log-likelihood also converges to constant $>0$.  This asymptotic behavior is not
observed once quadratic terms are included, because the sensitivity to
$C_{tG}$ is enhanced by $sv^2/\Lambda^4$ in high-energy bins, see
Eq.~\eqref{eq:mtt-dist} and Tab.~\ref{tab:he-scaling}. Combining all $t\bar t$ measurements the likelihood recovers a fairly symmetric form,
but with a distinct shift of the minimum towards small positive values
of $C_{tG}$.

For comparison, the asymptotic behavior in the linear fit is not
observed for an interfering four-quark operator like $C_{tu}^8$,
because it induces a significantly different shape in the kinematic
distributions compared to the SM, scaling as $s/\Lambda^2$. Large
values of the Wilson coefficients are therefore strongly disfavored by
at least one of the bin measurements that drive the log-likelikood towards zero. In the right panel of
Fig.~\ref{fig:ctG}, we show that $C_{tu}^8$ is well constrained by
measurements of normalized distributions. Total rates have
little impact on the fit results.

\subsection{Incoming up versus down quarks}
\label{sec:ttbar_u_vs_d}

The set of four-quark operators laid out in Eq.~\eqref{eq:ops_llrr} and
Eq.~\eqref{eq:ops_lr} span all possible assignments of the quark fields
to representations the SM symmetry groups:
\begin{enumerate}
\item chirality of the light quark and top quark currents;
\item left-handed currents: singlet or triplet under $SU(2)_L$;
\item right-handed currents: up- or down-type light quarks;
\item singlet or octet color contraction of the currents.
\end{enumerate}
Top pair production through strong interactions is not sensitive at parton level to the nature of the incoming quarks, \ie, questions 2 and 3. However, up-type and down-type quarks in the initial state are distinguished by the parton densities. The relative $u\bar{u}$ and
$d\bar{d}$ contributions to the $t\bar{t}$ final state are determined
by
\begin{align}
r(x) = \frac{f_u(x)f_{\bar u}(s/(xS))}{f_d(x)f_{\bar d}(s/(xS))}\,.
\end{align}
Here $f_p(x,s)$ denotes the usual parton distribution of parton $p$
with momentum fraction $x$ of the energy $\sqrt{s}/2$ in the proton.
$\sqrt{S}$ is the hadronic CM energy, and we suppressed the
factorization scale choice.  Around the valence quark maximum
$x\approx 0.1$ the ratio becomes $r \approx 2$. For most
observables used in our analysis, the ratio integrated over the
relevant phase-space region varies roughly in the range
\begin{align}\label{eq:lumi-ratio}
1.5\lesssim r \lesssim 3 \,.
\end{align}
In what follows we refer to $r$ as (roughly) the relative contribution up partonic up- and down-quark contributions to an observable. In what follows we discuss how the isospin of the incoming quarks can be disentangled in a minimal EFT analysis of $t\bar t$ production, neglecting quadratic EFT contributions and NLO QCD corrections.

Let us consider pairs of four-quark operators that are only distinguished by the nature of initial
quarks:  if the latter are right-handed, as in $O_{tu}^8$ and $O_{td}^8$, $t\bar{t}$ observables depend on the combination of Wilson
coefficients
\begin{align}\label{eq:weak-ud}
r\, C_{tu}^8 + C_{td}^8 \approx 2\, C_{tu}^8 + C_{td}^8\,.
\end{align}
If the initial quarks are left-handed, their nature is only distinguished by a singlet versus triplet $SU(2)$ structure, as in
$O_{Qq}^{1,8}$ and $O_{Qq}^{3,8}$.  In this case the typical combination is
\begin{align}\label{eq:weak-13}
\big(r + 1\big)C_{Qq}^{1,8} + \big(r - 1\big)C_{Qq}^{3,8} \approx 3\,C_{Qq}^{1,8} + C_{Qq}^{3,8}\,.
\end{align}
The numerical estimate $r\approx2$ holds for the bulk of the phase space in top pair production. On the other hand, measurements that select highly boosted tops can probe higher parton momentum fractions $x$ and  larger ratios $r$, thus constraining different directions in the EFT space. 

\begin{figure}[t]
\centering
\includegraphics[width=.49\textwidth]{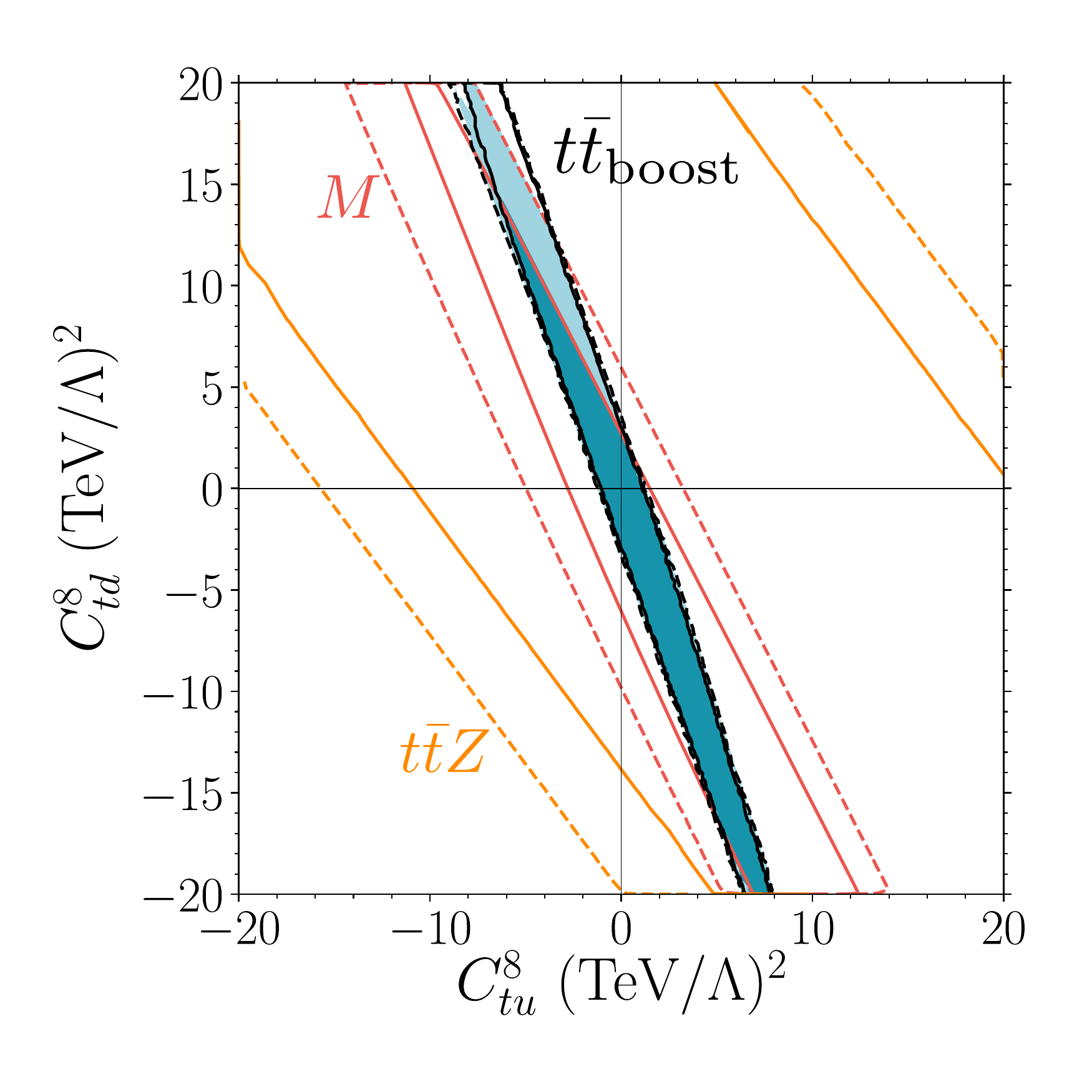}
\includegraphics[width=.49\textwidth]{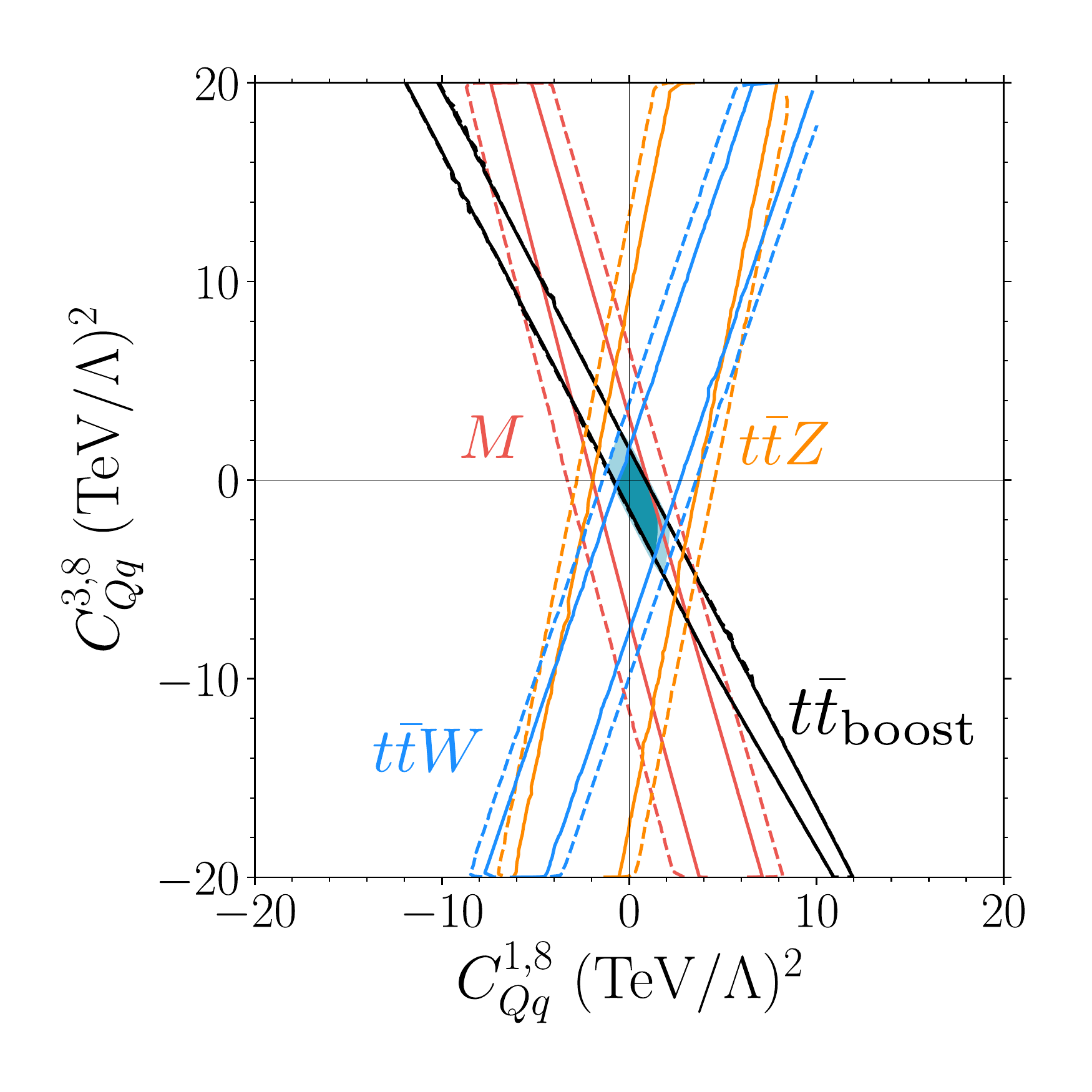}
\caption{Up-type versus down-type (left) and weak isospin (right) effects
  of four-quark operators from LO two-parameter fits to order
  $\Lambda^{-2}$. Solid and dashed lines mark
  the Gaussian equivalent of $\Delta\chi^2 = 1,4$ from fits to: set $M$ of $t\bar t$ observables (red, see Tab.~\ref{tab:data1}), highest-energy bins of a $t\bar t$ distribution in the boosted regime,  $t\bar{t}Z$ rates (orange), $t\bar{t}W$ rates (blue). The shaded areas show the combined fit.}
\label{fig:uvsd}
\end{figure}

To illustrate this effect, Figure~\ref{fig:uvsd}
shows bounds on these two pairs of operators obtained from two-dimensional
likelihood fit of top--anti-top observables to LHC data.

 The red contours use set 'M' of Table~\ref{tab:data1}, that contains rates and normalized $m_{t\bar t}$ distributions. These observables are most sensitive to quark-antiquark contributions around the maximum of the parton
distributions in $x$, where $r\approx 2$. They leave the directions
$(1,-2)$ for $(C_{tu}^8,C_{td}^8)$ and $(1,-3)$ for
$(C_{Qq}^{1,8},C_{Qq}^{3,8})$ essentially unbounded, as is expected
from the relations in Eqs.~\eqref{eq:weak-ud} and~\eqref{eq:weak-13}.

Boosted top pair production~\cite{Englert:2016aei} probes larger
momentum fractions $x$ and hence larger ratios $r$. The black contours in Fig.~\ref{fig:uvsd} show the likelihood obtained by fitting the last bins of a $p_{T,t}$ distribution in
the boosted regime, $p_{T,t} > 500\,\gev$. The blind
directions of this fit are tilted compared to the previous
analysis. They run roughly along $(1,-3)$ for $(C_{tu}^8,C_{td}^8)$ and along
$(1,-2)$ for $(C_{Qq}^{1,8},C_{Qq}^{3,8})$, which corresponds to
$r\approx 3$. Adding boosted top--anti-top observables thus breaks the blind directions in inclusive top--anti-top production, but only mildly.

To better resolve the weak gauge structure, we include $t\bar{t}Z$ and $t\bar{t}W$ production in the fit. As mentioned in Sec.~\ref{sec:eft_ttW_ttZ}, the radiation of the gauge boson changes the relative contributions of operators with different weak gauge structure. For operators with right-handed light quarks, only $t\bar{t}Z$ production is relevant. At $O(\Lambda^{-2})$, the contribution to the $t\bar{t}Z$ rate depends on the Wilson coefficients as
\begin{align}\label{eq:ttZ-RR}
\sigma_{t\bar{t}Z}^{\rm int} & = \Big(rC_{tu}^8 + C_{td}^8\Big)\sigma_{ff}  + \Big(r |g_{uZ}^R|^2\,C_{tu}^8 + |g_{dZ}^R|^2 C_{td}^8\Big)\sigma_{ii}\\\nonumber
& \quad + \Big(r g_{uZ}^R \,C_{tu}^8 + g_{dZ}^R C_{td}^8\Big)\sigma_{if}\,.
\end{align}
The three terms correspond to final--state radiation ($\sigma_{ff}$), initial--state radiation ($\sigma_{ii}$), and interference between initial-- and final--state radiation ($\sigma_{if}$) of the $Z$-boson. The quark couplings to the $Z$-boson are defined as $g_{uZ}^R = - \frac{2}{3}s_w^2$, $g_{dZ}^R = \frac{1}{3}s_w^2$, $g_{uZ}^L = \frac{1}{2} - \frac{2}{3}s_w^2$, and $g_{dZ}^L  = - \frac{1}{2} + \frac{1}{3}s_w^2$. The term $\sigma_{ff}$ includes contributions with $Z$ couplings to left- and right-handed top quarks. By comparing with Eq.~\eqref{eq:weak-ud}, we see that $t\bar{t}Z$ production probes a different direction in the $C_{tu}^8-C_{td}^8$ parameter space than inclusive $t\bar{t}$ production.

Operators with left-handed quarks and different weak isospin can be probed in both $t\bar{t}Z$ and $t\bar{t}W$ production. In $t\bar{t}Z$ production, they contribute at $O(\Lambda^{-2})$ as\footnote{Note that $\sigma_{ff}$, $\sigma_{ii}$ and $\sigma_{if}$ are generic symbols for the contributions to the total $t\bar t Z$ cross section, so their meaning is different in Eq.~\ref{eq:ttZ-RR} and in Eq.~\ref{eq:ttZ-LL}.}
\begin{align}\label{eq:ttZ-LL}
\sigma_{t\bar{t}Z}^{\rm int} & = \Big((r+1) C_{Qq}^{1,8} + (r-1) C_{Qq}^{3,8}\Big)\sigma_{ff}\\\nonumber
& \quad + \Big(\big(r |g_{uZ}^L|^2 + |g_{dZ}^L|^2\big) C_{Qq}^{1,8} + \big(r |g_{uZ}^L|^2 - |g_{dZ}^L|^2\big) C_{Qq}^{3,8} \Big)\sigma_{ii} \\\nonumber
& \quad + \Big(\big(r g_{uZ}^L + g_{dZ}^L\big) C_{Qq}^{1,8} + \big(r g_{uZ}^L - g_{dZ}^L\big) C_{Qq}^{3,8} \Big)\sigma_{if}\,.
\end{align}
In $t\bar{t}W$ production, the parton luminosity for operators with different weak isospin structure is the same, since all operators with left-handed light quarks contribute to the same partonic processes, dominantly $u\bar{d} \to t\bar{t}W^+$ and $d\bar{u} \to t\bar{t}W^-$, respectively. Associated $t\bar{t}W^+$ production probes the following direction in the $C_{Qq}^{1,8} - C_{Qq}^{3,8}$ plane at $O(\Lambda^{-2})$,
\begin{align}
\sigma_{t\bar{t}W^+}^{\rm int} & = \big(C_{Qq}^{1,8} + C_{Qq}^{3,8}\big)\sigma_{uu} + \big(C_{Qq}^{1,8} - C_{Qq}^{3,8}\big)\sigma_{dd} + C_{Qq}^{3,8}\sigma_{ud}\\\nonumber
& \approx C_{Qq}^{1,8}\big(\sigma_{uu} + \sigma_{dd}\big) + C_{Qq}^{3,8}\sigma_{ud}\,.
\end{align}
Here $\sigma_{uu}$ and $\sigma_{dd}$ denote cross section contributions where the $W^+$ boson is radiated off an incoming anti-down or up quark, which probes the operators $O_{Qq}^{1,8}$ and $O_{Qq}^{3,8}$ through their $(\bar{u}u)(\bar{t}t)$ and $(\bar{d}d)(\bar{t}t)$ contributions, respectively. In $\sigma_{ud}$ the $W^+$ is radiated off a anti-bottom quark in the final state, probing $O_{Qq}^{3,8}$ through its $(\bar{d}u)(\bar{t}b)$ contribution. The contribution of $O_{Qq}^{3,8}$ largely cancels between $\sigma_{uu}$ and $\sigma_{dd}$, so that the total cross section is sensitive to $O_{Qq}^{3,8}$ mostly through final state radiation. Very similar considerations hold for $t\bar{t}W^-$ production. In summary, $t\bar{t}W$ production probes a third direction in the $C_{Qq}^{1,8} - C_{Qq}^{3,8}$ plane, in addition to $t\bar{t}$ and $t\bar{t}Z$ production.

In Fig.~\ref{fig:uvsd}, we show the impact of cross section
measurements at 13 TeV for $t\bar{t}Z$ (orange) and $t\bar{t}W$
(blue). For the $RR$ operators $C_{tu}^8$ and $C_{td}^8$ (left panel),
$t\bar{t}Z$ production probes indeed a different direction than
inclusive $t\bar{t}$ production, leaving a band along $(1,-0.8)$
unconstrained. However, the sensitivity of the $t\bar{t}Z$ cross
section to $RR$ operators is much lower than in $t\bar{t}$
production. In the combined fit, shown as a blue area, the remaining
blind direction is thus aligned with boosted top pair
production. Differential $t\bar{t}Z$ distributions can help to resolve
this direction, featuring a better sensitivity to four-quark operators
at high energies, similar to $t\bar{t}$ production~\cite{CMS:2019too}.

The situation is different for $LL$ operators, as we show in the right
panel of Fig.~\ref{fig:uvsd}. Associated $t\bar{t}Z$ and $t\bar{t}W$
production probe similar directions in $(C_{Qq}^{1,8},C_{Qq}^{3,8})$,
leaving blind directions along roughly 
$(1,4.7)$ and $(1,2.8)$  respectively. Remarkably, the sensitivity of $t\bar{t}Z$ and $t\bar{t}W$
cross sections to $LL$ operators is comparable to that of differential
$t\bar{t}$ distributions. In $t\bar{t}W$ production, both SM and
dimension-6 contributions are induced by quark-antiquark
interactions. Compared to the SM rate, effects of $LL$ operators are
thus larger than in $t\bar{t}$ production, which is dominated by
gluon-gluon interactions. In $t\bar{t}Z$ production the sensitivity to
$LL$ operators is much larger than for $RR$ operators. This is due to
the different $Z$-couplings to left- and right-handed quarks,
$|g_{uZ}^L|/|g_{uZ}^R| \approx 2.4$ and $|g_{dZ}^L|/|g_{dZ}^R| \approx
5.8$, which affect the operator contributions, see
Eqs.~\eqref{eq:ttZ-RR} and \eqref{eq:ttZ-LL}. This makes $t\bar{t}W$
and $t\bar{t}Z$ production valuable probes of $LL$ four-quark
operators, complementary to $t\bar{t}$ production.

\subsection{Top chirality from charge asymmetry}
\label{sec:ttbar_L_vs_R_Ac}

One way to directly access the chiral structure of four-quark operators
is through observables like charge asymmetries, as discussed in
Sec.~\ref{sec:eft_tbar}. At the LHC it has been measured in terms
of absolute top and anti-top rapidities,
\begin{align}
A_C = \frac{\sigma(\Delta |y| > 0) - \sigma(\Delta |y| < 0)}{\sigma(\Delta |y| > 0) + \sigma(\Delta |y| < 0)} 
\quad \text{with} \quad  \Delta |y| = |y_t| - |y_{\bar t}|\,.
\end{align}
In QCD such an asymmetry arises only at NLO. In SMEFT, it
is induced at LO by four-quark contributions. For illustration, we consider the two operators $O_{Qq}^{1,8}$
and $O_{tq}^8$ with a left-handed light-quark current and
different chirality of the top current.  Since both operators are weak
singlets, there is no distinction between up and down quarks. Now the
chiral coefficients from Eq.~\eqref{eq:vv-aa} are given by
\begin{align}\label{eq:s-a-linear}
4C_{VV}^{q,8} = C_{Qq}^{1,8} + C_{tq}^8 = - 4C_{VA}^{q,8}\,,\qquad 4C_{AA}^{q,8} = C_{Qq}^{1,8} - C_{tq}^8 = - 4C_{AV}^{q,8}\,.
\end{align}
To leading order QCD, the charge asymmetry depends on the
corresponding Wilson coefficients as
\begin{align}
A_C = \frac{\sigma_{AA}\big(C_{Qq}^{1,8} - C_{tq}^8\big) }{\sigma_\text{SM} +  \sigma_{VV} \big(C_{Qq}^{1,8} + C_{tq}^8\big) }\, ,
\end{align}
Here $\sigma_\text{SM}$ is the SM $t\bar t$ rate,
$\sigma_{VV}$ and $\sigma_{AA}$ denote the 
contributions proportional to $4C_{VV}$ and $4C_{AA}$ (see
Eq.~\eqref{eq:qq_tt_analytic}), and the sum over all $q\bar{q}$ parton
contributions is implicit. This expression is easily inferred from Eq.~\eqref{eq:qq_tt_analytic}, observing that the charge asymmetry  probes the linear terms in $c_t$ in the partonic cross section. From the definition of $C_{AA}$ in Eq.~\eqref{eq:vv-aa}, we also see that $A_C$ is sensitive to $(LL-RL) + (RR-LR)$, thereby distinguishing between left- and right-handed top quarks.

\begin{figure}[t]\centering
\includegraphics[width=.49\textwidth]{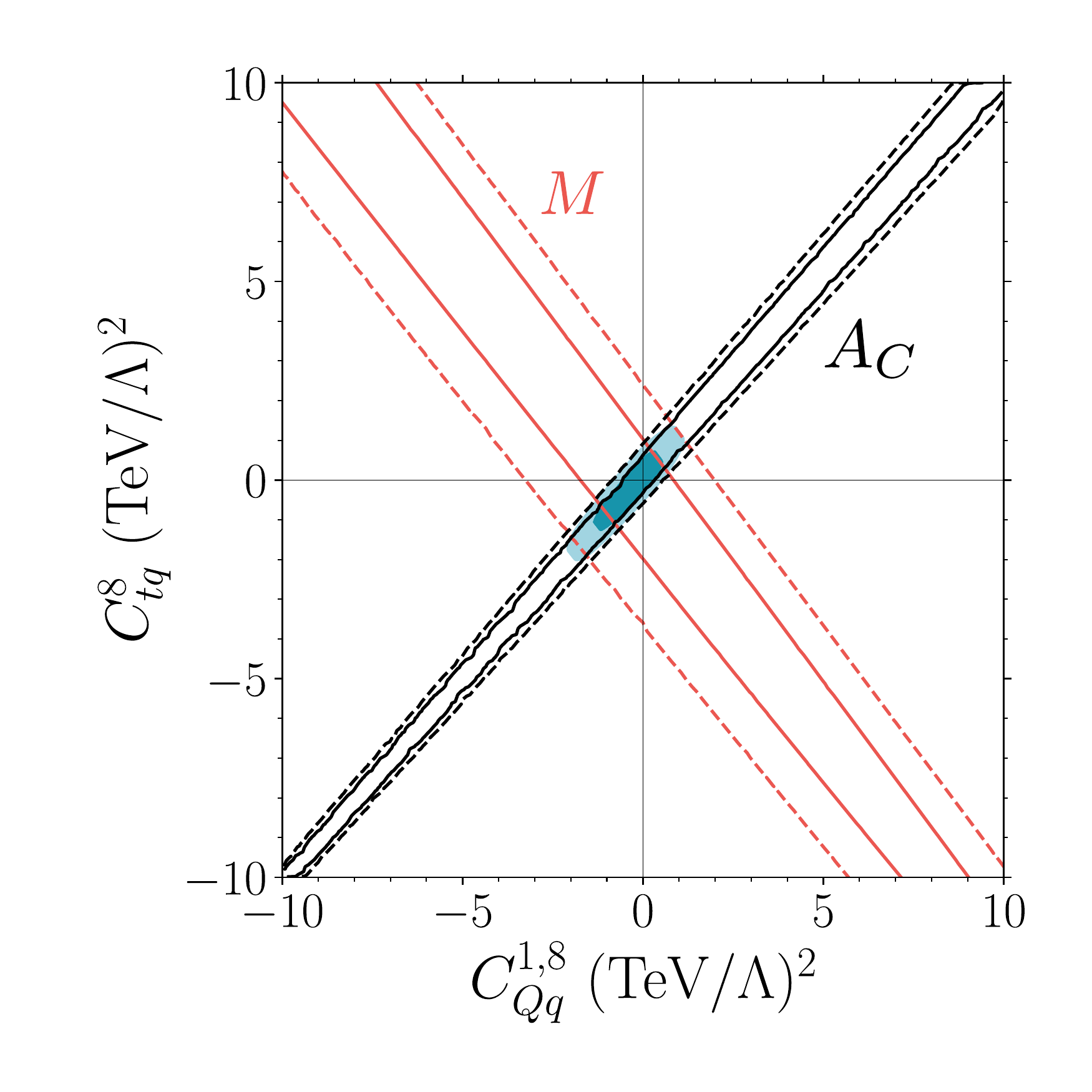}
\includegraphics[width=.49\textwidth]{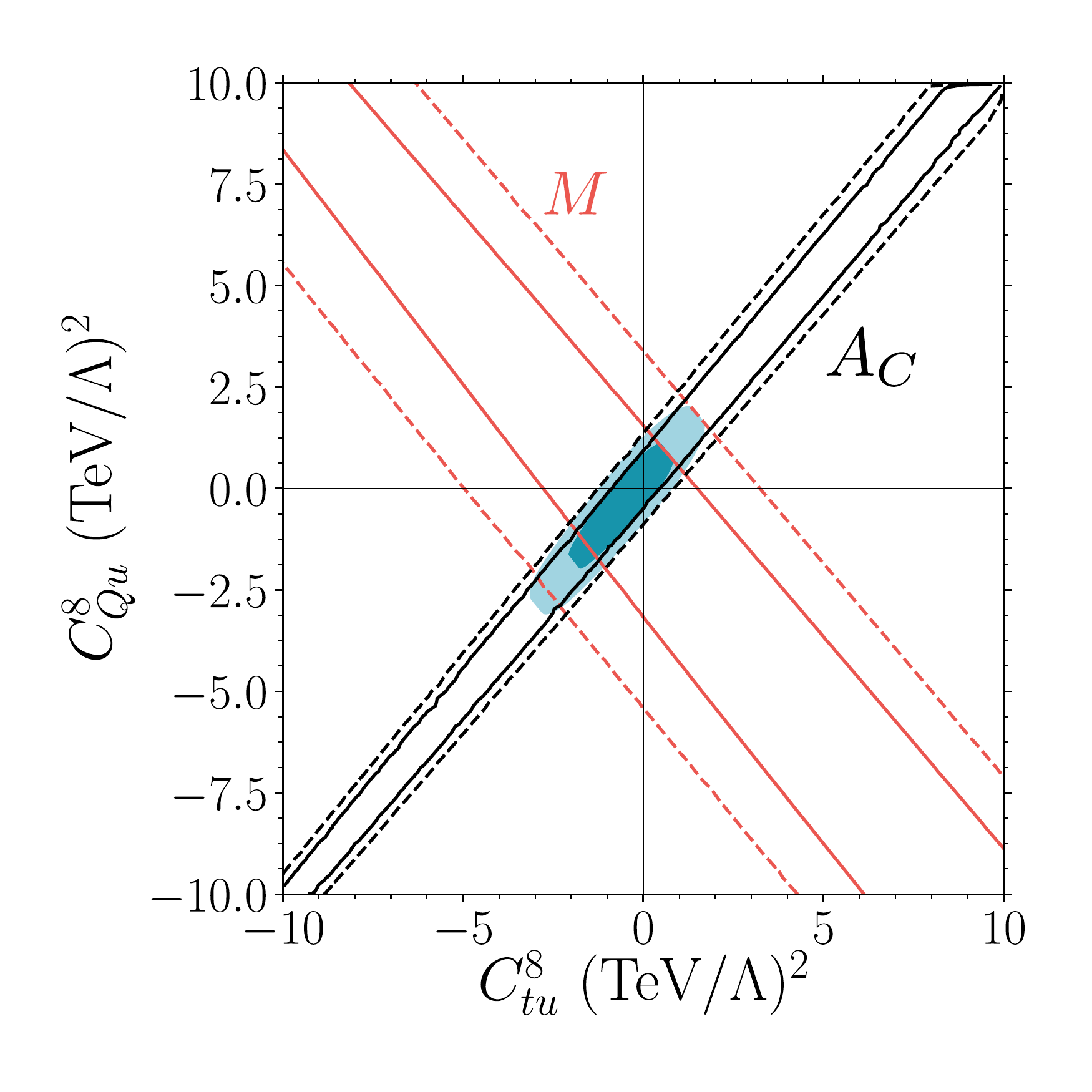}
\caption{Chirality effects of four-quark operators from LO two-parameter
  fits to order $\Lambda^{-2}$. Red lines use charge-symmetric
  observables (set M of Tab~\ref{tab:data1}) while black lines use asymmetries $A_C$.  The shaded areas show
  the combined fit.  Solid and dashed lines mark the Gaussian
  equivalent of $\Delta\chi^2 = 1,4$.}
\label{fig:LvsR}
\end{figure}

For the operator pair we have chosen, charge-symmetric observables probe the $(1,1)$ direction in $(C_{Qq}^{1,8},C_{tq}^8)$, which corresponds to a vector-like top coupling. The charge asymmetry is sensitive to the
$(1,-1)$ direction, which corresponds to an axial-vector-like top
coupling.
 The corresponding flat directions can be seen in the left panel of
Fig.~\ref{fig:LvsR}, where we show bounds on the Wilson coefficients
$(C_{Qq}^{1,8},\,C_{tq}^8)$ from a fit to measurements of total cross
sections and $m_{t\bar{t}}$ distributions labelled `M' in Tab.~\ref{tab:data1} (red lines) and of the charge
asymmetries $A_C$ (black lines).
The shaded blue region shows the combined fit with both
datasets, which probes both vector and
axial-vector currents with top quarks and breaks the respective blind
direction in $\sigma$ or $A_C$.

The same behavior applies to operators with right-handed initial quarks, like $O_{tu}^8$ and $O_{Qu}^8$. As shown in Fig.~\ref{fig:LvsR}, right, their effects on $\sigma$ and $A_C$ at
order $\Lambda^{-2}$ are the same as in Eq.~\eqref{eq:s-a-linear}, just
replacing $C_{Qq}^{1,8} \to C_{tu}^8$, $C_{tq}^8\to C_{Qu}^8$. 

\subsection{Top chirality from jet radiation}
\label{sec:ttbar_L_vs_R_NLO}

\begin{figure}[t]
\centering
\includegraphics[width=.48\textwidth]{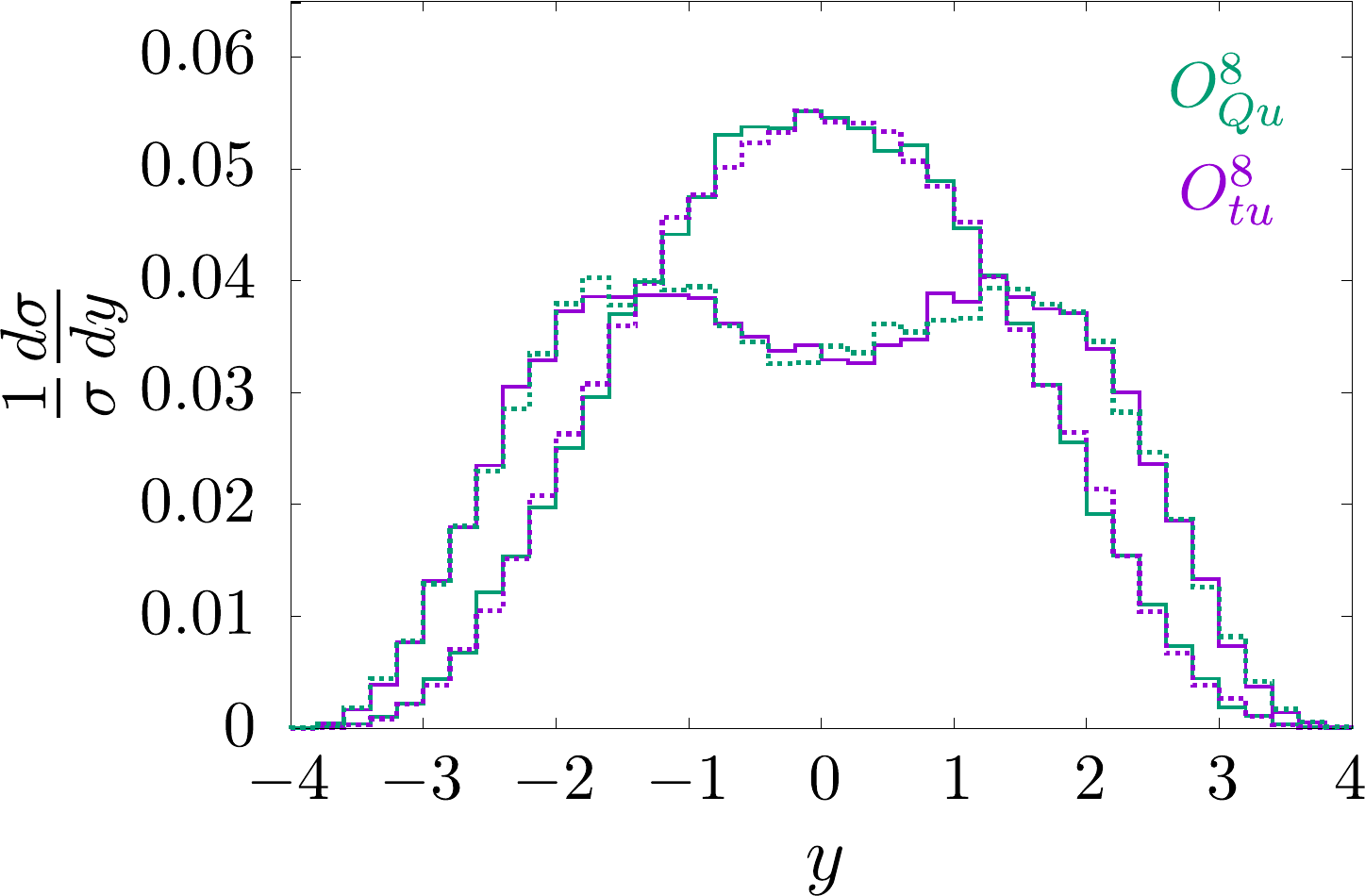}\hspace*{0.4cm}
\includegraphics[width=.48\textwidth]{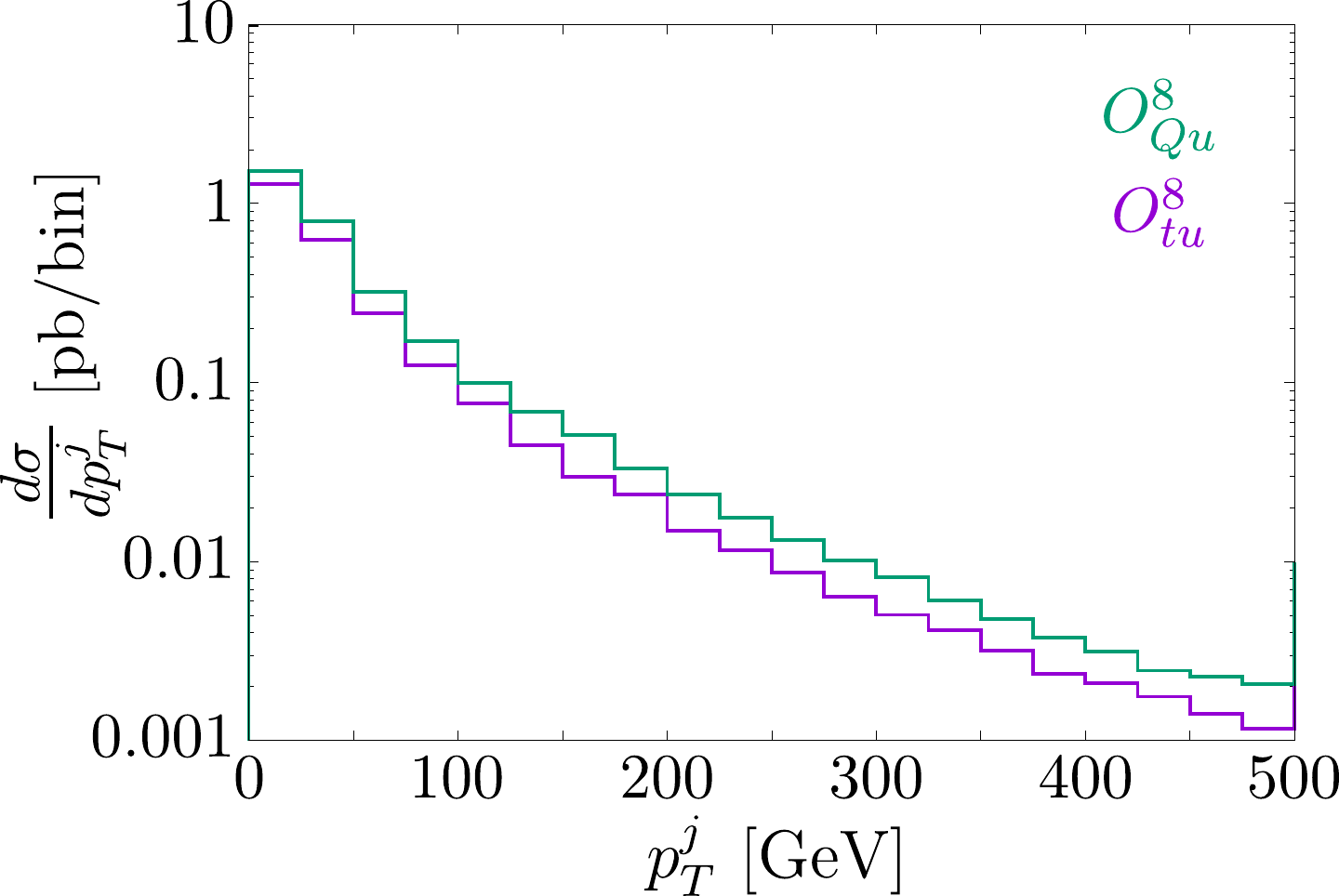}
\caption{Left: rapidity of top (plain curves) and anti-top (dashed curves) in $pp\to t\bar{t}$ for the SM-interference of the $LR$ (green) and $RR$ (purple) four-quark operators. Right: jet transverse momentum distribution in $pp\to t\bar{t}j$ for the SM-interference.}
\label{top-dist}
\end{figure}

As an alternative to the asymmetry in the previous section we can also
use patterns in QCD jet radiation to distinguish four-quark operators
with different chirality structures.  For instance the operators
$O^8_{tu}$ ($RR$) and $O^8_{Qu}$ ($LR$) differ only in the chirality of
the top quark. Their leading contribution to top pair production is the same for the inclusive rate and for any charge-symmetric observable, which probe $C_{VV}^{u,8} \propto C^8_{tu} + C^8_{Qu}$  and $|C_{V+A}^{u,8}|^2 \propto |C_{tu}^8|^2 + |C_{Qu}^8|^2$, cf. Eq.~\eqref{eq:vv-aa}. However, the two operators are distinguishable in top rapidity distributions, as shown in the left panel of
Fig.~\ref{top-dist}.

 Here $O^8_{tu}$ gives more forward or backward
tops, compared to $O^8_{Qu}$ which leads to more central tops. These
different rapidity distributions are directly related to the angular
distribution of the top quark in the CM frame of the
collision (cf. Eq.~\eqref{eq:qq_tt_analytic}),
\begin{align} 
 \frac{d \sigma(u\bar{u} \to t\bar{t})}{d \cos \theta_t} & \propto
 \left( 1+2 \beta_{t\bar{t}} \cos \theta_t  + 4 \overline{m}^2 +
 \beta^2_{t\bar{t}} \cos^2 \theta_t \right) C^8_{tu}\\\nonumber 
 & \quad +  \left( 1- 2
 \beta_{t\bar{t}} \cos \theta_t  + 4 \overline{m}^2 +
 \beta^2_{t\bar{t}} \cos^2 \theta_t \right) C^8_{Qu} \; , 
\end{align}
where $\theta_t$ is the angle between the incoming up quark and the top. In that sense the contribution of the $RR$ operator is `forward' whilst the $LR$ operator contributes as `backward'.

Combined with the color structure this directionality implies that an
additional jet can break the degeneracy of the two operators. In the
hard process $q \bar{q} \to t \bar{t}$ the triplet color charge flows
from the incoming quark to the top quark and from the anti-quark to
the anti-top. This leads to a stronger acceleration of color, and
consequently more QCD radiation, when the top is produced backwards compared to forwards in
the $q\bar{q}$ frame. The same effect can
be seen in the context of the top rapidity
asymmetry~\cite{Skands:2012mm}.  The additional radiation when the top
is backwards pushes the recoiling top--anti-top pair to higher
transverse momentum.  Indeed, in the right panel of
Fig.~\ref{top-dist} we find that $O^8_{Qu}$ gives a harder jet $p_T$
distribution than $O^8_{tu}$. The same effect can be seen in the invariant mass
distribution, where $O^8_{Qu}$ gives a harder $m_{t\bar{t}}$
distribution.

\begin{figure}[t]
\centering
\includegraphics[width=.48\textwidth]{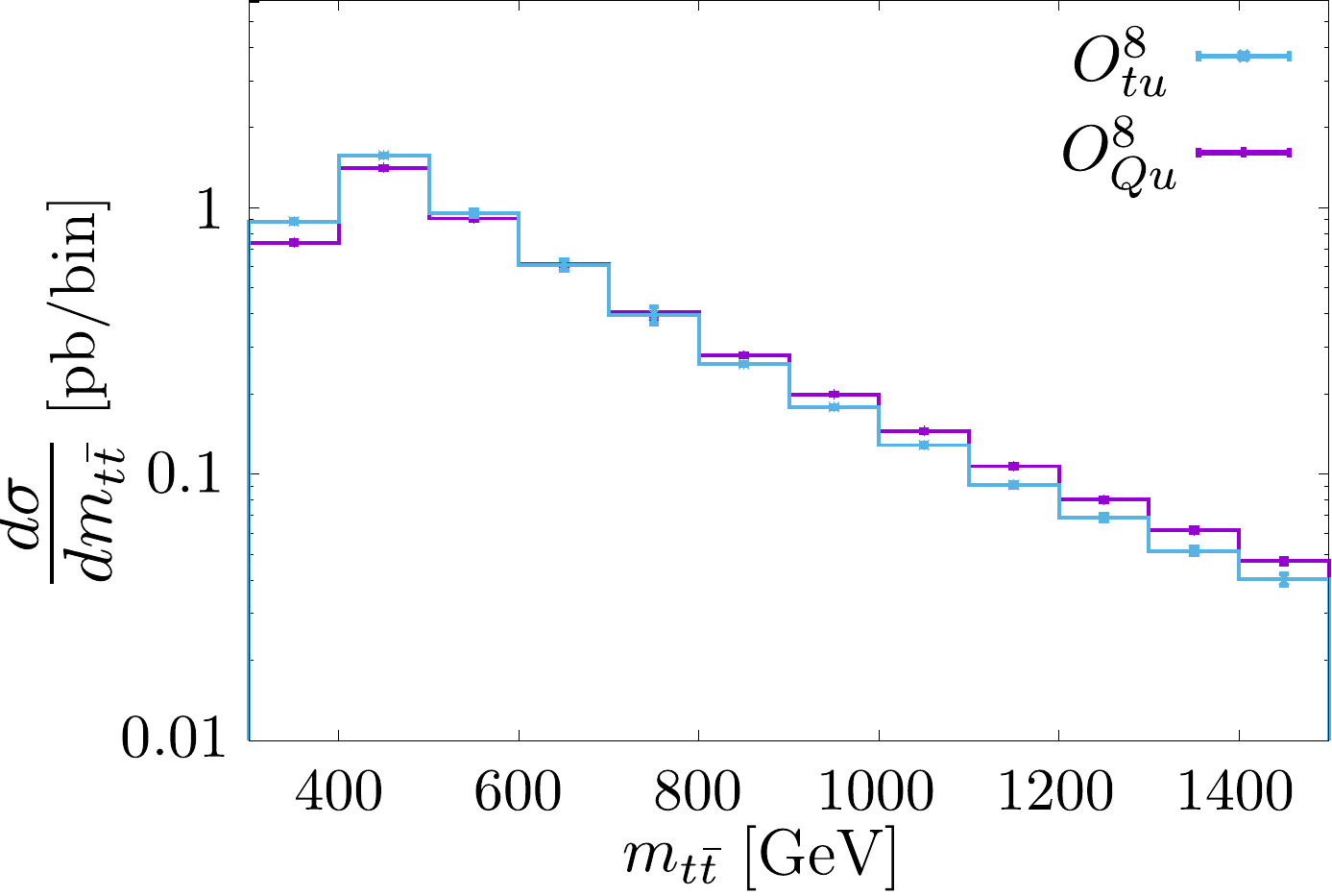}\hspace*{0.4cm}
\includegraphics[width=.5\textwidth]{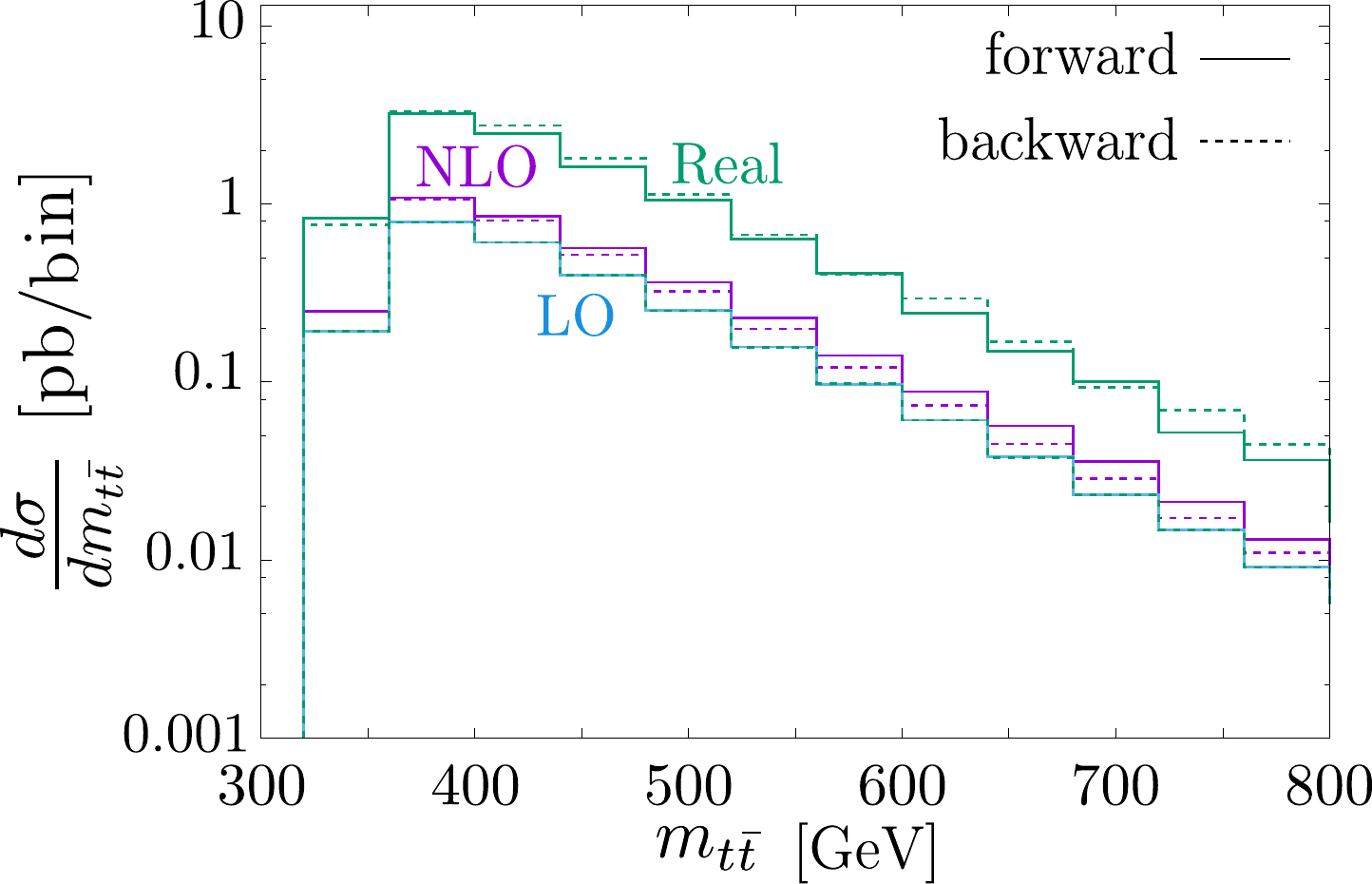}
\caption{Left: top pair invariant mass distribution in $pp\to t\bar{t}$ at NLO for the
  SM-interference of the $RR$ and $LR$ operators. Right: top pair
  invariant mass distribution in $p\bar{p} \to t\bar{t}$ at NLO in QCD. By `forward' we denote events with $y_t>0$ and by `backward'
  events with $y_t<0$.}
\label{nlo-mtt}
\end{figure}

The jet kinematics of the operator contributions illustrate the impact of NLO corrections in inclusive top--anti-top production. At NLO both the real and virtual corrections break the operator degeneracy in the $t\bar{t}$ distributions. The invariant mass distribution in $t\bar{t}$ production at NLO is shown in the left panel of Fig.~\ref{nlo-mtt}. Now the $RR$ operator $O^8_{tu}$ gives the harder distribution,
implying that the virtual corrections have a large effect in the
opposite direction of the real emission. The difference between the $LR$ and $RR$ operators at NLO reaches 20\% in the distributions.

To clarify the interplay between virtual and real corrections, we perform a comparison between forward and backward tops in QCD.  For a cleaner comparison, we use
$p\bar{p}$ collisions that are dominated by the $q\bar{q}$ partonic initial state. We define forward top quarks as emitted in the direction of the proton and use positive and negative rapidities to define forward and backward tops. In the right panel of Fig.~\ref{nlo-mtt} we show the NLO distributions in $p\bar{p}\to t\bar{t}$ separately for forward and backward tops. The results confirm that real radiation behaves differently from the total rate at NLO, given by the sum of Born, virtual and real corrections. This means that NLO QCD corrections break the degeneracy of
operators that occurs at LO. Our example demonstrates the potential of using NLO QCD corrections more generally to distinguish between operators.

\subsection{Quadratic terms and flat directions}
\label{sec:quadratic}

The dependence of the observables on effective operators changes significantly if we include 
 contributions to order $\Lambda^{-4}$.
 This is particularly true for four-quark operators that do not interfere with
the SM amplitude to leading order because of their color or helicity
structure. For these operators, quadratic contributions can be the leading effect in an observable. For operators that interfere with the SM quadratic terms can change the bounds from LHC measurements significantly, for instance in case of strong cancellations between linear and quadratic contributions or in case of limited sensitivity. A dominance of quadratic terms is thus per se not a problem with the convergence of the effective theory, but can be due to a distinctive physics pattern which suppresses the naively leading contribution. In general,
 an effective field theory approach is justified if a heavy particle can be decoupled for a given
observable. This does not necessarily imply a fast convergence of the EFT expansion. Since the LHC covers a wide range of scales for a wide
range of measurements, the justification of an EFT approach has to come
from the decoupling pattern of the underlying theory. An illustrative
extreme example is heavy quark effective theory which can be used in
hadronization calculations for the LHC even though the LHC energy is
clearly above the bottom mass. Another example are Higgs decays, for
which the Higgs mass defines an upper limit on the energy scale
independently of the Higgs production mode~\cite{Brehmer:2016nyr}.

As an illustration we look again at the operators $O_{Qq}^{1,8}$ and
$O_{Qq}^{3,8}$, for which the $t\bar{t}$ cross section and other charge-symmetric
observables depend on the Wilson coefficients as
\begin{align}
\sigma_{t\bar{t}} & = \sigma_\text{SM} + \sigma_{VV}^d \Big[r\big(C_{Qq}^{1,8} + C_{Qq}^{3,8}\big) + \big(C_{Qq}^{1,8} - C_{Qq}^{3,8}\big)\Big]  \notag \\
&  \qquad \quad \, + \sigma_{V+A}^d\Big[r \big(C_{Qq}^{1,8} + C_{Qq}^{3,8}\big)^2 + \big(C_{Qq}^{1,8} - C_{Qq}^{3,8}\big)^2 \Big]\,,
\label{eq:squared}
\end{align}
where $\sigma_{VV}^d$ and $\sigma_{V+A}^d$ are the
contributions from the partonic $d\bar{d} \to t\bar{t}$ process.  As
discussed in Sec.~\ref{sec:ttbar_u_vs_d}, the linear terms to order
$\Lambda^{-2}$ have a flat direction which can be resolved using the
kinematic variation of the parton densities.  From Fig.~\ref{fig:uvsd}
we learn that the linear fit leaves certain directions in the parameter space basically unexplored. This is particularly true for $RR$ operators, where bounds only appear around values of $C/\Lambda^2 \approx \pm 10/\rm TeV^2$. In this range  contributions from the squared dimension-6 amplitudes, \ie, the terms in the second line of Eq.~\eqref{eq:squared}, are numerically dominant.

Due to the presence of quadratic terms of order $\Lambda^{-4}$ any rate prediction $d \sigma$ is positive even for large Wilson coefficients. This implies that in a fit of the two-dimensional
parameter space $(C_{Qq}^{1,8},C_{Qq}^{3,8})$, we can set an upper
bound in any direction. From the second line of Eq.~\eqref{eq:squared}
we can immediately read off that there still exists a flat direction,
where the cross section remains constant for varying Wilson
coefficients. In contrast to the linearized case this flat direction
forms an ellipse, which we can collapse into any direction and indeed
derive a limit on the individual coefficients.

This argument also applies to more than two parameters. For
rate observables that are positive-definite, the $n-1$-dimensional hyper-surface in the parameter
space of $n$ Wilson coefficients is always a compact manifold, like a
hyper-ellipsoid.  Including the quadratic terms does not reduce the
dimension of the parameter space, it merely changes the topology of
the likelihood function describing the combinations of Wilson
coefficients that provide a certain level of agreement between
predictions and data. In particular, it does not break any blind
directions in the parameter space.

\begin{figure}[t]\centering
\includegraphics[width=.49\textwidth]{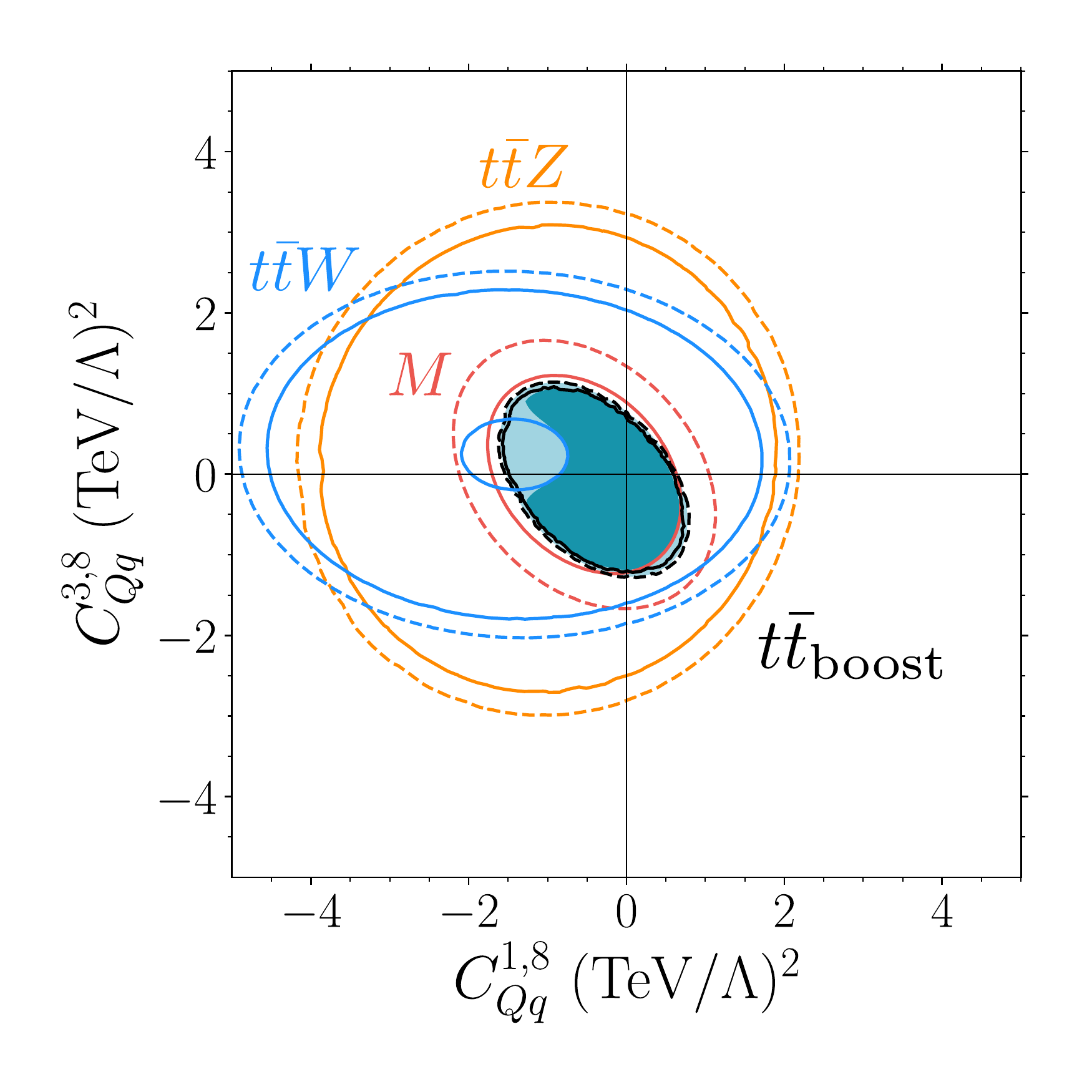}
\includegraphics[width=.49\textwidth]{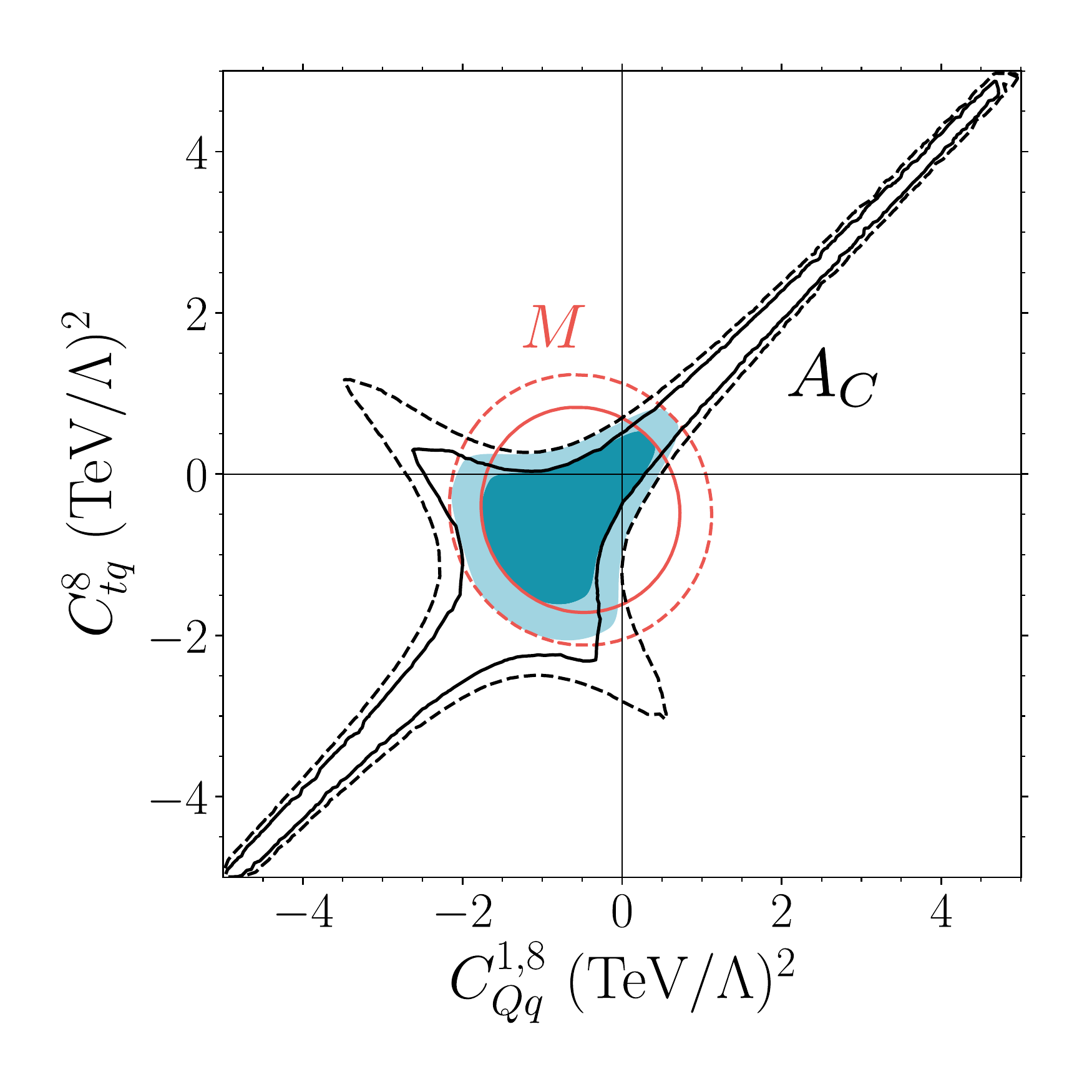}
\caption{Impact of the squared dimension-6 contribution on the fit
  result originally shown for the isospin distinction in
  Fig.~\ref{fig:uvsd} (right) and for the chirality distinction in
  Fig.~\ref{fig:LvsR} (left). The lines are based on the same $t\bar{t}$ data set as before, but the predictions now include SMEFT contributions to order $\Lambda^{-4}$ for $t\bar{t}$ (red), $t\bar{t}Z$ (orange) and $t\bar{t}W$ (blue). The black
  lines show the fit for symmetric observables in the boosted regime (left) and for the asymmetries $A_C$ (right) to order $\Lambda^{-4}$. Shaded areas show the combined fit to order $\Lambda^{-4}$. Solid and dashed lines mark the Gaussian
  equivalent of $\Delta\chi^2 = 1,4$.}
\label{fig:quadratics}
\end{figure}

In the left panel of Fig.~\ref{fig:quadratics} we show the same fit of
$(C_{Qq}^{1,8},C_{Qq}^{3,8})$ as in the right panel of
Fig.~\ref{fig:uvsd}, but including dimension-6 squared terms in the
predictions.
%
 %
 The elliptic shape of the bounds reflects the geometric
dependence of the observables on the two Wilson coefficients. This is
in contrast with the linear fit from Fig.~\ref{fig:uvsd}, where the
combined bound had a diamond shape. It is interesting to compare the
respective sensitivity of the linear and quadratic fits. For
$t\bar{t}$ production alone, quadratic contributions drastically
increase the sensitivity to the individual operators. Including
$t\bar{t}Z$ and $t\bar{t}W$ rates resolves the blind direction in top
pair production already at order $\Lambda^{-2}$. Quadratic terms from
associated production (orange and blue ellipses) do not alter the fit
results by much, since the combined fit result (blue area) is
dominated by the quadratic contributions in boosted $t\bar{t}$
observables (black ellipse). This illustrates nicely the interplay of
linear and quadratic contributions in a global fit. The bound on individual Wilson coefficients can be set either by quadratic terms in the dominant observable (for limited sensitivity) or by the interplay of linear terms in
several observables that probe different directions of the parameter
space (for high sensitivity). Which effect dominates depends on the overall sensitivity of the observables to operator contributions and on the precision of their measurement.

Beyond rates, quadratic terms do not necessarily lead to ellipses in a
fit. The charge asymmetry in $t\bar{t}$ production is an example where
quadratic terms modify the topology of the flat direction, but do not
allow to set bounds in all directions of the parameter space. The reason is that $A_C$ is not defined to be positive, and therefore negative squared contributions can generally appear. For the
chiral operators $O_{Qq}^{1,8}$ and $O_{tq}^8$, the cross section and
the asymmetry read
\begin{align}\label{eq:as-quad}
\sigma_{t\bar{t}} & = \sigma_\text{SM} + \sigma_{VV} \big(C_{Qq}^{1,8} + C_{tq}^8\big) + \sigma_{V+A} \big(|C_{Qq}^{1,8}|^2 + |C_{tq}^8|^2\big) +\sigma_{V-A} \,C_{Qq}^{1,8}C_{tq}^8 \,, \phantom{\Bigg]} \notag \\
A_C & = \frac{\sigma_\text{SM}^A + \sigma_{AA}\big(C_{Qq}^{1,8} - C_{tq}^8\big)  + \sigma_{VVAA} \big(|C_{Qq}^{1,8}|^2 - |C_{tq}^8|^2 \big)}{\sigma_{t\bar{t}}} \,.
\end{align}
A fit of charge-symmetric observables leads to a spherical bound in
$(C_{Qq}^{1,8},C_{tq}^{8})$, shown as red curves in the right panel of
Fig.~\ref{fig:quadratics}. For the charge asymmetry the isocurves are
hyperbolas with asymptotes along the directions $(1,1)$ and
$(1,-1)$. The fit results reflect this shape in the black curves. They
leave the direction $(1,1)$ unconstrained. The fact that the direction
$(1,-1)$ is bounded is due to the combination of asymmetry
measurements with different best-fit points.

\section{Single top analysis}
\label{sec:global_single}

In addition to the top pair observables described in the previous
section our global top analysis also includes single top
production. Some Feynman diagrams for the different processes are
shown in Fig.~\ref{fig:singletop_diagrams}. The structure of the
single top sector is very similar to classic global SMEFT analyses in
that the operators listed in Tab.~\ref{tab:wilson-contributions} have
distinctive observable effects and can be probed with the sizeable
number of different measurements listed in Tab.~\ref{tab:data2}. Flat
directions are not an issue in this sector, but it is interesting to
test if there exist correlations in the bounds on the individual
operators.

\begin{figure}[t]
  \centering
  \includegraphics[width=0.325\linewidth]{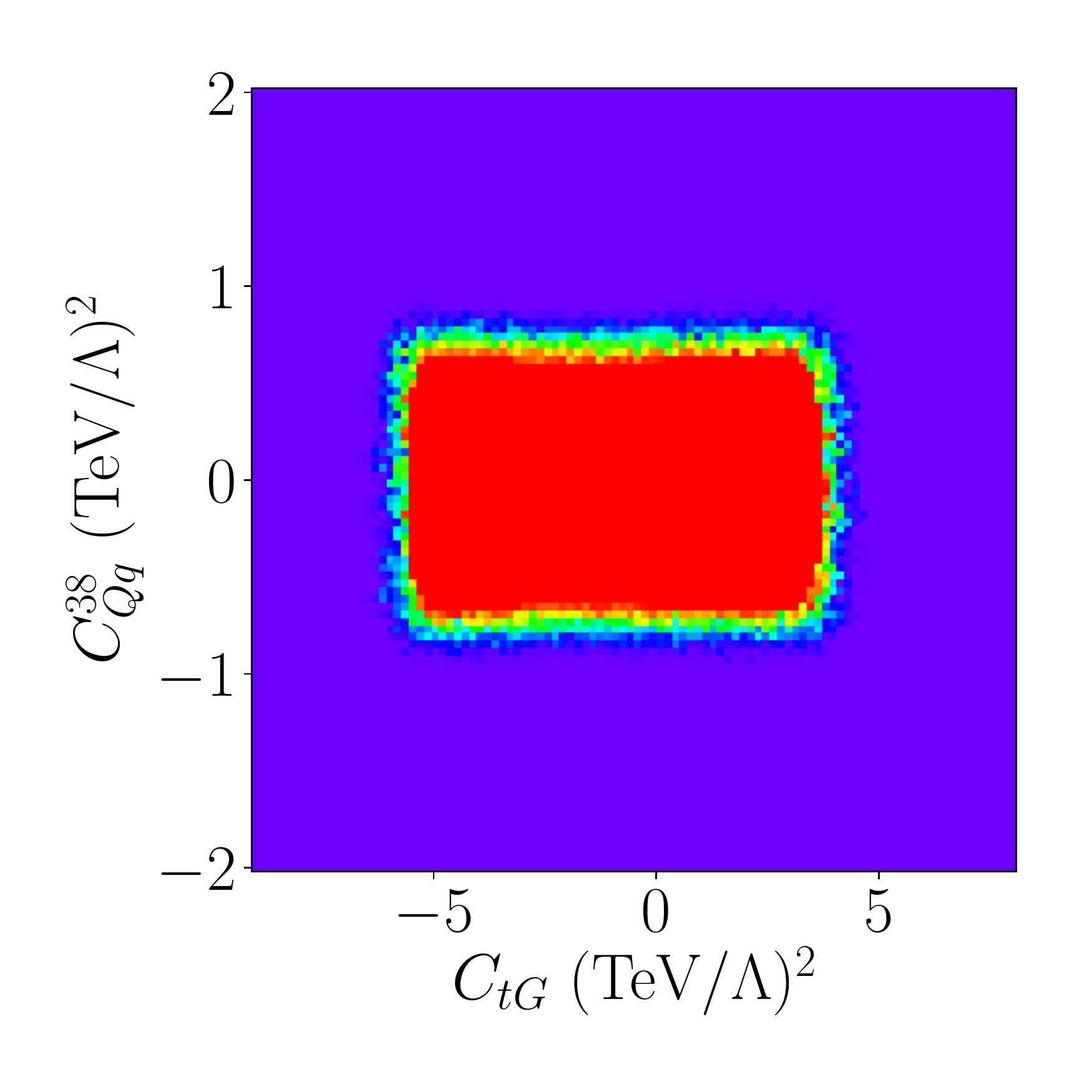}
  \includegraphics[width=0.325\linewidth]{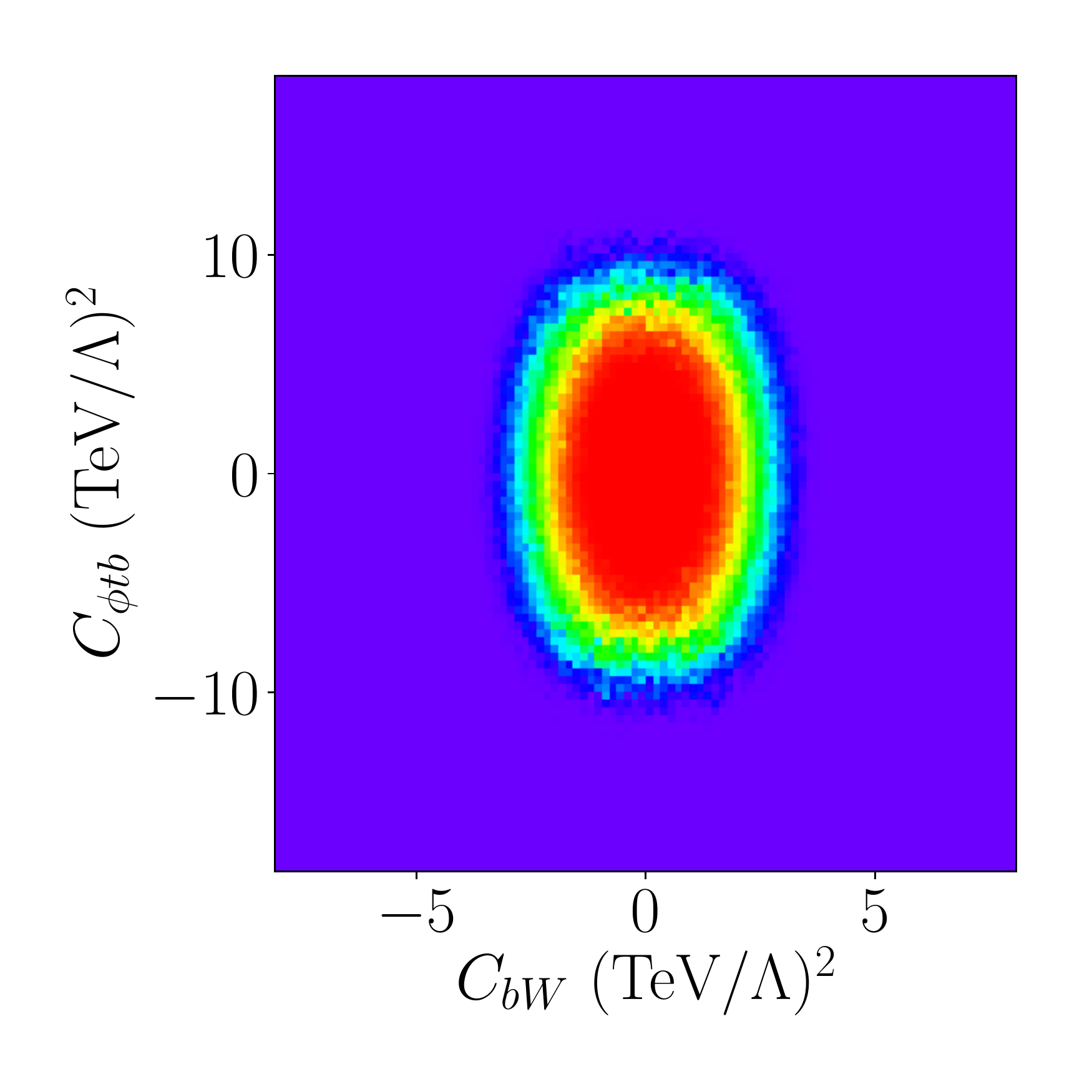}
  \includegraphics[width=0.325\linewidth]{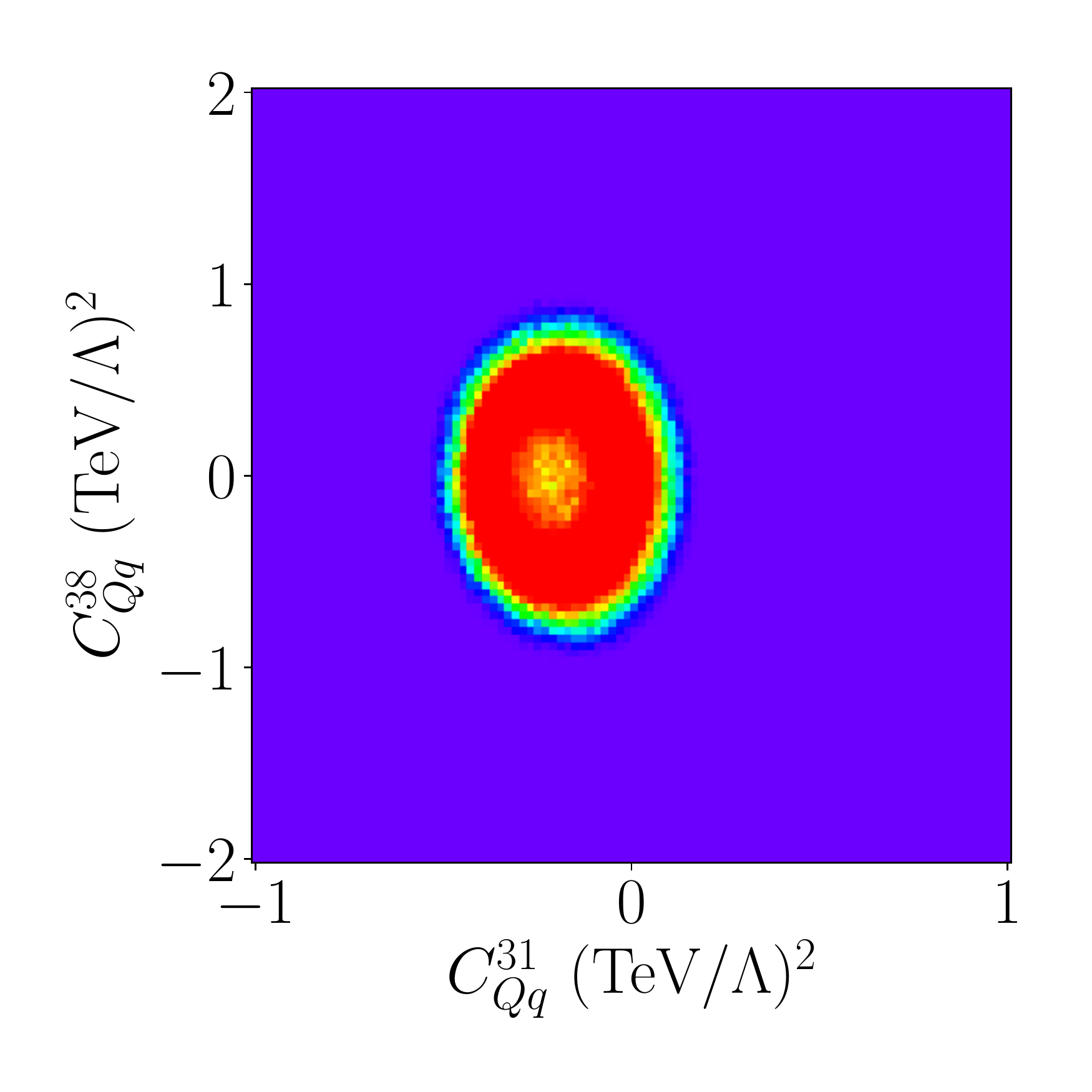} \\
  \includegraphics[width=0.325\linewidth]{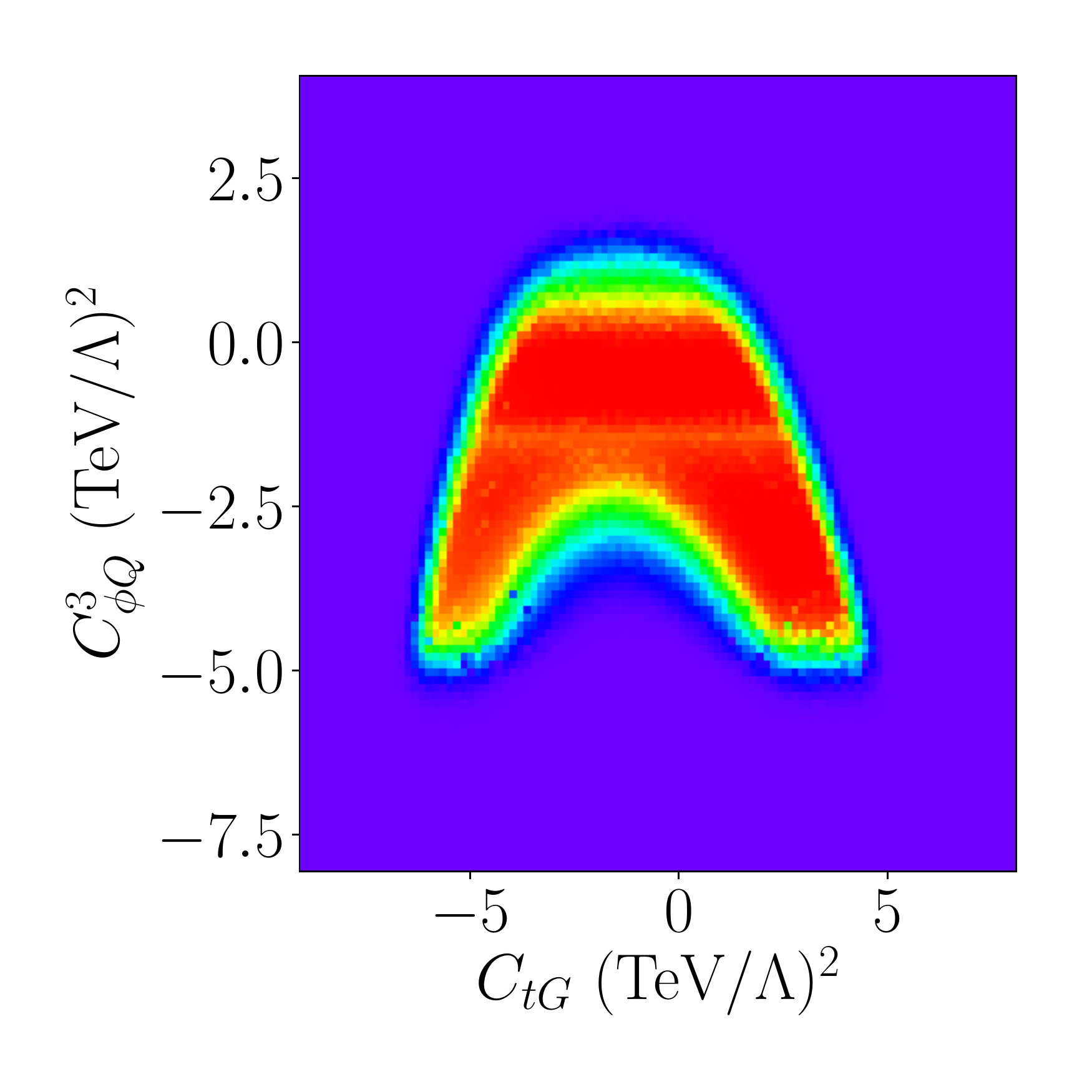}
  \includegraphics[width=0.325\linewidth]{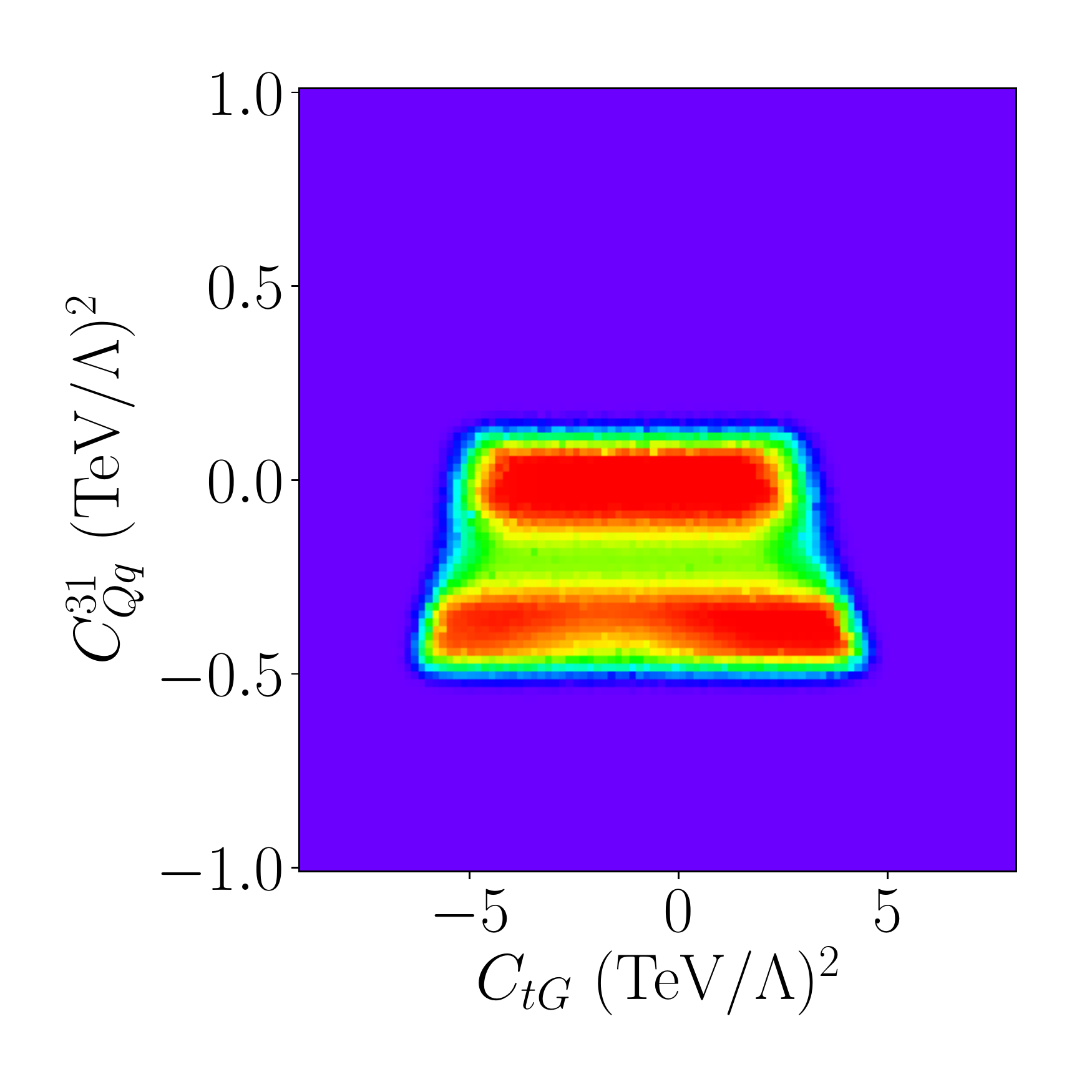}
  \includegraphics[width=0.325\linewidth]{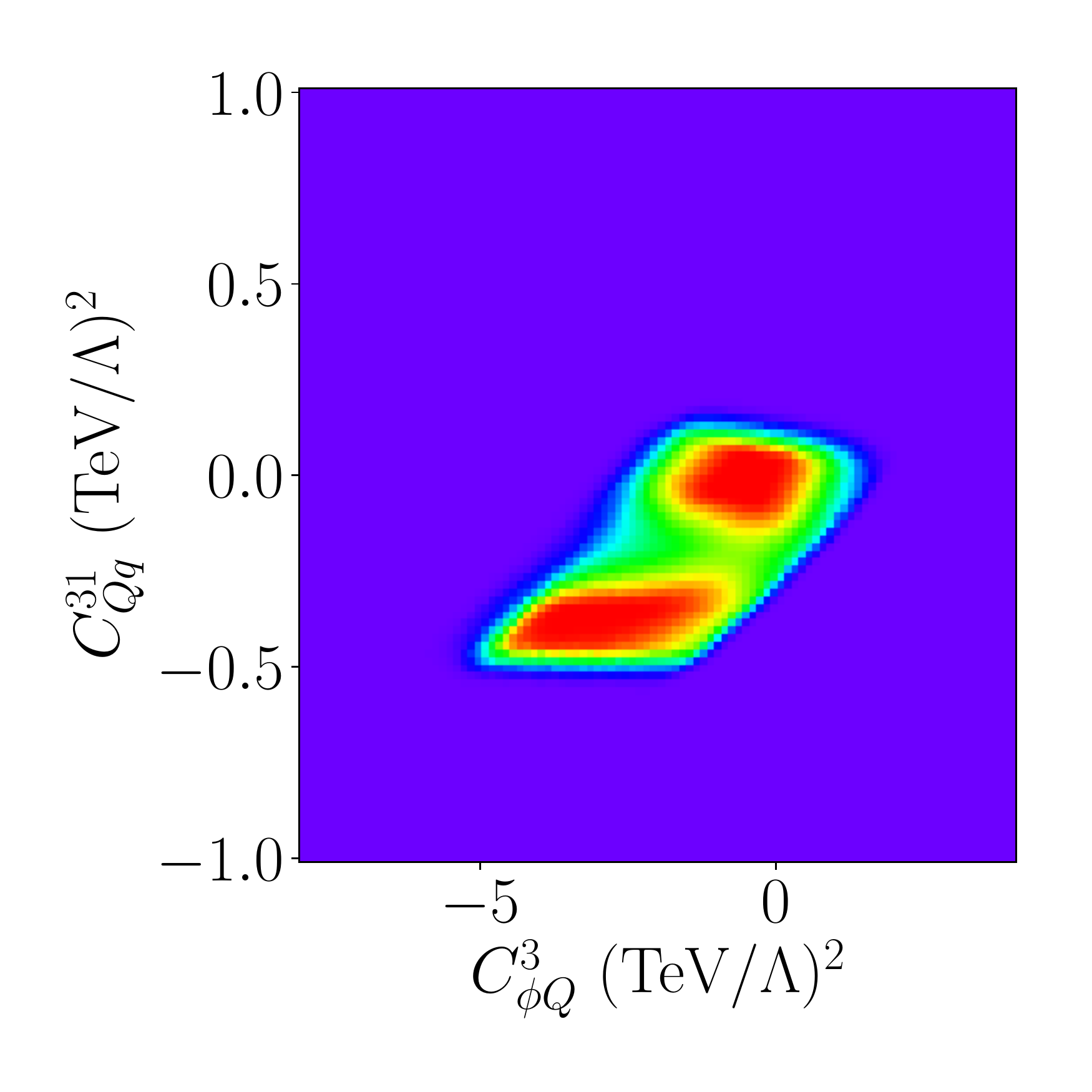} 
   \caption{Upper: examples for correlated 2-dimensional profile
     likelihoods of operators in a global fit to the single top
     data. Lower: correlated profile likelihoods for a three-parameter
     fit of the same data.}
\label{fig:SingleTCorrelations}
\end{figure}

We evaluate all two-operator correlations based on two-dimensional
profile likelihoods and find three distinct patterns shown in the
upper row of Fig.~\ref{fig:SingleTCorrelations}. First, a box shape
like for $C_{tG}$ and $C_{Qq}^{3,8}$ appears if two Wilson
coefficients are bounded by two separate sets of observables. Next, an
elliptic disk like the one between $C_{bW}$ and $C_{\phi tb}$ appears
if two operators contribute quadratically to the same
observable. Finally, a shifted circle like in the $C_{Qq}^{3,1} -
C_{Qq}^{3,8}$ plane appears if two operators contribute to the same
observables, but one of them linearly ($C_{Qq}^{3,1}$) and the other
one only quadratically ($C_{Qq}^{3,8}$). For this pattern the SM value
cannot be at the center of the circle. 

One of the few noteworthy correlations in the single top fit is the
inverted heart shape in the $C_{tG}-C_{\phi Q}^3$ plane shown in the
lower left panel of Fig.~\ref{fig:SingleTCorrelations}.  It can be
understood as the interplay of the three operators $C_{tG}$, $C_{\phi
  Q}^3$, and $C_{Qq}^{3,1}$ with at least two measurements. The only
single top measurement sensitive to $O_{tG}$ is $tW$ production. Using
its rate to constrain $C_{tG}$ and $C_{\phi Q}^3$ we find an elliptic
correlation centered at negative values of $C_{\phi Q}^3$. When we add
the strong constraints on $C_{\phi Q}^3$ from $t$-channel production
the bottom part of the ellipsis gets removed. Finally, once we add
$C_{Qq}^{3,1}$ to the fit we find that $O_{\phi Q}^3$ and
$O_{Qq}^{3,1}$ are slightly correlated and hence more negative values
of $O_{\phi Q}^3$ become consistent with data. In the lower panels of
Fig.~\ref{fig:SingleTCorrelations} we project the 3-dimensional
profile likelihood  from a 3-parameter fit along each of the three directions.  In the left
panel we see a very faint barrier for $C_{\phi Q}^3/\Lambda^2 \approx 1.5 \tev^{-2}$. It
corresponds to the two disconnected regions, one for $C_{Qq}^{3,1}/\Lambda^2
\approx 0 $ and one for $C_{Qq}^{3,1}/\Lambda^2 \approx 0.4 \tev^{-2}$, which we see
clearly in the central and right panels.
In the global single top fit, once all observables are included, only the region for $C_{\phi Q}^3/\Lambda^2\approx 0 $ remains, while the other region becomes disfavored.

\begin{figure}[t]
  \centering
  \includegraphics[width=0.95\linewidth]{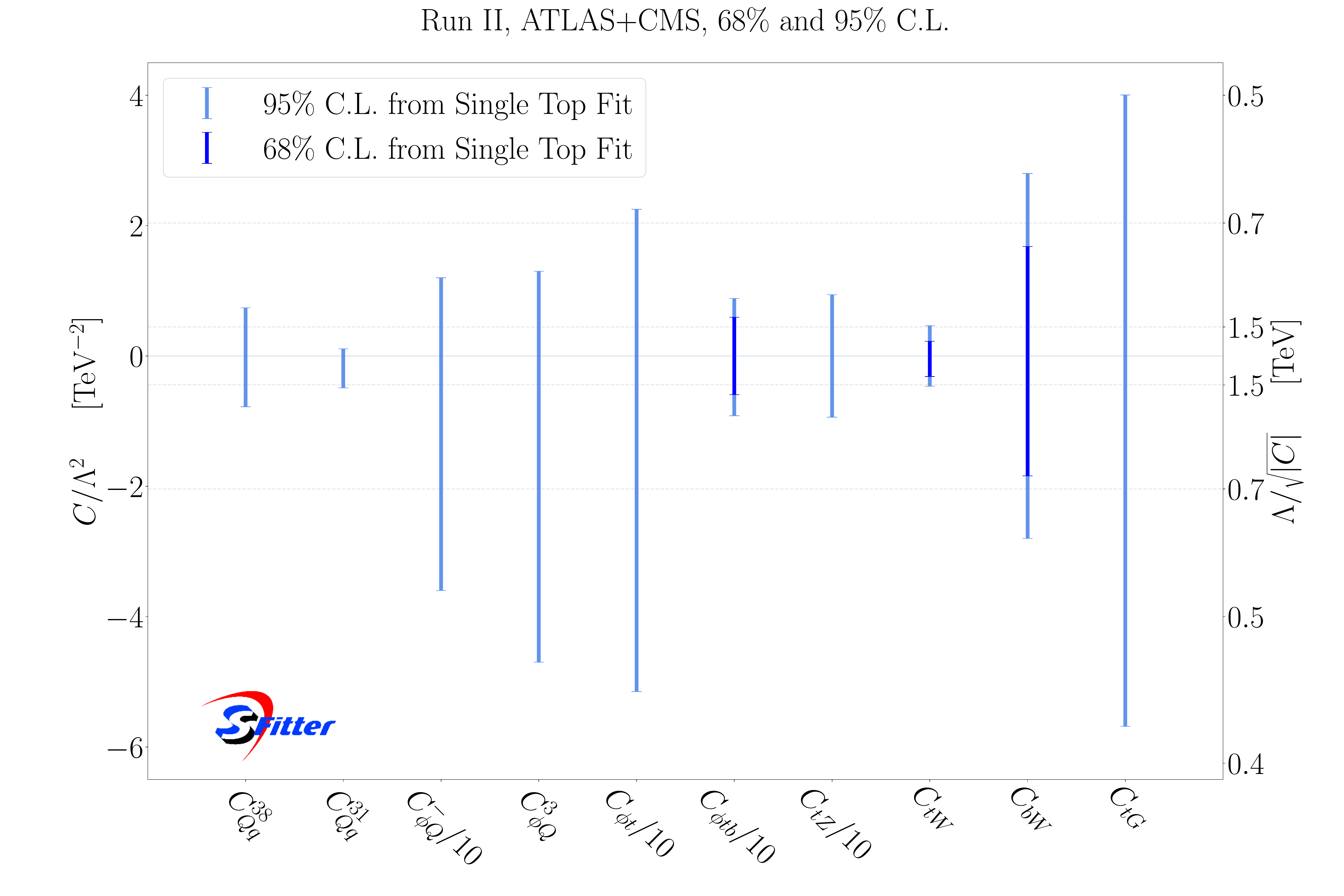}
  \caption{95\% and 68\% CL bounds for the global fit to the single
    top data set from Tab.~\ref{tab:data2}. Whenever the 68\% CL is
    not shown it falls into the flat profile likelihood regime
    reflecting dominant theoretical uncertainties.}
  \label{fig:GlobalSingle}
\end{figure}
 
Given the smooth behavior of the multi-dimensional likelihood we can
perform a global fit of the single top sector, including the
$W$ helicity fractions in top decay and associated $tV$
production. The one-dimensional profile likelihoods are shown in
Fig.~\ref{fig:GlobalSingle}.  The only non-standard aspect in these
results is that we cannot define meaningful 68\% CL limits for some of
the operators. This happens when a flat core of the profile likelihood
covers more than 68\% of the integral and there exists no unique
definition of a range. We observe this for all operators except for
$O_{\phi tb}$, $O_{tW}$, and $O_{bW}$, implying that for all other
operators the theory uncertainty is large compared to the experimental
statistics and systematics.

One aspect which sticks out in the global fit is the low sensitivity
to $O_{\phi Q}^3$, compared to $O_{Qq}^{3,1}$ and $O_{tW}$. All three
operators interfere with the SM amplitude in $t$-channel single top
production, but for $O_{\phi Q}^3$ the effect is numerically
smaller by about a factor three. As discussed in
Sec.~\ref{sec:eft_single}, $O_{\phi Q}^3$ only rescales the SM
contribution, while $O_{Qq}^{3,1}$ changes the kinematics in
$t$-channel production, see Tab~\ref{tab:he-scaling}. The operator
$O_{tW}$ is best constrained by the $W$ helicity fractions in top
decay, see Eq.~\eqref{eq:w-hel}, which are very sensitive to this
operator.

The bounds on $C_{Qq}^{3,8}$, $C_{\phi tb}$ and $C_{bW}$ are symmetric
around zero, since the corresponding operators contribute to single
top observables only at order $\Lambda^{-4}$,
cf. Tab~\ref{tab:wilson-contributions}. The coefficients $C_{\phi Q}^-$,
$C_{tZ}$ and $C_{\phi t}$ are bound by $tZ$ production. Due to the
limited experimental precision, the bounds on these operators are very
loose. Also here the SM-interference plays a role, leading to asymmetric bounds for $C_{\phi Q}^-$ and $C_{\phi t}$. The sensitivity to $O_{\phi t}$ is especially poor because its
contribution to $tZ$ production is suppressed, see
Sec.~\ref{sec:eft_single}. This will change once we include the
better-measured $t\bar{t}Z$ channel in the global fit.

\section{Global top analysis}
\label{sec:global_comb}
In the final step we add all top pair measurements from
Tab.~\ref{tab:data1} to our single top fit based on the measurements
in Tab.~\ref{tab:data2} and presented in
Sec.~\ref{sec:global_single}. On the parameter side we add the large
number of four-quark operators, which roughly doubles the number of
model parameters. For the measurements we not only include top pair
production, but also associated $t\bar{t}W$ and $t\bar{t}Z$ rate
measurements. They constrain some of the electroweak top operators in
single top production and four-quark operators in top pair production,
thus linking both sectors in the global fit.

First, we briefly comment on 2-dimensional correlations in the
complete fit. The box-shaped correlations for separate operators and
separate measurements, filled ellipses for more than one operator
affecting a measurement, and shifted circles from linear contributions
to compact flat directions which we observed in the single top fit (Fig.~\ref{fig:SingleTCorrelations}) also appear in the global fit.

\begin{figure}[t]
  \centering
  \includegraphics[width=0.325\linewidth]{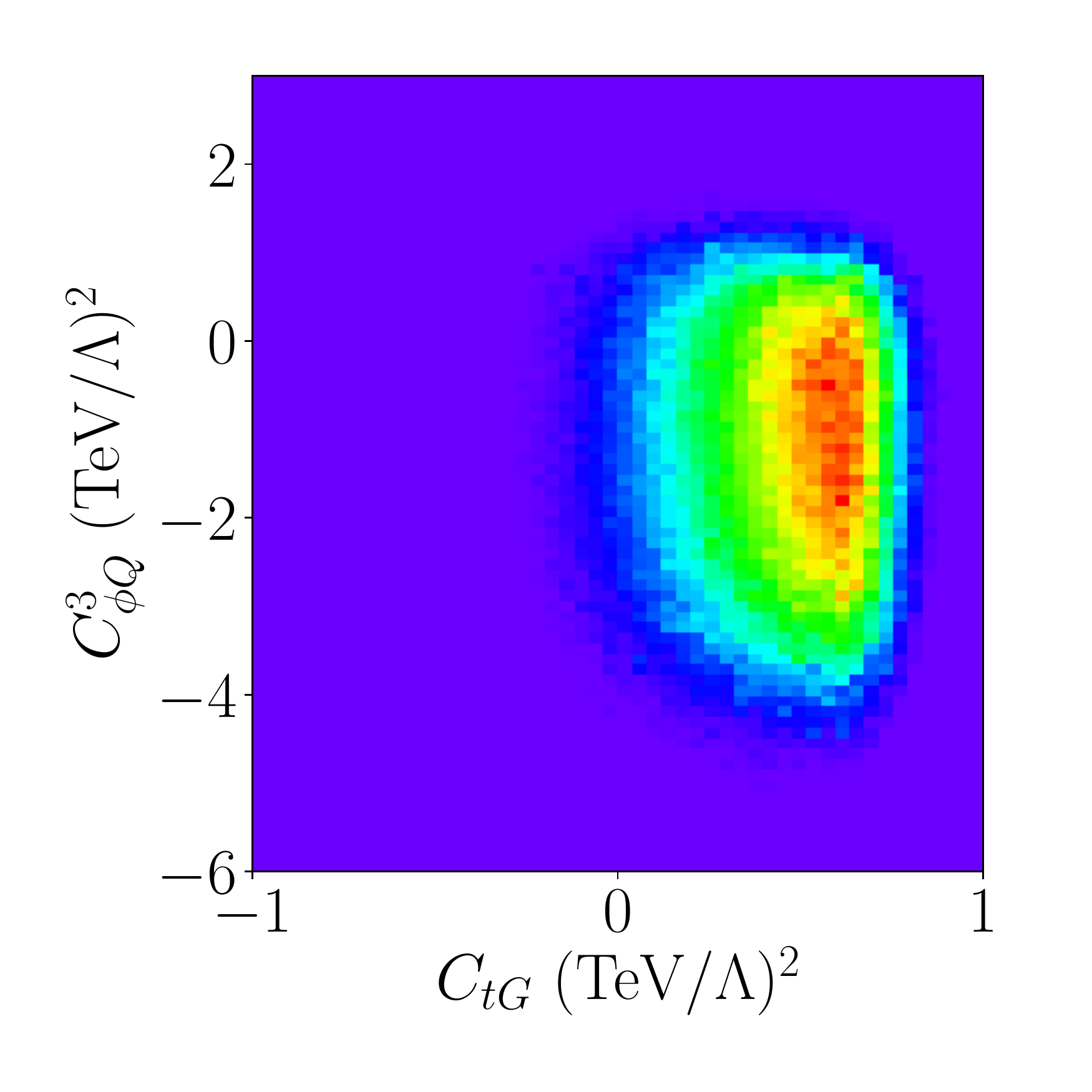}
  \includegraphics[width=0.325\linewidth]{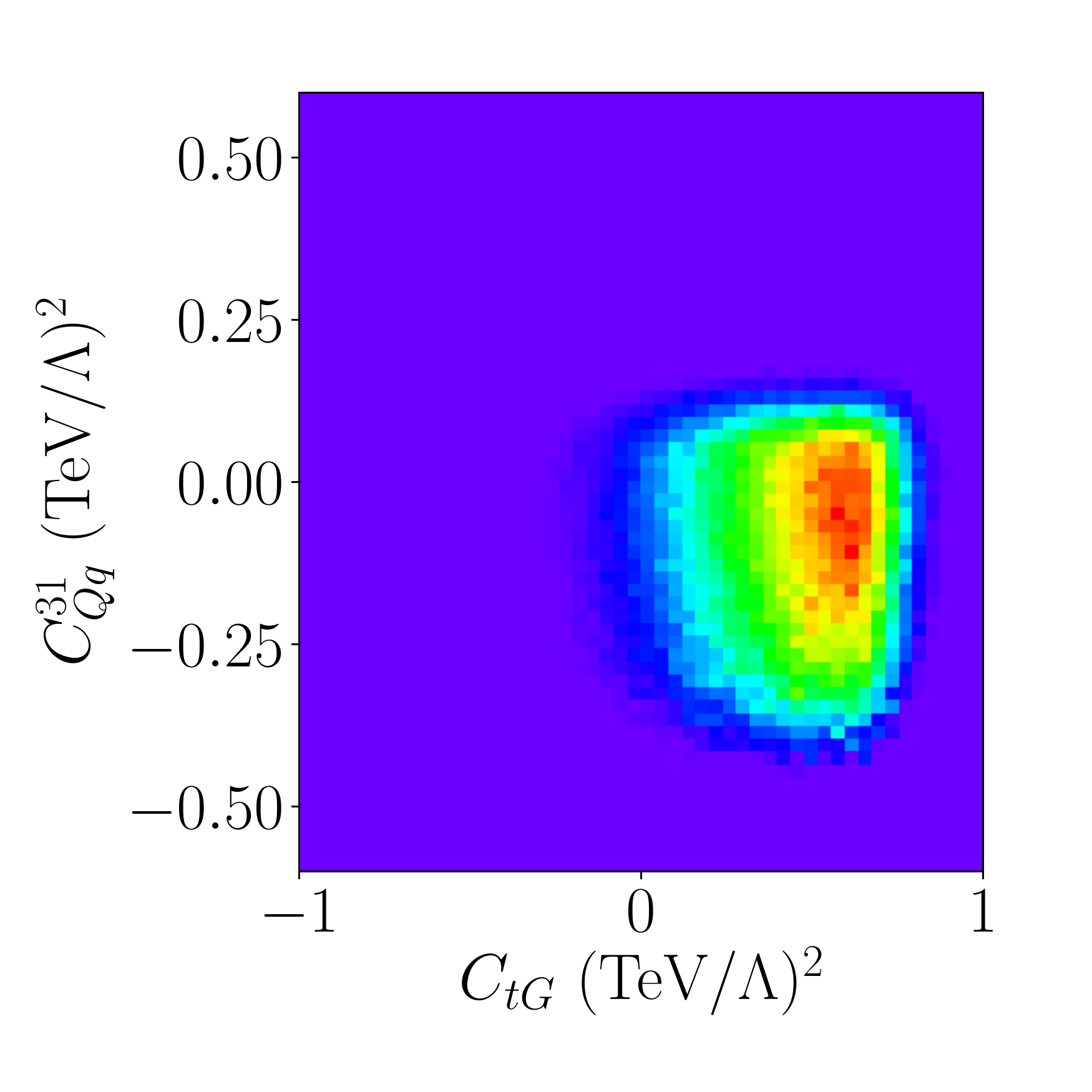}
  \includegraphics[width=0.325\linewidth]{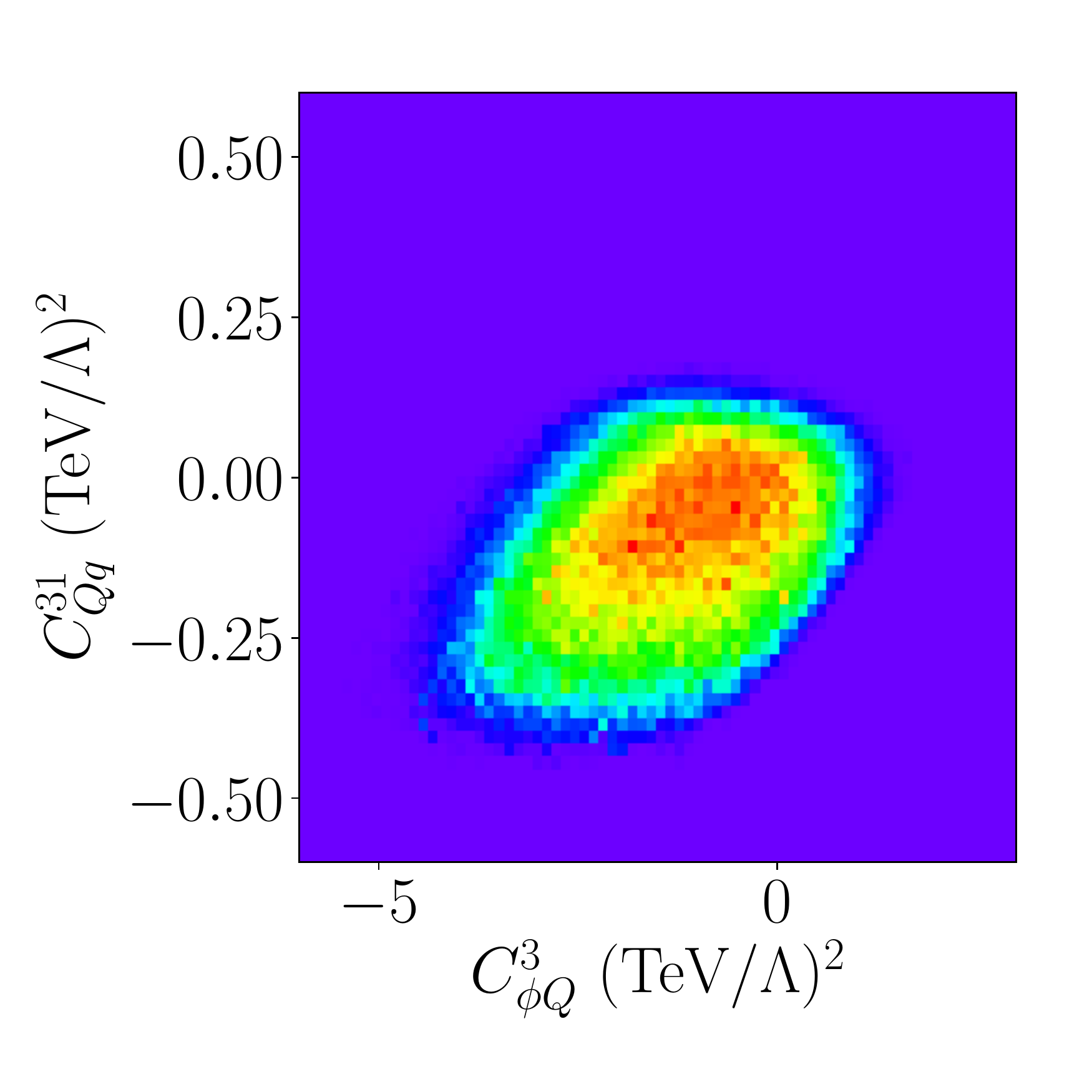}
   \caption{Examples of correlated 2-dimensional profile likelihoods
     from the global fit, showing the same operators as in the lower
     panels of Fig.~\ref{fig:SingleTCorrelations}.}
\label{fig:Correlations}
\end{figure}

Non-trivial correlations as between $C_{tG}$, $C_{\phi Q}^3$, and $C_{Qq}^{3,1}$ vanish once we include the
full data set, see Fig.~\ref{fig:Correlations}.
 The reason is that $C_{tG}$ and $C_{Qq}^{3,1}$ are
strongly constrained individually by top pair production. For the weak-triplet operators $O_{Qq}^{3,1}$ and $O_{Qq}^{3,8}$ the bounds are the same in the single top fit and the global fit, see also Fig~\ref{fig:3fits}. Single top production is indeed more sensitive to these four-quark operators than top pair production.

\begin{figure}[t]
  \centering
  \includegraphics[width=1.0\linewidth]{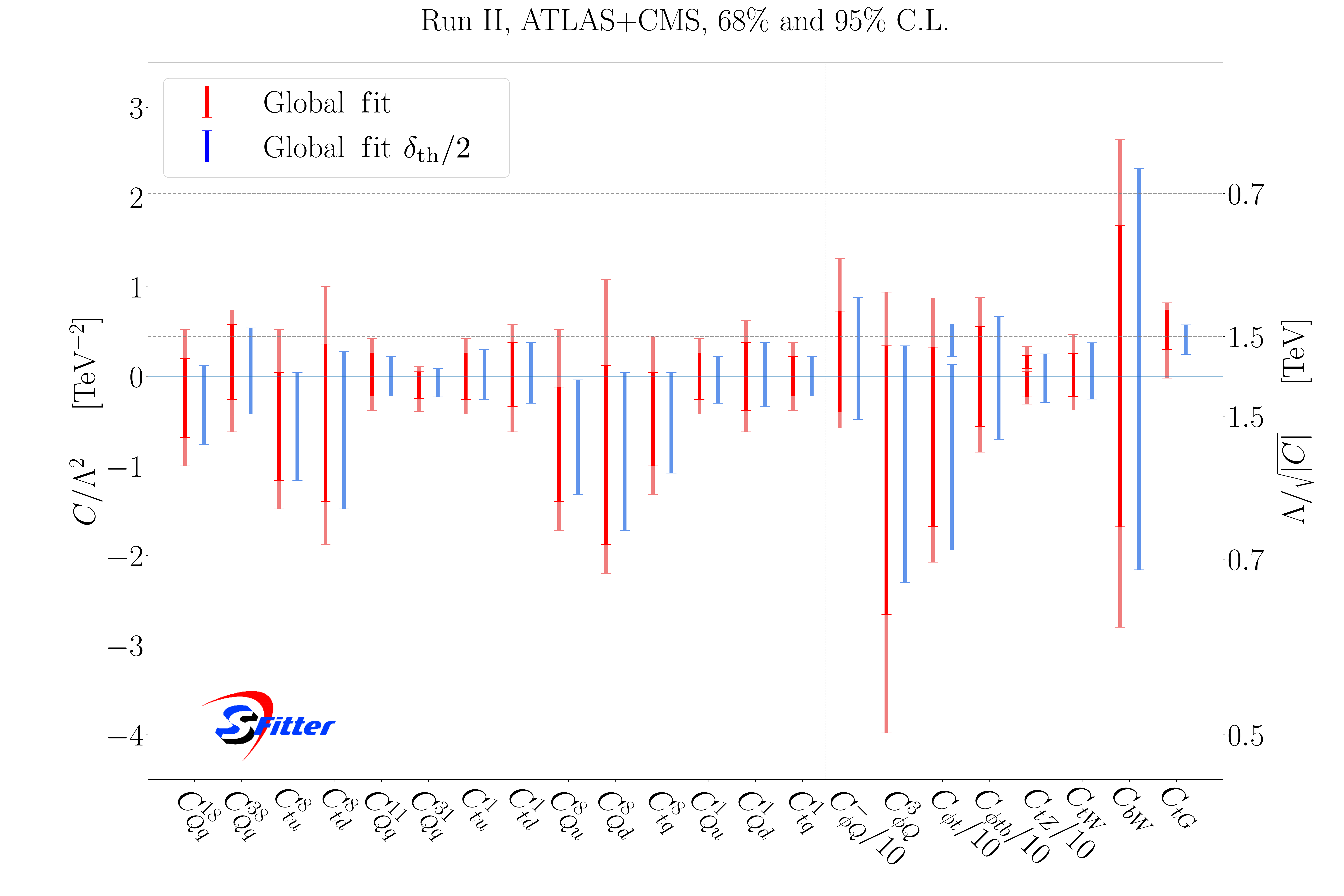}
  \caption{95\% and 68\% CL bounds on top operators from a global fit to the full
    data set from Tabs.~\ref{tab:data1} and \ref{tab:data2}. We show the results including all uncertainties (red) and
    with theoretical uncertainties reduced by a factor of two, $\delta_{\rm th}/2$ (blue).}
  \label{fig:Global}
\end{figure}

In Fig.~\ref{fig:Global} we show the profile likelihoods for each of
the top-related effective operators. On the $x$-axis we start with the
diagonal $LL$ and $RR$ four-quark operators listed in
Eq.~\eqref{eq:ops_llrr}, continue with the $LR$ and $RL$ four-quark
operators from Eq.~\eqref{eq:ops_lr}, and finally include the bosonic
operators from Eq.~\eqref{eq:tboson}. For each operator the red bars
indicate the final result at 68\% and 95\% CL. These confidence levels
are compact intervals defined by the area under the profile likelihood
curve, where in addition we require the likelihood values on each side
to be equal. For a Gaussian distribution we expect the 95\% error bar
to be symmetric around the best-fit value and twice as wide as the symmetric 68\% error bar. For some of the Wilson coefficients, non-Gaussian effects occur, which are mainly due to theoretical uncertainties treated as flat likelihoods.

In general, the four-quark operators are extremely well constrained with
limits in the range of $\Lambda/\sqrt{C} \approx 1-2$~TeV. For all weak-singlet four-quark operators, the sensitivity is dominated by high-energy bins in $t\bar{t}$ distributions. For $LL$ operators, we made this observation earlier in the left panel of Fig.~\ref{fig:quadratics}. As
discussed in Sec.~\ref{sec:quadratic}, for most operators the
well-defined limits on each of the four-quark operators rest entirely on
the quadratic contributions to the observables. For color-singlet operators, which contribute to top pair production only at order $\Lambda^{-4}$, the bounds are fully determined by quadratic contributions and symmetric around zero. The only asymmetric limit on a color-singlet operator appears for $O_{Qq}^{3,1}$ through a linear contribution to single top production.

Color-octet operators have asymmetric error bars due to their interference with QCD in top pair production. This interference is also the reason for the correlation patterns of shifted circles in Fig.~\ref{fig:quadratics}. The bounds on color-octet operators thus rely on the interplay between contributions of order $\Lambda^{-2}$ and $\Lambda^{-4}$, where the inclusion of both terms is particularly important.  In Fig.~\ref{fig:Global}, the error bars for color-singlet operators appear much smaller than for color-octet operators. This is due to the fact that top--anti-top observables always probe the combination $(C^8)^2 + \frac{9}{2} (C^1)^2$ at order $\Lambda^{-4}$, see Eq.~\eqref{eq:qq_tt_analytic}. Top--anti-top observables therefore constrain the quantities $C^8$ and $(C^8)^2 + \frac{9}{2} (C^1)^2$ at LO, which allows us to distinguish color-singlet from color-octet structures in kinematic distributions. The color combination is also changed at NLO in QCD, which offers the possibility to determine the color structure of operators from jet radiation.

Looking at the quark chirality, we observe that the bounds on operators with left- and right-handed tops are similar in strength. Charge-symmetric $t\bar{t}$ observables do not distinguish between these operators at high energies, see Eqs.~\eqref{eq:vv-aa} and \eqref{eq:qq_tt_analytic}. The charge asymmetry is sensitive to the top chirality, see Eq.~\eqref{eq:as-quad}, but still leads to equal bounds on the magnitude of $LL$ and $RL$ operators due to its small SM contribution.

Regarding different light quark flavors, operators with up quarks are better constrained than operators with down quarks. This reflects the parton content of the proton, which leads to an enhanced sensitivity of $t\bar{t}$ observables to up-quark operators over down-quark operators, see Eq.~\eqref{eq:weak-ud}.

Let us now turn our attention to the bosonic operators. The strongest bounds are obtained for the dipole operators $O_{tG}$ and $O_{tW}$. For $O_{tW}$ the bound does not change compared to the single top fit (see also Fig~\ref{fig:3fits}), because it is dominated by the precise measurements of $W$ helicities in top decays. From our global fit, we obtain at 95\% CL
\begin{align}
\frac{\Lambda}{\sqrt{C_{tW}}} \in [-0.38,0.47] \tev\,.
\end{align}
For $O_{tG}$ the best global limit at 95\% CL is obtained from top--anti-top production,
\begin{align}
\frac{\Lambda}{\sqrt{C_{tG}}} \in
[-0.02,0.82]  \tev\,.
\end{align}
This bound is consistent with the estimate of Eq.~\eqref{eq:limit_estimate} and much stronger than the bound from associated $tW$ production in Fig.~\ref{fig:GlobalSingle}. Most of the remaining bosonic operators are better constrained in the global fit than in the single top fit (see Fig~\ref{fig:3fits}). This shows the impact of the
$t\bar{t}Z$ cross section measurements in the global fit. For the operator $O_{\phi tb}$, which does not contribute to $t\bar{t}Z$ or $t\bar{t}W$ production, the sensitivity remains very low.

In our fit, theory uncertainties affect the relation between the (rate) measurements and the Wilson coefficients. Since we treat these uncertainties as flat errors in our statistical, they lead to plateaus in the center of the likelihood distributions and to some of the non-Gaussian effects. To study the relative impact of theoretical and experimental uncertainties on the fit results, we have performed a global fit with theory uncertainties divided by a factor of two. The 95\% CL results are shown as blue bars in Fig.~\ref{fig:Global}. We find that throughout the global analysis these theory uncertainties are limiting factors in interpreting Run-II results already now. Significant improvements are crucial to fully exploit the potential of Run-III measurements.

\section{Conclusions}
\label{sec:conclusions}
We have presented a comprehensive analysis of the LHC Run~II data in the
top sector. We use NLO simulations in \textsc{Madgraph5\_aMC@NLO} and the
\textsc{SFitter} framework to constrain the Wilson coefficients of 22
dimension-6 operators.  The bulk of the measurements involve final states with a top
pair, including kinematic distributions, the charge
asymmetry, and associated top pair production with a weak boson. In
addition, we include different single top channels and $W$ helicity
measurements in top decays. The measurements we use are based on up to $139~\ifb$
of integrated luminosity.

The main challenge of this global analysis is the large number of
four-quark operators in top pair production, whose contribution to the QCD process are largely degenerate. We have discussed several ways
of breaking this degeneracy, including parton luminosity effects, the
charge asymmetry, jet radiation patterns, and associated production
with weak bosons. 
We have also discussed the impact of dimension-6 squared terms on the fit results, and their role in lifting this degeneracy.

Altogether, we derive limits in the range of $\Lambda/\sqrt{C} =
0.35 - 2 \tev$ for the different Wilson coefficients from a profile
likelihood. The strongest limit is on the anomalous top coupling to
the gluon, driven by the QCD production rate. Similarly strong limits
apply to several four-quark operators, stemming mostly from normalized kinematic distributions. The top dipole interaction with the $W$ boson is also strongly constrained by the precisely known $W$ helicity fractions in top decays. Other operators with weak bosons are much less constrained,
because they only occur in electroweak top processes with a limited sensitivity in total rates. Differential distributions in electroweak top production, as well as precision observables in electroweak and flavor physics can help to increase the sensitivity. For all operators we find that the theory uncertainties linking 
measurements to Wilson coefficients in the Lagrangian are
becoming the limiting factor in interpreting LHC results.

\begin{center} \textbf{Acknowledgments} \end{center}

\noindent We thank Dirk Zerwas for many helpful discussions and
experimentalist's advice. The research of SB, TP and SW is supported
by the Deutsche Forschungsgemeinschaft (DFG, German Research
Foundation) under grant no. 396021762 - TRR 257. SW acknowledges
funding by the Carl Zeiss foundation through an endowed junior
professorship (\emph{Junior-Stiftungsprofessur}). The authors
acknowledge support by the state of Baden-Württemberg through bwHPC
and the German Research Foundation (DFG) through grant no INST
39/963-1 FUGG (bwForCluster NEMO). EV is supported by a Marie Skłodowska-Curie Individual Fellowship of the European Commission’s
Horizon 2020 Programme under contract number 704187.
CZ is supported by IHEP
under Contract No.~Y7515540U1.
FM was partly supported by F.R.S.-FNRS under the “Excellence of Science – EOS” – be.h project n. 30820817

\appendix

\section{Operator relations}
\label{app:operators}
In this appendix we list the relations between the relevant operators in our analysis and the operators in the Warsaw basis, following the notation of Refs.~\cite{Grzadkowski:2010es,Hartland:2019bjb}. Using the $SU(2)$ and $SU(3)$ identities
\begin{align*}
\tau_{ij}^I\tau_{kl}^I = -\delta_{ij}\delta_{kl} + 2\delta_{il}\delta_{jk},\qquad T_{ab}^A T_{cd}^A = -\frac{1}{6}\delta_{ab}\delta_{cd} + \frac{1}{2}\delta_{ad}\delta_{bc},
\end{align*}
and the Fierz identities for anti-commutating fermion fields,
\begin{align*}
 (\bar{q}\gamma^\mu q)(\bar{Q}\gamma_\mu Q) =  (\bar{q}\gamma^\mu Q)(\bar{Q}\gamma_\mu q),\qquad (\bar{u}\gamma^\mu u)(\bar{t}\gamma_\mu t) =  (\bar{u}\gamma^\mu t)(\bar{t}\gamma_\mu u),
\end{align*}
we derive the following relations:
\begin{itemize}
\item four-quark operators with $LL$ and $RR$ chiral structure ($i=1,2$),
\begin{align*}
O_{Qq}^{1,8} & \equiv (\bar{Q}\gamma_\mu T^A Q)(\bar{q}_i\gamma^\mu T^A q_i) & = & -\frac{1}{6}\mathcal{O}_{qq}^{1(33ii)} + \frac{1}{4}\mathcal{O}_{qq}^{1(3ii3)} + \frac{1}{4}\mathcal{O}_{qq}^{3(3ii3)}\\\nonumber
O_{Qq}^{3,8} & \equiv (\bar{Q}\gamma_\mu T^A\tau^I Q)(\bar{q}_i\gamma^\mu T^A \tau^I q_i) & = & -\frac{1}{6}\mathcal{O}_{qq}^{3(33ii)} + \frac{3}{4}\mathcal{O}_{qq}^{1(3ii3)} - \frac{1}{4}\mathcal{O}_{qq}^{3(3ii3)}\\\nonumber
O_{Qq}^{1,1} & \equiv (\bar{Q}\gamma_\mu Q)(\bar{q}_i\gamma^\mu q_i) & = & \ \mathcal{O}_{qq}^{1(33ii)}\\\nonumber
O_{Qq}^{3,1} & \equiv (\bar{Q}\gamma_\mu\tau^I Q)(\bar{q}_i\gamma^\mu\tau^I q_i) & = & \ \mathcal{O}_{qq}^{3(33ii)}\\\nonumber
O_{tu}^8 & \equiv (\bar{t}\gamma_\mu T^A t)(\bar{u}_i\gamma^\mu T^A u_i) & = & -\frac{1}{6}\mathcal{O}_{uu}^{(33ii)} + \frac{1}{2}\mathcal{O}_{uu}^{(3ii3)}\\\nonumber
O_{tu}^1 & \equiv (\bar{t}\gamma_\mu t)(\bar{u}_i\gamma^\mu u_i) & = & \ \mathcal{O}_{uu}^{(33ii)}\\\nonumber
O_{td}^8 & \equiv (\bar{t}\gamma^\mu T^A t)(\bar{d}_i\gamma_\mu T^A d_i) & = & \ \mathcal{O}_{ud}^{8(33ii)}\\\nonumber
O_{td}^1 & \equiv (\bar{t}\gamma^\mu t)(\bar{d}_i\gamma_\mu d_i) & = & \ \mathcal{O}_{ud}^{1(33ii)}\,,
\end{align*}
\item four-quark operators with $LR$ and $RL$ chiral structure
\begin{align*}
O_{Qu}^8 & \equiv (\bar{Q}\gamma^\mu T^A Q)(\bar{u}_i\gamma_\mu T^A u_i) & = & \ \mathcal{O}_{qu}^{8(33ii)} &
\qquad O_{Qu}^1 & \equiv (\bar{Q}\gamma^\mu Q)(\bar{u}_i\gamma_\mu u_i) & = & \ \mathcal{O}_{qu}^{1(33ii)}\\\nonumber
O_{Qd}^8 & \equiv (\bar{Q}\gamma^\mu T^A Q)(\bar{d}_i\gamma_\mu T^A d_i) & = & \ \mathcal{O}_{qd}^{8(33ii)} & O_{Qd}^1 & \equiv (\bar{Q}\gamma^\mu Q)(\bar{d}_i\gamma_\mu d_i) & = & \ \mathcal{O}_{qd}^{1(33ii)}\\\nonumber
O_{tq}^8 & \equiv (\bar{q}_i\gamma^\mu T^A q_i)(\bar{t}\gamma_\mu T^A t) & = & \ \mathcal{O}_{qu}^{8(ii33)} & O_{tq}^1 & \equiv (\bar{q}_i\gamma^\mu q_i)(\bar{t}\gamma_\mu t) & = & \ \mathcal{O}_{qu}^{1(ii33)}\,,
\end{align*}
\item operators with two heavy quarks and bosonic fields
\begin{align*}
O_{\phi Q}^1 & \equiv (\phi^\dagger \,i\stackrel{\longleftrightarrow}{D_\mu} \phi)(\bar{Q}\gamma^{\mu}Q) & = & \ \mathcal{O}_{\phi q}^{1(33)} & \qquad ^\ddagger O_{tB} & \equiv (\bar{Q}\sigma^{\mu\nu} t)\,\widetilde{\phi}\,B_{\mu\nu} & = & \ ^\ddagger \mathcal{O}_{uB}^{(33)} \\\nonumber
O_{\phi Q}^3 & \equiv (\phi^\dagger \,i\stackrel{\longleftrightarrow}{D_\mu^I} \phi)(\bar{Q}\gamma^{\mu}\tau^I Q) & = & \ \mathcal{O}_{\phi q}^{3(33)} & \qquad ^\ddagger O_{tW} & \equiv (\bar{Q}\sigma^{\mu\nu} t)\,\tau^I\widetilde{\phi}\,W_{\mu\nu}^I & = & \ ^\ddagger \mathcal{O}_{uW}^{(33)} \\\nonumber
O_{\phi t} & \equiv (\phi^\dagger \,i\stackrel{\longleftrightarrow}{D_\mu} \phi)(\bar{t}\gamma^{\mu}t) & = & \ \mathcal{O}_{\phi u}^{(33)} & \qquad ^\ddagger O_{bW} & \equiv (\bar{Q}\sigma^{\mu\nu} b)\,\tau^I\phi \,W_{\mu\nu}^I & = & \ ^\ddagger \mathcal{O}_{dW}^{(33)} \\\nonumber
^\ddagger O_{\phi t b} & \equiv (\widetilde{\phi}^\dagger iD_\mu \phi)(\bar{t}\gamma^{\mu}b) & = & \ ^\ddagger \mathcal{O}_{\phi ud}^{(33)} & \qquad ^\ddagger O_{tG} & \equiv (\bar{Q}\sigma^{\mu\nu} T^A t)\,\widetilde{\phi}\,G_{\mu\nu}^A & = & \ ^\ddagger \mathcal{O}_{uG}^{(33)} \,,
\end{align*}
\end{itemize}
with the Higgs field $\phi = (0,\frac{1}{\sqrt{2}}(v+h))^\top$ in unitary gauge, $\widetilde{\phi} = i\sigma_2\,\phi^\ast$ and the covariant derivative
\begin{align}
D_\mu  = \partial_\mu - i \frac{e}{2s_W} A_{\mu}^I\tau^I - i \frac{e}{c_W}B_{\mu}Y,\qquad D_{\mu}^I  = \tau^I D_\mu,\qquad \tau^I = \sigma_I.
\end{align}
The relations between the corresponding Wilson coefficients $C_i$ and $\mathcal{C}_i$ can be obtained by requiring that both bases lead to the same terms in the effective Lagrangian~\cite{AguilarSaavedra:2018nen},
\begin{align}
\lag_\text{eff} = \sum_a \left(\frac{C_a}{\Lambda^2}\,^\ddagger O_a + \text{h.c.} \right) 
               + \sum_b \frac{C_b}{\Lambda^2}\,O_b =  \sum_c \left(\frac{\mathcal{C}_c}{\Lambda^2}\,^\ddagger \mathcal{O}_c + \text{h.c.} \right) 
               + \sum_d \frac{\mathcal{C}_d}{\Lambda^2}\,\mathcal{O}_d.
\end{align}
After electroweak symmetry breaking, the effective interactions of the physical weak gauge bosons are described by linear combinations of the operators in the unbroken phase. In unitary gauge, the relations read
\begin{align}
\begin{pmatrix}
O_{\phi Q}^{1}\\
O_{\phi Q}^{3}
\end{pmatrix} & = \begin{pmatrix}
1 & 1 & 0 & 0\\
-1 & 1 & 1 & 1\\
\end{pmatrix} \begin{pmatrix}
- \frac{e}{2s_Wc_W}\left(\overline{t}\gamma^\mu t_L\right)Z_\mu(v+h)^2\\ - \frac{e}{2s_Wc_W}\left(\overline{b}\gamma^\mu b_L\right)Z_\mu (v+h)^2\\ \frac{e}{\sqrt{2}s_W}\left(\overline{t}\gamma^\mu b_L\right)W_\mu^+(v+h)^2\\ \frac{e}{\sqrt{2}s_W}\left(\overline{b}\gamma^\mu t_L\right)W_\mu^-(v+h)^2\\
\end{pmatrix}, \\\nonumber
\begin{pmatrix}
^\ddagger O_{tB}\\
^\ddagger O_{tW}
\end{pmatrix} & = \begin{pmatrix}
c_W & - s_W & 0\\
s_W & c_W & 1\\
\end{pmatrix} \begin{pmatrix}
  \frac{1}{\sqrt{2}}\left(\overline{t}\sigma^{\mu\nu} t_R\right)A_{\mu\nu}(v+h)\\  \frac{1}{\sqrt{2}}\left(\overline{t}\sigma^{\mu\nu} t_R\right)Z_{\mu\nu}(v+h)\\   \left(\overline{b}\sigma^{\mu\nu} t_R\right)W_{\mu\nu}^-(v+h)
\end{pmatrix}\,,\\\nonumber
^\ddagger O_{bW} & = \left[ - \tfrac{1}{\sqrt{2}}\,\overline{b}\,\sigma^{\mu\nu} b_R \big(c_w Z_{\mu\nu} + s_w A_{\mu\nu}\big) + \bar{t}\,\sigma^{\mu\nu} b_R\,W_{\mu\nu}^+\right](v+h)\,.
\end{align}

\section{Numerical bounds on operators}
Here we list the limits on the 22 Wilson coefficients, obtained from fits to different data sets. Tab.~\ref{tab:dataGlobal} shows the results of our global fit, Tab.~\ref{tab:dataSingleT} corresponds to our single top fit, and Tab.~\ref{tab:dataTTBar} shows a fit of observables in top pair production only.

We also show a comparison of the bounds obtained from fits to top-pair production, single top production, and from the full global fit in Fig.~\ref{fig:3fits}.\\

\begin{table}[h!]\centering
\renewcommand{\arraystretch}{1.2}
\setlength{\tabcolsep}{4mm}
\begin{small} \begin{tabular}{c|*2{>{$}c<{$}}}
\toprule
Operator & 68\%\, \text{CL} &  95\%\, \text{CL}\\\midrule[1pt]
$C_{tG}$ & -  & [ -5.68 , 4.00 ] \\
$C_{Qq}^{38}$ & - & [ -0.78 , 0.74 ] \\
$C_{Qq}^{31}$ & - & [ -0.49 , 0.11 ] \\
$C_{bW}$ & [ -1.84 , 1.68 ]  & [ -2.80 , 2.80 ] \\
$C_{tW}$ & [ -0.32 , 0.23 ]  & [ -0.47 , 0.47 ] \\
$C_{tZ}$ & -  & [ -9.40 , 9.40 ] \\
$C_{\phi t}$ & - & [ -51.50 , 22.50 ] \\
$C_{\phi t b }$ & [ -5.94 , 5.94 ]  & [ -9.18 , 8.82 ] \\
$C_{\phi  Q}^{3}$ & - & [ -4.70 , 1.30 ] \\
$C_{\phi  Q}^{-}$ & - & [ -36.00 , 12.00 ] \\
\bottomrule[1pt]
\end{tabular} \end{small}
\caption{The 68\% and 95\% confidence levels for single top fit,
  corresponding to Fig.~\ref{fig:GlobalSingle}.}
\label{tab:dataSingleT}
\end{table}

\begin{table}[h!]\centering
\setlength{\tabcolsep}{4mm}
\renewcommand{\arraystretch}{1.2}
\begin{small} \begin{tabular}{c|*2{>{$}c<{$}}}
\toprule
Operator & 68\%\, \text{CL} &  95\%\, \text{CL}\\\midrule[1pt]
$C_{tG}$ & [ 0.30 , 0.74 ]  & [ -0.03 , 0.82 ] \\
$C_{Qq}^{18}$ & [ -0.79 , 0.15 ]  & [ -1.11 , 0.49 ] \\
$C_{Qq}^{38}$ & [ -0.49 , 0.73 ]  & [ -0.84 , 1.16 ] \\
$C_{tq}^8$ & [ -1.21 , -0.09 ]  & [ -1.37 , 0.47 ] \\
$C_{Qu}^8$ & [ -1.51 , -0.09 ]  & [ -1.91 , 0.44 ] \\
$C_{Qd}^8$ & [ -2.09 , 0.15 ]  & [ -2.44 , 1.24 ] \\
$C_{tu}^8$ & [ -1.16 , 0.15 ]  & [ -1.48 , 0.65 ] \\
$C_{td}^8$ & [ -1.40 , 0.52 ]  & [ -1.93 , 1.16 ] \\
$C_{Qq}^{11}$ & [ -0.38 , 0.09 ]  & [ -0.47 , 0.30 ] \\
$C_{Qq}^{31}$ & [ -0.18 , 0.29 ]  & [ -0.34 , 0.42 ] \\
$C_{tq}^{1}$ & [ -0.27 , 0.21 ]  & [ -0.39 , 0.37 ] \\
$C_{Qu}^{1}$ & [ -0.47 , 0.09 ]  & [ -0.62 , 0.27 ] \\
$C_{Qd}^{1}$ & [ -0.41 , 0.37 ]  & [ -0.66 , 0.58 ] \\
$C_{tu}^{1}$ & [ -0.35 , 0.15 ]  & [ -0.47 , 0.34 ] \\
$C_{td}^{1}$ & [ -0.41 , 0.35 ]  & [ -0.58 , 0.63 ] \\
\bottomrule[1pt]
\end{tabular} \end{small}
\caption{The 68\% and 95\% confidence levels for the top pair production fit.}
\label{tab:dataTTBar}
\end{table}

\begin{table}[h!]\centering
\renewcommand{\arraystretch}{1.2}
\setlength{\tabcolsep}{4mm}
\begin{small} \begin{tabular}{c|*2{>{$}c<{$}}|>{$}c<{$}}
\toprule
Operator & 68\%\, \text{CL} &  95\%\, \text{CL} &  95\%\, \text{CL}, \delta_{\rm th}/2\\\midrule[1pt]
$C_{tG}$ & [ 0.30 , 0.74 ]        & [ -0.02 , 0.82 ]                      & [ 0.24 , 0.57 ] \\
$C_{Qq}^{18}$ & [ -0.68 , 0.20 ]  & [ -1.00 , 0.52 ]                      & [ -0.76 , 0.12 ] \\
$C_{Qq}^{38}$ & [ -0.26 , 0.58 ]  & [ -0.62 , 0.74 ]                      & [ -0.42 , 0.54 ] \\
$C_{tq}^8$ & [ -1.00 , 0.04 ]     & [ -1.32 , 0.44 ]                      & [ -1.08 , 0.04 ] \\
$C_{Qu}^8$ & [ -1.40 , -0.12 ]    & [ -1.72 , 0.52 ]                      & [ -1.32 , -0.04 ]\\
$C_{Qd}^8$ & [ -1.88 , 0.12 ]     & [ -2.20 , 1.08 ]                      & [ -1.72 , 0.04 ] \\
$C_{tu}^8$ & [ -1.16 , 0.04 ]     & [ -1.48 , 0.52 ]                      & [ -1.16 , 0.04 ] \\
$C_{td}^8$ & [ -1.40 , 0.36 ]     & [ -1.88 , 1.00 ]                      & [ -1.48 , 0.28 ] \\
$C_{Qq}^{11}$ & [ -0.22 , 0.26 ]  & [ -0.38 , 0.42 ]                      & [ -0.22 , 0.22 ] \\
$C_{Qq}^{31}$ & [ -0.25 , 0.05 ]  & [ -0.39 , 0.11 ]                      & [ -0.23 , 0.09 ] \\
$C_{tq}^{1}$ & [ -0.22 , 0.22 ]   & [ -0.38 , 0.38 ]                      & [ -0.22 , 0.22 ] \\
$C_{Qu}^{1}$ & [ -0.26 , 0.26 ]   & [ -0.42 , 0.42 ]                      & [ -0.30 , 0.22 ] \\
$C_{Qd}^{1}$ & [ -0.38 , 0.38 ]   & [ -0.62 , 0.62 ]                      & [ -0.34 , 0.38 ] \\
$C_{tu}^{1}$ & [ -0.26 , 0.26 ]   & [ -0.42 , 0.42 ]                      & [ -0.26 , 0.30 ] \\
$C_{td}^{1}$ & [ -0.34 , 0.38 ]   & [ -0.62 , 0.58 ]                      & [ -0.30 , 0.38 ] \\
$C_{bW}$ & [ -1.68 , 1.68 ]       & [  -2.80 , 2.64 ]                     & [ -2.16 , 2.32 ] \\
$C_{tW}$ & [ -0.23 , 0.26 ]       & [ -0.38 , 0.47 ]                      & [ -0.26 , 0.38 ] \\
$C_{tZ}$ & [ -2.30 , 2.30 ]  & [ -3.10 , 3.30 ]           & [ -2.90 , 2.50 ]^* \\
$C_{\phi t}$ & [ -16.75 , 3.25 ]     & [ -20.75 , 8.75 ]                  &[ -19.38 , 5.83 ]^* \\
$C_{\phi t b }$ & [ -5.58 , 5.58 ]   & [ -8.46 , 8.82 ]                   & [ -7.02 , 6.66 ]\\
$C_{\phi  Q}^{3}$ & [ -2.66 , 0.34 ] & [ -3.98 , 0.94 ]                   & [ -2.30 , 0.34 ]\\
$C_{\phi  Q}^{-}$ & [ -3.98 , 7.28 ] & [ -5.78 , 13.12 ]                  & [ -4.80 , 8.80 ]\\
\bottomrule[1pt]
\end{tabular} \end{small}
\caption{The 68\% and 95\% confidence levels for the full global fit,
  corresponding to Fig.~\ref{fig:Global}. The asterisk marks
  non-Gaussian effects for which we quote conservative envelopes of
  the likelihood. The label $\delta_{\rm th}/2$ stands for the fit with halved theoretical uncertainties.}
\label{tab:dataGlobal}
\end{table}

\begin{figure}[h!]\centering
\includegraphics[width=\linewidth]{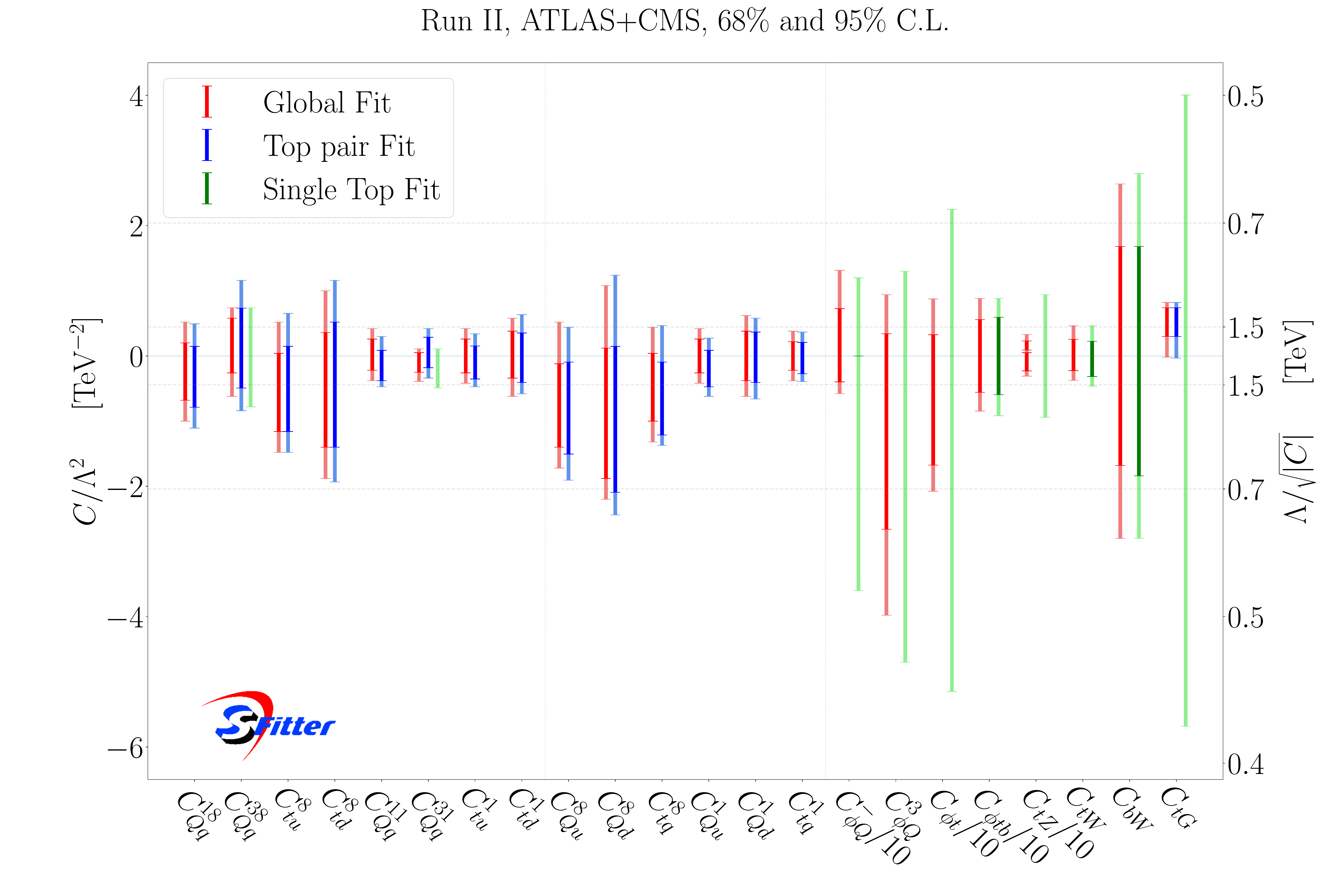}
\caption{
95\% and 68\% CL bounds on top operators global fits to top pair production measurements (blue), single top (green) and to the full
    data set from Tabs.~\ref{tab:data1} and\ref{tab:data2} (red).
}\label{fig:3fits}
\end{figure}

\newpage


\providecommand{\href}[2]{#2}\begingroup\raggedright\endgroup

\end{document}